\documentclass[a4paper,fleqn,usenatbib]{mnras}

\usepackage{newtxtext,newtxmath}

\usepackage[T1]{fontenc}
\usepackage{ae,aecompl}
\usepackage{pdflscape}

\usepackage{graphicx}	
\usepackage{amsmath}	
\usepackage{amssymb}	
\usepackage{multicol, wrapfig}

\title[Extreme variables from UKIDSS - II]{Extreme infrared variables from UKIDSS - II. an end-of-survey 
catalogue of eruptive YSOs and unusual stars}

\author[P. W. Lucas et al.]{
P. W. Lucas,$^{1}$\thanks{E-mail: p.w.lucas@herts.ac.uk (PWL)}
L. C. Smith$^{1}$
C. Contreras Pe\~{n}a$^{2}$, Dirk Froebrich$^{3}$, Janet E. Drew$^{1}$,
\newauthor M.S.N. Kumar$^{1}$, J. Borissova$^{4,5}$, D. Minniti$^{6,5,7}$, R. Kurtev$^{4,5}$, M. Mongui\'{o}$^{1}$\\
$^{1}$Centre for Astrophysics, University of Hertfordshire, College Lane, Hatfield, AL10 9AB, UK\\
$^{2}$School of Physics, University of Exeter, Stocker Road, Exeter, EX4 4QL, UK\\
$^{3}$Centre for Astrophysics and Planetary Science, University of Kent, Canterbury CT2 7NH, UK\\
$^{4}$Instituto de F\'{i}sica y Astronom\'{i}a, Universidad de Valpara\'{i}so, ave. Gran Breta\~{n}a, 1111, Casilla 5030, Valpara\'{i}so, Chile\\
$^{5}$Millennium Institute of Astrophysics, Av. Vicuna Mackenna 4860, 782-0436, Macul, Santiago, Chile\\
$^{6}$Departamento de Ciencias Fisicas, Universidad Andres Bello, Republica 220, Santiago, Chile\\
$^{7}$Vatican Observatory, V00120 Vatican City State, Italy
}

\date{Accepted XXX. Received YYY; in original form ZZZ}

\pubyear{2017}

\begin{document}
\label{firstpage}
\pagerange{\pageref{firstpage}--\pageref{lastpage}}
\maketitle

\begin{abstract}
We present a catalogue of 618 high amplitude infrared variable stars ($1 < \Delta K < 5$~mag) detected by 
the two widely separated epochs of 2.2~$\mu$m data in the UKIDSS Galactic plane survey, from searches
covering $\sim$1470~deg$^2$. Most were discovered by a search of all fields at $30<l<230^{\circ}$. 
Sources include new dusty Mira variables, three new CV candidates, a blazar and a peculiar source that may 
be an interacting binary system. However, $\sim$60\% are YSOs,
based on spatial association with star forming regions at distances ranging from 300~pc to over 10~kpc. 
This confirms our initial result in Contreras Pe\~{n}a et al. (Paper I) that YSOs dominate the high amplitude infrared 
variable sky in the Galactic disc. It is also supported by recently published VVV results at $295<l<350^{\circ}$. The spectral 
energy distributions of the YSOs indicate class I or flat spectrum systems in most cases, as in the VVV sample.  
A large number of variable YSOs are associated with the Cygnus X complex and other groups are associated 
with the North America/Pelican nebula, the Gemini OB1 molecular cloud, the Rosette complex, the Cone nebula, the 
W51 star forming region and the S86 and S236 HII regions. Most of the YSO variability is likely due to 
variable/episodic accretion on timescales of years, albeit usually less extreme than classical FUors and EXors. 
Luminosities at the 2010 WISE epoch range from $\sim$0.1~$L_{\sun}$ to 10$^3 L_{\sun}$ but only rarely exceed 
$10^{2.5} L_{\sun}$. 
\end{abstract}


\section{Introduction}

The advent of the wide field infrared telescopes in recent years has enabled the first 
panoramic explorations of the variable sky in the near infrared waveband. 
The UKIDSS Galactic Plane Survey (UGPS, \citealt{lucas08}; \citealt{lawrence07}) surveyed 
the northern and equatorial Galactic plane at two epochs, while in the southern
hemisphere, the VVV survey (VISTA Variables in the Via Lactea, \citealt{minniti10})
surveyed 560~deg$^2$ of the Galactic bulge and disc at several dozen epochs from 2010--2015.
An important subject for such surveys is the phenomenon of eruptive variability in YSOs
in which the variability is driven by episodic accretion. Given that YSOs suffer increasingly
high extinction at earlier stages of pre-main sequence (pre-MS) evolution, infrared observation is 
needed to place the study of variable accretion on a firm observational footing. Equally important
is the possibility of making significant unexpected discoveries in other areas of astronomy.

Episodic accretion is a serious unresolved matter in star formation (reviewed by \citealt{audard14}), owing to the effect that 
it may have on masses and ages inferred from Hertzsprung-Russell (HR) diagrams for pre-main sequence (pre-MS) clusters. 
This may help to resolve the luminosity problem (\citealt{kenyon90}; \citealt{enoch09}; 
\citealt{caratti12}) which is that typical luminosities of class I YSOs in nearby clusters 
are lower than expected for stars that should lie on Hayashi tracks above the main sequence. Variable
accretion may also explain the scatter of pre-MS stars about any given isochrone in HR diagrams, e.g. 
\citet{mayne08}, \citet{weights09}. Some theoretical work has suggested that
episodic accretion has a lingering effect on the location of pre-MS stars on the HR diagram 
(\citealt*{baraffe09}, \citealt*{baraffe12}) while another study
indicates that there is little effect on low mass stars \citep*{hosokawa11}.
Aside from the effect on stellar luminosity and radius, the effect of variable accretion on planet 
formation is worthy of investigation \citep{cieza16}. 

While low level accretion variations are thought to
be understood in terms of the role of the stellar magnetosphere in generating multiple, often unstable
accretion flows (e.g. \citealt*{romanova08}), theories of episodic accretion are quite diverse. E.g.
\citet*{zhu09} proposed an imbalance between mass transport rates determined by gravitional instability
in the outer disc and magneto-rotational instability in the inner disc. \citet{bonnell92} proposed that
outbursts are triggered by a binary companion passing through the perioastron of an eccentric orbit, 
a model that may apply to the short period eruptive variable LRLL 54361 \citep{muzerolle13}.
\citet{vorobyov15} proposed that outer disc fragmentation due to gravitational instability can cause 
fragments to spiral inward to the protostar via exchange of angular momentum with spiral arms and
other fragments in the disc, the resulting outbursts being most common in the class I stage of evolution.

The UGPS surveyed 1868~deg$^2$ of the northern and equatorial Galactic plane in the $J$, $H$ and $K$ 
passbands from 2005--2013. The survey used the Wide Field Camera (WFCAM, \citealt{casali07}) mounted on 
the 3.8-m United Kingdom Infrared Telescope. With the aims of detecting rarely seen high amplitude infrared 
variable stars and measuring proper motions, a second epoch of $K$ photometry was obtained for most 
of the UGPS area with a time baseline of 1.8 to 8 yrs. 
New high proper motion discoveries were presented by \citet{smith14b}.
Initial discoveries from the variable star search were presented in Paper I of this series 
\citep{cp14} which described 45 stars from the 5th and 7th UGPS data releases with
variations above 1~mag in $K$.
These near infrared surveys have been complemented by two-epoch mid-infrared searches of data
from the {\it Spitzer} Space Telescope \citep{werner04} and the 
Wide-field Infrared Survey Explorer \citep{wright10}, see \citet*{scholz13}; \citet{antoniucci14}.

In Paper I we found that most of the 45 high amplitude infrared variable stars detected in our
search of 155~deg$^2$ of the Galactic plane were YSOs. This suggested that highly variable YSOs 
dominate the near infrared variable sky on Galactic disc sightlines, a result that had not been predicted. 
The space density of these sources also appeared to be higher than that of highly variable asymptotic 
giant branch stars (AGB stars). However, most of the variable YSOs were located in just two nearby star 
forming complexes (the Serpens OB2 region and parts of Cygnus X) so it was possible that the results
would not be replicated in a pan-Galactic search. A search for high amplitude variables in 118~deg$^2$ of the VVV 
Galactic disc dataset confirmed the dominance of YSOs on mid-plane sightlines in the inner Galaxy (longitudes
$295<l<350^{\circ}$), see \citet{cp17a,cp17b}, hereafter CP17a, CP17b.
That study found that the variable YSOs had a variety of light curve types but systems dominated by short term variations
generally varied by $<$1.5~mag in $K$, whereas higher amplitudes were more often associated with outbursts with 
durations of a few years, most often seen amongst class I and flat-spectrum YSOs.
A similar rarity of high amplitude short term eruptions is seen in optical studies of T Tauri stars and 
Herbig Ae/Be stars, see \citealt{herbst94}; \citealt{herbst99}; \citealt{findeisen13}, \citealt{cody14}; \citealt{stauffer14}. 
\citet{cody17} found a few examples of short duration bursts (hours to days) with optical amplitudes slightly over 1~mag 
and one case of a 2~mag burst, providing the exception that proves the rule. High amplitude short term optical changes are 
more often reductions in flux below the usual level: the UXor phenomenon, usually but not always attributed to extinction 
(\citealt{calvet04}; \citealt{herbst99}; \citealt{findeisen13}; \citealt{cody14}). Such extinction-driven variability will of course 
have a much lower amplitude in the infrared than the optical, though the YSOVAR survey data indicate that
variable reddening can dominate the low-amplitude variability observed on short timescales even at
mid-infrared wavelengths, e.g. \citet{wolk15}.

The VVV study of CP17a greatly increased the available sample of candidate eruptive
variable YSOs. The mean amplitude of $\sim$1.7~mag in $K_s$ was lower than is traditionally
associated with eruptive variables in the FUor and EXor subclasses but the continuous distribution of
amplitudes seemed to argue against setting an arbitrary threshold. We adopted a 1~mag threshold in
Paper I and CP17a on the grounds that this is the level above which short term variations become rather
rare, such that variability is more often long term, i.e. timescales of years rather than days. 
A limitation of the VVV sample is that it is dominated by sources at distances, $d>2$~kpc, often suffering high 
extinction. This inhibits spatially resolved follow up and optical investigation and selects against less luminous
sources.

In this work we present the results of a search for high amplitude variable stars in all UGPS fields with 
two epochs of $K$ photometry at longitudes $l>30^{\circ}$. This allows us to confirm the prevalence
of high amplitude YSOs across a much broader range of Galactic sightlines and detect low luminosity 
variables in nearby, well-studied star forming regions. The non-YSOs in the catalogue are likely to be of interest
in a variety of sub-fields of astronomy. We also include the results of our previous 
UGPS searches in our catalogue for ease of reference. In section 2 we describe our search methods and present the 
catalogue, which includes information gleaned from public multiwaveband datasets and the literature.  A third epoch of near 
infrared photometry from 2MASS \citep{skrutskie06} is provided where available. In section 3 we provide an overview of the 
properties of catalogue members. We summarise the few previously known variable stars, describe the identification of likely YSOs 
and give a breakdown of non-YSOs into various categories. In section 4 we discuss the properties of likely YSOs in detail, including 
a description of the groups of highly variable YSOs found in several well-studied star forming regions. In section 5 we briefly 
discuss a few of the highest amplitude variables, the nearest pre-MS variables and a single object with unique properties. A 
summary of our main findings is presented in section 6.

\section{Variable Search}

The variable stars in our catalogue were found in several searches of the UGPS data as the survey 
progressed, all of which were minor variations on the same strategy. A minimum variation $\Delta K = 1$~mag
was adopted, as in Paper I. Below this level short term variability becomes much more common in YSOs, see
e.g. \citet*{carpenter01}, Paper I and section 4.1.3. A brightness threshold $K<16$ was required in at 
least one of the two epochs. All of the many candidates identified by catalogue-based selections were
inspected visually to identify genuine variable stars, using the FITS images publicly available at the WFCAM Science
Archive. We summarize the searches here and give full details in Appendix~\ref{A} in order to not encumber the main 
text with excessive detail.

First, the 45 variables discovered in Paper I were found by searching the 5th and 7th UGPS Data Releases (DR5,
DR7) available in the WFCAM Science Archive (WSA), these containing data for 156~deg$^2$ of two-epoch sky. An 
additional search of the 8th Data Release is described in the PhD thesis of \citet{cp15}. It used
the same method, adding 26 variables in a further 104~deg$^2$ of two-epoch sky. These searches used
SQL queries of the public data releases and they were inefficient: only 2\% of candidates passed visual
inspection because most of the available two-epoch sky was in fairly crowded parts of the 1st Galactic 
quadrant, where blends increase the number of false positives despite the quality cuts. Many false positive
candidates were also created by bad pixels. 

\begin{figure*}
	\includegraphics[width=\textwidth]{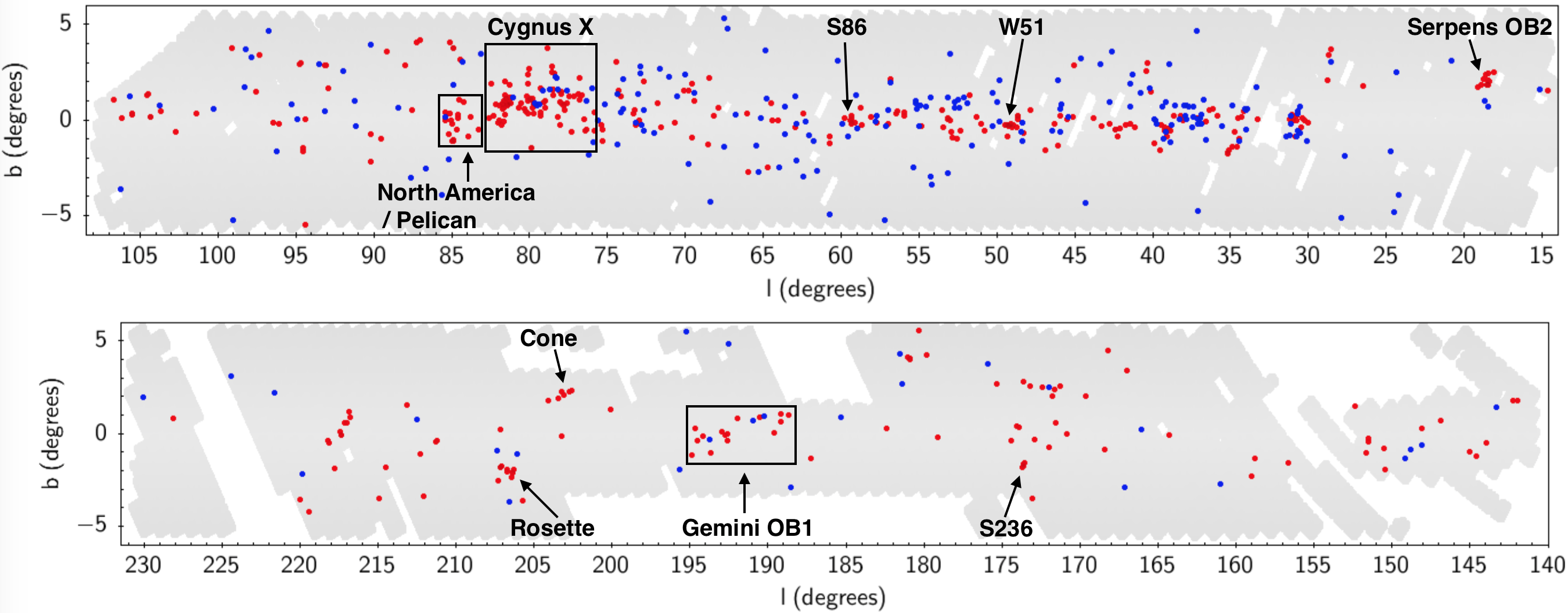}
	\vspace{-4mm}
    \caption{Locations of the 618 variable stars in Galactic coordinates. Likely YSOs (red dots) are identified
    by spatial association with SFRs, see section 3.3. Other sources are shown in blue and the UGPS two-epoch 
    coverage is shown as a grey background. Groups of likely YSOs in well-studied SFRs are indicated (see section 4.2) 
    though some are unresolved in this large scale view. At $l<30^{\circ}$ only part of the surveyed area has been searched, 
    see Paper I. The vertical scale is slightly exaggerated for clarity.}
    \label{fig:cov}  
\end{figure*}

\begin{figure}
	\includegraphics[width=\columnwidth]{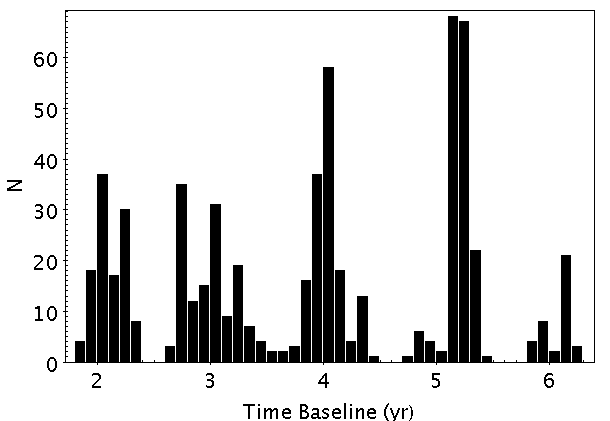}
    \caption{Distribution of time baselines for the 618 variable stars in the catalogue.}
    \label{fig:hist_baseline}  
\end{figure}

Later searches were undertaken using the publicly available FITS catalogues for each field. All of these are now available 
for the full survey area, though not all the 2nd epoch data are included in the most recent SQL data release (DR10). Using this 
approach we initially performed a very efficient search of the less crowded star fields at $l>60^{\circ}$, using a method that
minimised the number of candidates caused by bad pixels. After completion of the survey we extended the search to 
include both previously unobserved fields at $l>60^{\circ}$ and more crowded fields at $30<l<60^{\circ}$. This final search 
also included some candidates flagged as having possible bad pixels in the aperture and some candidates arising from the third 
epoch of data available for a small proportion of fields due to doubts about data quality at an earlier epoch. Only a small number 
of bona fide variables were found amongst using third epoch data or sources with possible bad pixels (the latter are flagged in the 
catalogue but are thought to have reliable photometry). Unlike our early searches of the the SQL data releases, the searches of the 
FITS catalogues include sources bright enough to have a "possible saturation" warning in the SQL releases. Stars with $K>11$ 
generally appear to have reliable photometry (see Appendix~\ref{A}). The catalogue does not include any sources with 
$K<11$ at both epochs and we flag all photometry that might be influenced by saturation.

We decided to not to investigate FITS catalogue-based candidates at $l<30^{\circ}$ because $\sim$2500 candidates were found,
of which under 5\% would be real (and limited time remained for this project). We note that UGPS fields at $l<20^{\circ}$ are included 
in the new extension of the multi-epoch VVV survey, VVVX, so these fields may be more readily searched at a later date. 
An area of $\sim$75~deg$^{2}$ at $15<l<30^{\circ}$ located off the mid-plane was included in the early SQL searches of DR5 
and DR7, see Paper I. 





In summary, we have searched all of the UGPS area that may be readily searched for high amplitude 
variables, finding a total of 618 variables in $\sim$1470~deg$^{2}$ of sky, see figure \ref{fig:cov}. 
The FITS catalogue searches yielded 547 of these, all at $l>30^{\circ}$, and the remaining 71 were found by the earlier 
SQL-based searches. The time baseline of the two UGPS epochs for these sources ranges from 1.8 to 6.3~yrs, as illustrated in 
figure \ref{fig:hist_baseline}. These variables may be termed extreme because only 1 star in $\sim$105 000 was observed 
to vary in this manner, based on a count of 591 variables at $l>30^{\circ}$ from 62 million stars that passed all our algorithmic 
selections save variability. Additional variables could be found by relaxing our quality cuts (see appendix A) or by matching 
primary and secondary detections of the same source in the small overlap regions in adjacent fields via the SQL tables in the 
forthcoming final UKIDSS data release (which will include all the two-epoch data). However, such searches 
would yield a very small proportion of discoveries from a lot of candidates.

The completeness and depth of our search declines unavoidably with decreasing longitude, as 
increasing source confusion causes a growing proportion of sources to fail our quality cuts or escape detection
entirely. Our quality cuts would be expected to remove $\sim$50\% of normal stars but we note that nearby class I
YSOs are often seen as spatially resolved nebulae that would typically not satisfy our requirements for a stellar
image profile and low ellipticity (necessary to reduce candidates caused by blends). A recent UKIRT-based study
of H$_2$ jet-driving YSOs in the Cygnus X region (Makin \& Froebrich submitted) has found a number of highly
variable YSOs missed by our search, principally sources that failed the image profile or ellipticity cuts at one of the 
two $K$ epochs. Consequently, our quality cuts appear to lead to completeness of only $\sim$30\% for jet-driving sources
in the Cygnus X region at $d$=1.4~kpc, though the issue would be less significant for more distant YSOs or more 
evolved systems, given that YSOs with 2.12~$\mu$m jets are generally class I systems \citep{davis09}.
As noted in Paper I, simulated searches of the VVV light curves for the 816 variables in 
CP17a showed that UKIDSS two epoch searches would be expected to recover only 25\% of 
those variables as high amplitude systems. The VVV sampling is itself quite sparse, detecting $\sim$80\% of 
highly variable YSOs with stellar image profiles (see CP17a). The overwhelming majority of highly variable YSOs and
other highly variable stars therefore remain to be discovered in the northern and equatorial Galactic plane.

\subsection{The Catalogue}

In Table 1 we present the catalogue of 618 high amplitude infrared variables, including photometry from optical to far- infrared public surveys. 
The first 10 rows are shown in the printed version of the paper and a description of all the columns is given in Appendix \ref{B}. There we also 
describe our quality control on the list-driven photometry from the WISE All-Sky database (following a similar procedure to \citealt{koenig14}) 
and details of our multi-waveband cross-matching procedure. The contents of most columns are obvious from the column label but here we note 
that $"K_c"$ is a UGPS $K$ magnitude that is always contemporaneous with the UGPS $J$ and $H$ photometry and $"K_o"$ is the other UGPS 
$K$ measurement. Also, $\Delta K$ is the amplitude of variation based on the two UGPS fluxes and $\Delta K_{all}$ is the amplitude after taking 
account of 2MASS $K_s$ photometry, where available (neglecting the slight difference in the bandpass).

\section{Overview of catalogue members}

\subsection{Amplitude distribution}

\begin{figure}
	\includegraphics[width=\columnwidth]{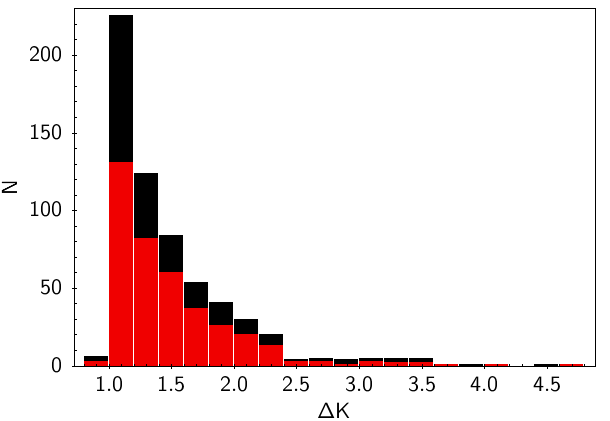}
	\includegraphics[width=\columnwidth]{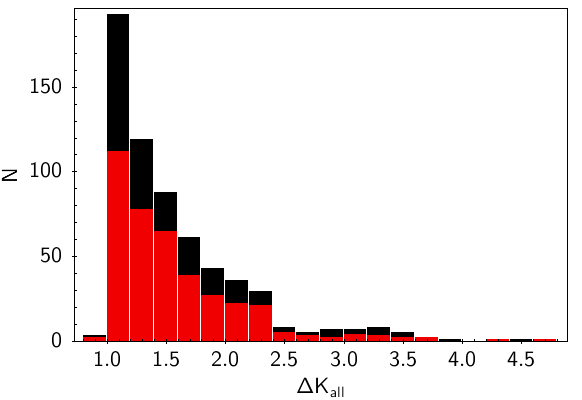}
    \caption{Distribution of amplitudes for the 618 variable stars in the catalogue. {\it (upper panel):} two-epoch amplitude $\Delta K$ from UGPS. {\it (lower panel):} 
    amplitude $\Delta K_{all}$, which includes a 3rd epoch of photometry in $K_s$ from 2MASS, where available. The subset classified as likely YSOs is shown in red.
    A few sources have $\Delta K$=0.98--1.00 due to small calibration adjustments between the FITS catalogue photometry and DR10, see Appendix \ref{B}.}
    \label{fig:hist_ampl}  
\end{figure}

Histograms of the variability amplitude are shown in figure \ref{fig:hist_ampl}. The distribution of $\Delta K$ falls steeply from 1 to 2.5 mag but a low level tail extends up to 5~mag. The distribution of $\Delta K_{all}$, which includes the 3rd epoch of $K_{s}$ photometry from 2MASS available for 43\% of the catalogue, has a slight shift to higher amplitudes as one would expect. The tail with $\Delta K_{all}>2.5$ includes 43 sources. The proportion of sources with $\Delta K>1.5$ is 34\%, falling to 13\% with $\Delta K>2$. For the subset with a 3rd epoch 
from 2MASS, 43\% have $\Delta K_{all}>1.5$ and 21\% have $\Delta K_{all}>2$. The likely YSOs (shaded in red) have a very similar distribution of $\Delta K$ to the full sample.

While it might be tempting to ascribe the drop above 2.5~mag to a physical difference, we note that the VVV sample of 816 high amplitude variables (with much better sampled light curves) does not show this feature but rather a fairly smooth decline with increasing amplitude. Inspection of the distribution for UGPS (using different binning choices and trying a vertical log scale) suggests that the edge at 2.5~mag is due to a chance under-density at 2.4--2.5 mag. The more significant feature appears to be a change from a linear decline in $log$$\left( N \right)$ with increasing amplitude to an approximately flat distribution at higher amplitudes, as distinct from a decline to $N=0$. Since a variety of light curve types are seen in high amplitude YSOs (CP17a) the UGPS amplitude distribution cannot be
simply interpreted.

\subsection{Known variable stars and other identifications}

To identify known variable sources we performed searches of the SIMBAD database, supplemented by checks of the General Catalogue of Variable Stars \citep{samus10}, the GAIA Photometric Science Alerts database, the OGLE-III catalogue of variable stars, the K2 archive of the {\it Kepler} satellite and the {\it CoRoT} catalogues available at the VizieR database. In addition we searched the \citet{witham08} catalogue of IPHAS H$\alpha$ emitters, the \citet{viironen09} catalogue of IPHAS PN candidates and the full list of individual source catalogues used by \citet{acero15} for automatic source association with the {\it FERMI} catalogue of high energy sources.


\begin{landscape}
 \begin{table}
  \caption{Catalogue of 618 high amplitude infrared variable stars from UGPS}
  \label{tab:landscape}
  \begin{tabular}{rrlrrrrrrrrrrrr}
    \hline
  \multicolumn{1}{|c|}{No.} &
  \multicolumn{1}{c|}{UGPS Designation} &
  \multicolumn{1}{c|}{Other Name} &
  \multicolumn{1}{c|}{RA} &
  \multicolumn{1}{c|}{Dec} &
  \multicolumn{1}{c|}{$l$} &
  \multicolumn{1}{c|}{$b$} &
  \multicolumn{1}{c|}{$J$} &
  \multicolumn{1}{c|}{$J_{err}$} &
  \multicolumn{1}{c|}{$H$} &
  \multicolumn{1}{c|}{$H_{err}$} &
  \multicolumn{1}{c|}{$K_{c}$} &
  \multicolumn{1}{c|}{$K_{c,err}$} &
  \multicolumn{1}{c|}{$K_{o}$} &
  \multicolumn{1}{c|}{$K_{o,err}$} \\
  & & &  \multicolumn{1}{c|}{deg} & \multicolumn{1}{c|}{deg} & \multicolumn{1}{c|}{deg} & \multicolumn{1}{c|}{deg} &  &  &  & & & & & \\
    \hline
1 & UGPS J032726.66+584508.5 &{\footnotesize [GMM2009] AFGL490 25}& 51.8611 & 58.7524 & 141.9958 & 1.7798 & & & 17.95 & 0.05 & 14.25 & 0.02 & 12.98 & 0.02 \\
2 & UGPS J032904.43+583617.6 &{\footnotesize [GMM2009] AFGL490 337}& 52.2685 & 58.6049 & 142.2539 & 1.7771 & 16.90 & 0.02 & 15.63 & 0.02 & 14.67 & 0.02 & 15.88 & 0.03 \\
3 & UGPS J032946.55+554538.1 &  & 52.4440 & 55.7606 & 143.9428 & -0.5137 & 19.03 & 0.09 & 16.48 & 0.02 & 14.01 & 0.02 & 12.12 & 0.02  \\
4 & UGPS J033048.77+544852.8 &  & 52.7032 & 54.8147 & 144.6023 & -1.2077 & 18.55 & 0.05 & 16.86 & 0.02 & 15.88 & 0.03 & 14.86 & 0.02  \\
5 & UGPS J033352.60+574259.2 &  & 53.4692 & 57.7165 & 143.2822 & 1.4112 & 18.41 & 0.05 & 16.84 & 0.02 & 15.50 & 0.02 & 16.82 & 0.07  \\
6 & UGPS J033410.59+544603.0 &  & 53.5442 & 54.7675 & 145.0261 & -0.9674 & 15.65 & 0.02 & 14.85 & 0.02 & 14.49 & 0.02 & 13.36 & 0.02  \\
7 & UGPS J035121.92+545646.8 &  & 57.8413 & 54.9463 & 146.8871 & 0.6713 & 17.93 & 0.03 & 15.94 & 0.02 & 14.31 & 0.02 & 15.61 & 0.02  \\
8 & UGPS J035220.42+531052.9 &  & 58.0851 & 53.1814 & 148.1109 & -0.6085 & 19.23 & 0.07 & 18.19 & 0.05 & 17.44 & 0.08 & 15.37 & 0.02  \\
9 & UGPS J035442.91+515708.3 &  & 58.6788 & 51.9523 & 149.1704 & -1.3286 & 16.68 & 0.02 & 16.00 & 0.02 & 15.31 & 0.02 & 13.97 & 0.02 \\
10 & UGPS J035454.09+523007.8 &  & 58.7254 & 52.5022 & 148.8411 & -0.8873 & 17.72 & 0.03 & 16.64 & 0.02 & 15.91 & 0.03 & 17.41 & 0.09 \\
    ...\\
    \hline
  \end{tabular}
  \\
   \begin{tabular}{rrrccrrrrrll}
    \hline
  \multicolumn{1}{c|}{$K_{c}$} &
  \multicolumn{1}{c|}{MJD a} &
  \multicolumn{1}{c|}{MJD b} &
  \multicolumn{1}{|c|}{Saturation} &
  \multicolumn{1}{|c|}{Bad pixel} &
  \multicolumn{1}{c|}{$\Delta K$} &
  \multicolumn{1}{c|}{$\Delta K_{all}$} &
  \multicolumn{1}{c|}{Source} &
  \multicolumn{1}{c|}{$d$} &
  \multicolumn{1}{c|}{Ref.} &
  \multicolumn{1}{c|}{Association} &
  \multicolumn{1}{c|}{Ref. for} \\
  \multicolumn{1}{|c|}{Epoch} &  \multicolumn{1}{|c|}{days} &   \multicolumn{1}{|c|}{days} &  \multicolumn{1}{|c|}{flag} &  \multicolumn{1}{|c|}{flag} &  &  & Type & kpc & \multicolumn{1}{c|}{for $d$} &  &  \multicolumn{1}{c|}{Association}  \\
 \hline
 a & 53673.52904 & 54742.63889 & 0000 &  & 1.27 & 2.28 & YSO & 0.9 & 1 & {\footnotesize AFGL490 SFR.} & 2,3\\
b & 53673.40762 & 54757.62116 & 0000 &  & 1.21 & 1.21 & YSO & 0.9 & 1 & {\footnotesize AFGL490 SFR.} & 2,3 \\
a & 55483.56806 & 56293.37766 & 0000 &  & 1.89 & 1.89 & YSO &  &  & {\footnotesize New pre-MS cluster, no.1, centred on IRAS 03262+5536.}\\
a & 55490.4603 & 56293.40218 & 0000 &  & 1.02 & 1.02 & YSO &  &  & {\footnotesize New pre-MS cluster, no.2, matching IRAS 03270+5438.}\\
a & 55482.54308 & 56291.23234 & 0000 &  & 1.32 & 1.32 &  &  &  &\\
a & 55490.51391 & 56293.40543 & 0000 &  & 1.13 & 1.29 & YSO &  &  &\\ 
a & 55490.55902 & 56293.42058 & 0000 &  & 1.30 & 1.30 & YSO &  &  &\\
a & 55508.56666 & 56584.51157 & 0000 &  & 2.07 & 2.07 &  &  &  &\\
a & 55476.64744 & 56601.43958 & 0000 &  & 1.34 & 1.34 &  &  &  &\\ 
a & 55476.64744 & 56601.43958 & 0000 &  & 1.51 & 1.51 &  &  &  &\\
    ...\\
    \hline
  \end{tabular}
 \\
  \begin{tabular}{lrrrrrrrrrrrrrrrrrrrr}
    \hline
  \multicolumn{1}{c|}{Spatial} &
  \multicolumn{1}{c|}{SED} &
  \multicolumn{1}{|c|}{$\alpha$} &
   \multicolumn{1}{c|}{$r$} &
  \multicolumn{1}{c|}{$i$} &
  \multicolumn{1}{c|}{H$\alpha$} & Optical &
  \multicolumn{1}{c|}{$J$} &
  \multicolumn{1}{c|}{$H$} &
  \multicolumn{1}{c|}{$K_{s}$} &
  \multicolumn{1}{c|}{W1} &
  \multicolumn{1}{c|}{W2} &
  \multicolumn{1}{c|}{W3} &
  \multicolumn{1}{c|}{W4} &
  \multicolumn{1}{c|}{[24]} &
  \multicolumn{1}{c|}{I1} &
  \multicolumn{1}{c|}{I2} &
  \multicolumn{1}{c|}{I3} & 
  \multicolumn{1}{c|}{I4} &
  \multicolumn{1}{c|}{PACS [70]}\\
\multicolumn{1}{c|}{Group} & \multicolumn{1}{c|}{Class} & & & & & \multicolumn{1}{c|}{MJD} & \multicolumn{3}{c|}{\footnotesize (2MASS)} &  &  &  &   &   &  &  &   & & \multicolumn{1}{c|}{mJy}  \\
 \hline
  & 0.94 & I & & & & & &  & 15.26 & 11.52 & 9.32 & 6.37 & 3.75 & & 10.72 & 9.35 &  &    \\
  &  & & & & & & 17.00 & 15.53 & 14.76 & 13.90 & 13.08 &  &  & & 13.79 & 13.15 &  &    \\
  & 0.37 & I & & & & & & 15.40 & 13.48 & 11.84 & 10.09 & 7.10 & 4.98 & & 11.20 & 10.05 &  &  \\
  &  &  & &  &  &  &  &  &  &  &  &  &  &  &  &  & & \\
 &  &  & &  & & & & &  &  &  &  &  & & 14.51 & 13.66 &  &   \\
 & 0.19 & Flat & 19.32 & 17.85 & 18.81 & 53655.0 & 15.25 & 14.12 & 13.22 & 13.96 & 13.55 & 9.99 & 6.54 &  & \\
 & 0.04 & Flat & & & & & 17.17 & 15.94 & 14.88 & 13.22 & 11.97 & 9.43 & 7.40 & & 12.42 & 11.60 &   \\
&  &  &  & &  & & & & & 14.64 & 14.24 & 11.55 &  & & 14.56 & 14.23 &  & \\
&  & & 19.71 & 18.0 & 19.06 & 54066.0 & 16.32 & 15.94 & 15.24 & 13.84 & 12.94 & 11.15 &  &  &   \\
&  & &  &  &  &  &  &  &  &  &  &  &  &  &  &  & & \\
    ...\\
    \hline
  \end{tabular}\\
 Note: Only the first 10 rows of the table are shown here, split into three sections. The full table is available in the online supplementary information.
 \end{table}
 \end{landscape}
 
\begin{landscape}
References for Table 1:
\small (1) \citet*{testi98}
(2) \citet{gutermuth09}
(3) \citet{masiunas12}
(4) \citet*{wouterloot93}
(5) \citet*{froebrich07}
(6) \citet*{camargo15}
(7) \citet{lumsden13}
(8) \citet*{cpm89}
(9) \citet{walmsley75}
(10) \citet{difrancesco08}
(11) \citet{lundquist14}
(12) \citet{balanutsa14}
(13) \citet*{solin12}
(14) \citet{buckner13}
(15) \citet{dobashi11}
(16) \citet{prisinzano11}
(17) \citet{wang09}
(18) \citet{casoli86}
(19) \citet{kronberger06}
(20) \citet{lim15}
(21) \citet{liu14}
(22) \citet{evans81}
(23) \citet{gyulbudaghian11}
(24) \citet{lada03}
(25) \citet*{carpenter95b}
(26) \citet*{moffat79}
(27) \citet*{reich97}
(28) \citet{magnier99}
(29) \citet{kawamura98}
(30) \citet{dunham10}
(31) \citet*{bdb03}
(32) \citet{chavarria08}
(33) \citet*{leistra06}
(34) \citet{bds03}
(35) \citet{reid09}
(36) \citet{dutra01}
(37) \citet{shimoikura13}
(38) \citet{ackermann15}
(39) \citet{perez87}
(40) \citet{romanzuniga08}
(41) \citet{wang08}
(42) \citet{phelps97}
(43) \citet{cambresy13}
(44) \citet{poulton08}
(45) \citet{kato01}
(46) \citet{kamezaki14}
(47) \citet{reipurth04}
(48) \citet{venuti14}
(49) \citet{dahm05}
(50) \citet*{lee96}
(51) \citet{lennon90}
(52) \citet{puga09}
(53) \citet*{horner97}
(54) \citet*{lee91}
(55) \citet{elia13}
(56) \citet{difrancesco08}
(57) \citet*{may97}
(58) \citet{cp15}
(59) \citet*{kaminski07}
{60} \citet{hog00}
(61) \citet{cp14}
(62) \citet{cohen80}
(63) \citet{forbes00}
(64) \citet{jimenez06}
(65) \citet*{radhakrishnan72}
(66) \citet{mallick13}
(67) \citet{georgelin70}
(68) \citet{ageorges97}
(69) \citet{hashimoto94}
(70) \citet{robitaille08}
(71) \citet{engels83}
(72) \citet{eden12}
(73) \citet{anderson09}
(74) \citet{veneziani13}
(75) \citet*{teyssier02}
(76) \citet{kuchar94}
(77) \citet{morales13}
(78) \citet{billot11}
(79) \citet{nakano03}
(80) \citet{peretto09}
(81) \citet{anderson12}
(82) \citet{simpson12}
(83) \citet{watson03}
(84) \citet{chengalur93}
(85) \citet{avalos09}
(86) \citet*{fontani10}
(87) \citet{schlingman11}
(88) \citet*{ramirez14}
(89) \citet{alexander12}
(90) \citet*{wam82}
(91) \citet{rathborne10}
(92) \citet*{ragan09}
(93) \citet{solomon87}
(94) \citet{faustini09}
(95) \citet*{lockman96}
(96) \citet{abj09}
(97) \citet*{bronfman96}
(98) \citet{zhang09}
(99) \citet{rygl14}
(100) \citet*{mauerhan11}
(101) \citet{wood89}
(102) \citet{alexander13}
(103) \citet{ellsworth13}
(104) \citet{eden13}
(105) \citet{wienen12}
(106) \citet{dirienzo12}
(107) \citet{rosolowsky10}
(108) \citet{codella95}
(109) \citet*{williams04}
(110) \citet{simon06}
(111) \citet{churchwell06}
(112) \citet{simon01}
(113) \citet{lewis90}
(114) \citet{kolpak03}
(115) \citet{herbig83}
(116) \citet{mercer05}
(117) \citet{nagayama11}
(118) \citet{parsons11}
(119) \citet{kang09}
(120) \citet{cyganowski08}
(121) \citet*{kwok97}
(122) \citet{lundquist15}
(123) \citet*{bania12}
(124) \citet{viironen09}
(125) \citet{stephenson89}
(126) \citet{billot10}
(127) \citet*{bica08}
(128) \citet{riaz12}
(129) \citet{chapin08}
(130) \citet{xu09}
(131) \citet{corradi11}
(132) \citet{wu12}
(133) \citet{belczynski00}
(134) \citet{raj08}
(135) \citet{kharchenko13}
(136) \citet{yun95}
(137) \citet{balog02}
(138) \citet*{eder88}
(139) \citet{guenther12}
(140) \citet*{lin12}
(141) \citet{rygl10}
(142) \citet*{kurtz94}
(143) \citet{urquhart11}
(144) \citet{tadross09}
(145) \citet{hennemann08}
(146) \citet{tej07}
(147) \citet*{delgado97}
(148) \citet*{massey95}
(149) \citet{straizys14}
(150) \citet*{dickel69}
(151) \citet{kryukova14}
(152) \citet{leduigou02}
(153) \citet{downes66}
(154) \citet{schneider07}
(155) \citet{rygl12}
(156) \citet{maiz15}
(157) \citet*{maia16}
(158) \citet{schneider06}
(159) \citet{dobashi94}
(160) \citet{beerer10}
(161) \citet{rivilla14}
(162) \citet{marston04}
(163) \citet{laugalys02}
(164) \citet{laugalys06}
(165) \citet{guieu09}
(166) \citet{rebull11}
(167) \citet{scholz13}
(168) \citet{cambresy02}
(169) \citet{yung14}
(170) \citet{dickinson91}
(172) \citet*{arvidsson09}
(173) \citet{dame85}
(174) \citet{bernes77}
(175) \citet{samus10}
(176) \citet{getman12}
(177) \citet{contreras02}
(178) \citet{kuchar97}
(179) \citet{harvey08}
(180) \citet{jilinski03}
(181) \citet{liu12}
(182) \citet{kerton02}
(183) \citet{patriarchi01}
(184) \citet{kerton03}
(185) \citet*{kumar06}
(186) \citet{bruch87}
(187) \citet{saral17}
(188) \citet{marton17}\\
\end{landscape}

\begin{figure*}
	\vspace*{-2.5cm} \includegraphics[width=0.96\textwidth]{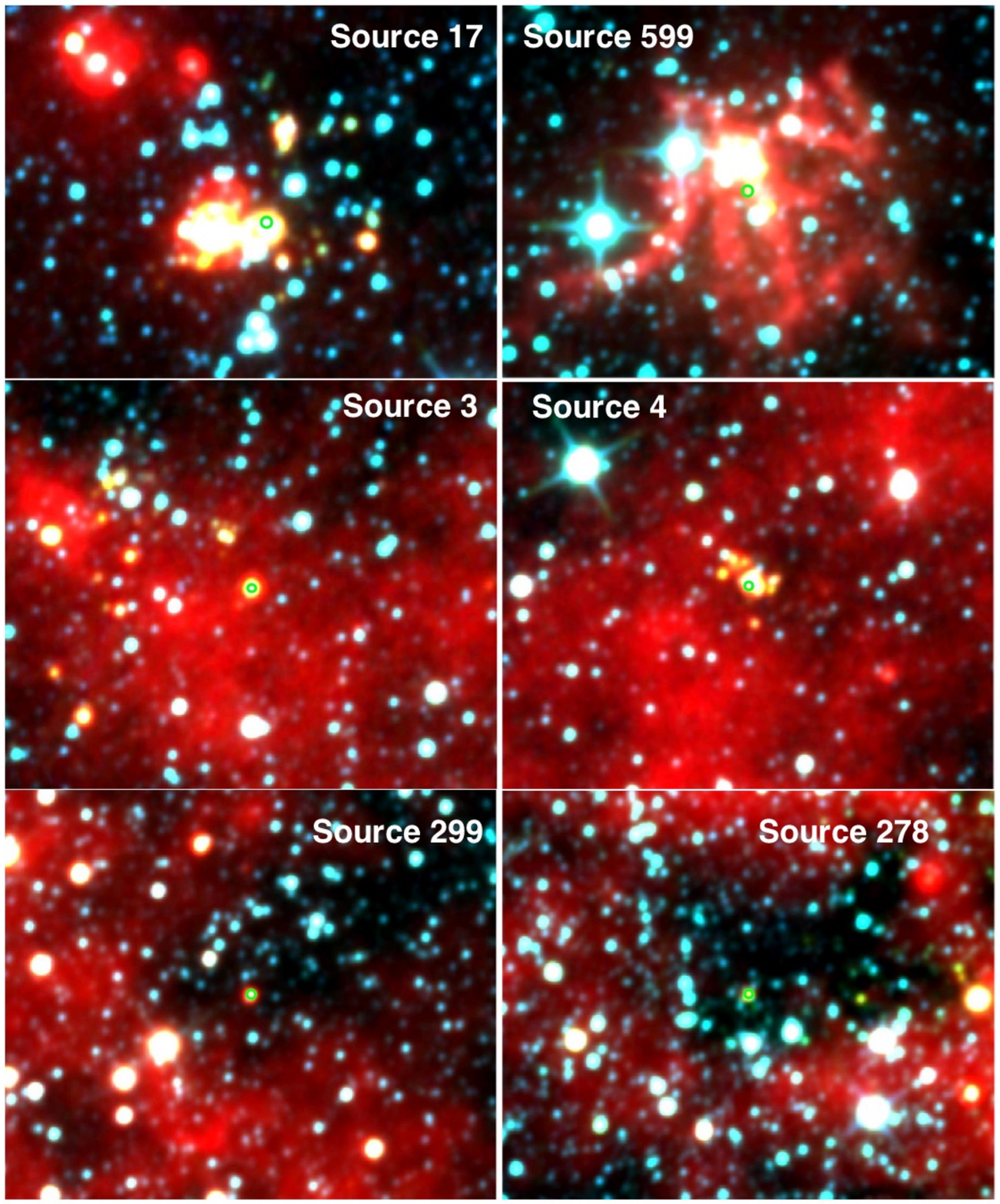}
    \vspace*{-2cm} \caption{Examples of YSO/non-YSO classification using WISE three colour images: W1 (3.3~$\mu$m, blue), W2 (4.6~$\mu$m, green), W3 (12~$\mu$m, red). 
    Normal stars usually appear blue whereas YSOs
    often appear red, yellow or green. Each image is $10\arcmin$ across with north at the top and east at left. The variable stars are marked with a green
    circle at the centre of each panel, though not all are clearly seen in WISE. The variables in the top and middle panels are classified as likely YSOs but 
    those in the lower panels are not. Source 17 is in the clearly apparent pre-MS cluster [FSR2007] 0671 \citep{froebrich07} so the WISE
    image was not required. Source 599 is in a previously uncatalogued SFR and likely HII region associated with IRAS 21401+5228, 
    see section 4.3. Source 3 is at the edge of a small uncatalogued group or cluster of pre-MS stars centred on IRAS 03262+5536. Source 4 is in a marginally resolved 
    uncatalogued group of YSOs. Sources 299 and 278 are in inner Galaxy mid-plane fields where YSO identification is more difficult. The field around 
    source 299 contains several bright stars that have red edges but most of these are probably distant AGB stars (see text). The source 278 field contains
    several rather faint green sources in the dark area at right of centre. This is likely an IRDC but it is unclear whether the green stars are YSOs or background 
    stars.}
    \label{fig:yso_select}
\end{figure*}

\noindent
Only 22 of the 618 variable stars were previously known as variable stars before our UGPS searches. Most of these, 15/22, are Long Period Variables on the Asymptotic Giant Branch (AGB), these typically appearing as isolated bright red stars in the WISE three colour images. A smaller group, 5/22, are cataclysmic variables (CVs, either dwarf novae or novae). Nova Cyg 2008 (source 391) has the 2nd highest amplitude in our catalogue, with $\Delta K=4.40$. Most of the variable AGB stars and AGB star candidates (11/15) are OH/IR stars or OH/IR candidates but two are classified as carbon stars (source 335=IRAS 19304+2529 and source 591=RAFGL 5625) and one (source 389=IRAS 19558+3333) is  an 8.4~GHz radio source likely to be a D-type symbiotic star (\citealt{belczynski00}; \citealt{seaquist94}), a system containing a Mira variable and a compact object with a circumstellar disc. 
Source 301 (IRAS 19210+1448) is classified as a likely O-rich AGB star \citep{kwok97}; this seems probable because it is a very bright infrared source (W4 = -2.4 in the WISE 22~$\mu$m passband) that satisfies our AGB star selection  (see section 3.4.1), even though it is on the outskirts of the W51 star formation region in a field containing several YSOs from the catalogue of \citet{kang09}. For some of the AGB stars it is unclear from the literature whether variability had been verified in past studies but typically the "probability of variability" column in the IRAS catalogue of point sources \citep{beichman88} has a value above 90\%.

The two remaining known variables are source 366, the binary central star of planetary nebula IPHASX~J194359.5+170901 (= the Necklace nebula, see \citet{corradi11} and source 551 (= 2MASS J20500940+4426522), a candidate eruptive variable YSO in the North America/Pelican nebula from the two epoch mid-infrared search of \citet{scholz13}. The latter has received little attention since its discovery. We discuss it further in section 4.2.2.

Apart from known variable stars, 81 sources were listed in SIMBAD as known YSOs or YSO candidates. The number of YSOs in the catalogue is actually much higher, see below. Five
sources in the Cygnus X region are identified in Makin \& Froebrich (submitted) as YSOs driving molecular outflows, see section 4.2.1.
In addition, four catalogue members (sources 41, 103, 326 and 555) are emission line stars included in a list of candidate PNe from the IPHAS survey \citep{viironen09}. However, it is uncertain
whether any of these actually are PNe. Sources 103 and 555 are located in SFRs, have low PN scores (0.38, 0.36 respectively, as defined in \citealt{viironen09}) and have red SEDs consistent with our YSO interpretation. Sources 41 and 326 have PN scores of 2.02 and 1.01, respectively, which are more promising indications of PN status. However, source 41 is one of several
red or green sources visible in the WISE three colour image and its near infrared colours are consistent with either a classical T Tauri star or a PN. The YSO interpretation seems more
likely, given the presence of adjacent YSO candidates and because such objects are more common than PNe.
Source 326 is in a field with less evidence for star forming activity and the near infrared colours would be unusual for a YSO ($J-H=0.27$, $H-K=0.61$). However, it has a very high amplitude 
($\Delta K_{all}$ = 2.96), having faded monotonically from $K_s$=13.19 in June 2000 (2MASS data) to $K$=14.79 in September 2005 and $K$=16.15 in July 2010. Such a high infrared amplitude would be extreme for the usual PN mechanism of a binary central star in which a hot white dwarf component irradiates one hemisphere of the companion star (as in the Necklace nebula). The nature of this H$\alpha$-emitting source is therefore unclear.  (We note that sources 41 and 326 lack WISE W3 detections so it is not possible to distinguish them from YSOs in the W1-W2 vs. W2-W3 plane in the manner indicated by \citet{koenig14}, in figure 10 of that work.)

One other catalogue member, source 83, was identified as a blazar of type II by \citet{ackermann15}. This variable is the gamma ray source 3FGL J0631.2+2019 from the Fermi Large Area Telescope Third Source Catalog \citep{acero15}.


\begin{figure*}
	\hspace*{0cm} \includegraphics[width=\textwidth,angle=0]{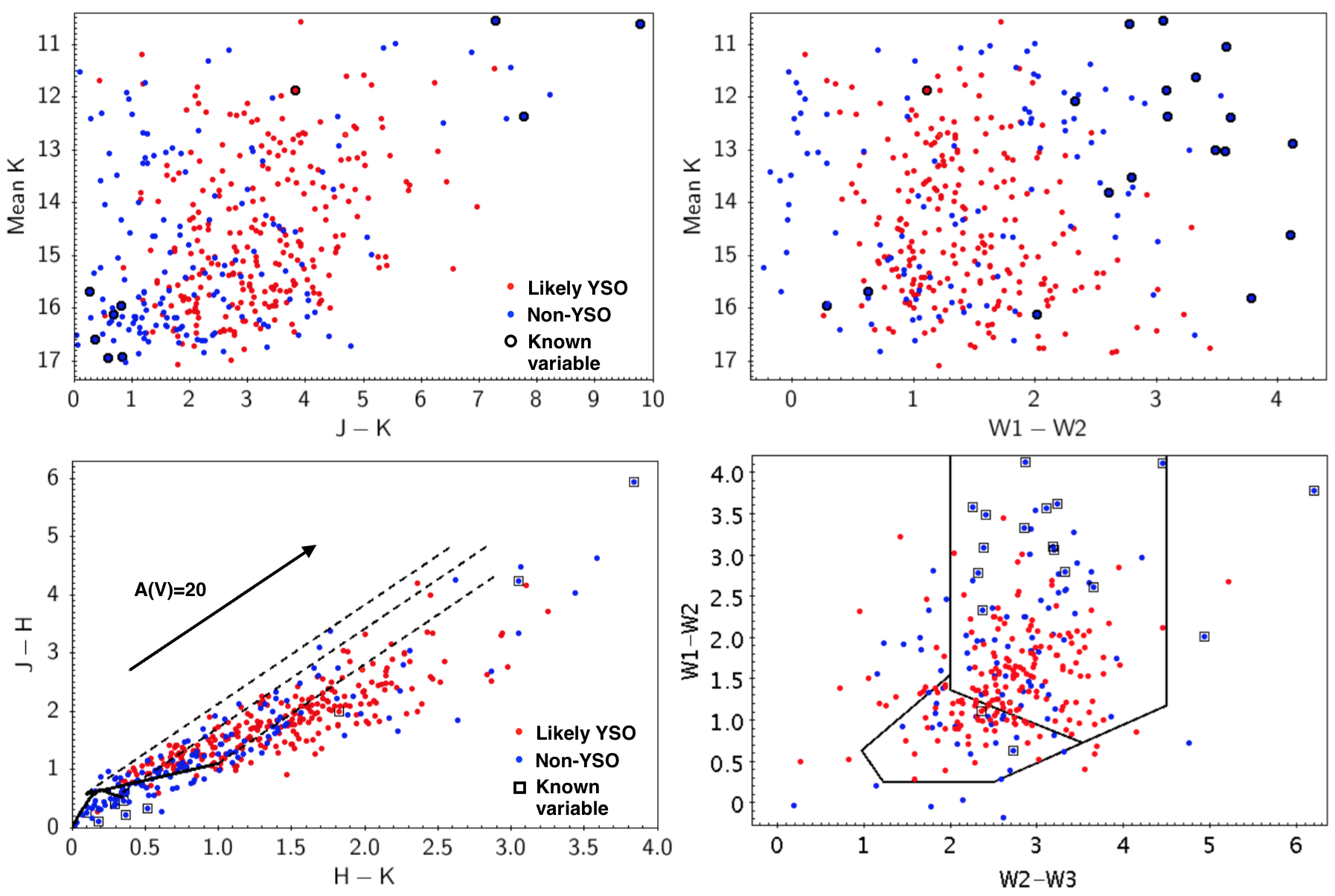}
	\vspace*{-4mm}
     \caption{CMDs and two colour diagram for likely YSOs (red points) and non-YSOs (blue points). Known variables are circled in black circles in the upper panels
     and black squares in the lower panels.
     {\it (upper left):} mean UGPS $K$ vs. $J-K$. {\it (upper right):} mean UGPS $K$ vs. W1-W2. {\it (lower left):} near infrared two colour diagram, with dashed lines illustrating 
     the reddening paths from the classical T Tauri locus (solid line) and the main sequence (curve). Reddened main sequence stars should lie between the left and central 
     dashed lines, whereas classical T Tauri stars may lie between the left and right dashed lines. Protostars and other stars with large amounts of circumstellar matter may 
     lie to the right of all the dashed lines. {\it (lower right):} W1-W2 vs. W3-W4, with solid lines enclosing the typical locations of class I YSOs (upper part) and class II YSOs 
     (lower part), see \citet{koenig14}. The locations of the likely YSOs in the lower panels support their position-based classification. Known variables are mostly dusty AGB stars 
     (sources redder than most YSOs, often undetected in $J$) or CVs (faint blue sources, often undetected in W1 and W2).}
    \label{fig:X}
\end{figure*}

\subsection{Proportion of YSOs}

We classify the majority of the individual variable stars (390/618 or 63\%) as likely YSOs. This was initially based on 395 sources (64\%) identified as YSOs in the literature or having spatial associations with SFRs, determined with the SIMBAD database and simple inspection of the WISE mid-infrared 3-colour images available at the Infrared Processing and Analysis Centre (IPAC). Subsequently, we re-classified four sources as likely AGB stars projected against SFRs and one source as an evolved star of uncertain nature (see section 3.4.1). We searched for indications of star formation within a $5\arcmin$ radius of each variable. This radius was found in Paper~I to be large enough to include most sources associated with pre-MS clusters but small enough to avoid including a high proportion of chance associations. The SIMBAD search was then expanded to a $10\arcmin$ radius to aid YSO identification and distance estimates but matches at 5--10$\arcmin$ were treated with caution and associated distances are given in the catalogue only if it seems likely that there is a genuine association with a spatially extended SFR.

\begin{figure*}
	\hspace*{-1cm} \includegraphics[width=0.8\textwidth]{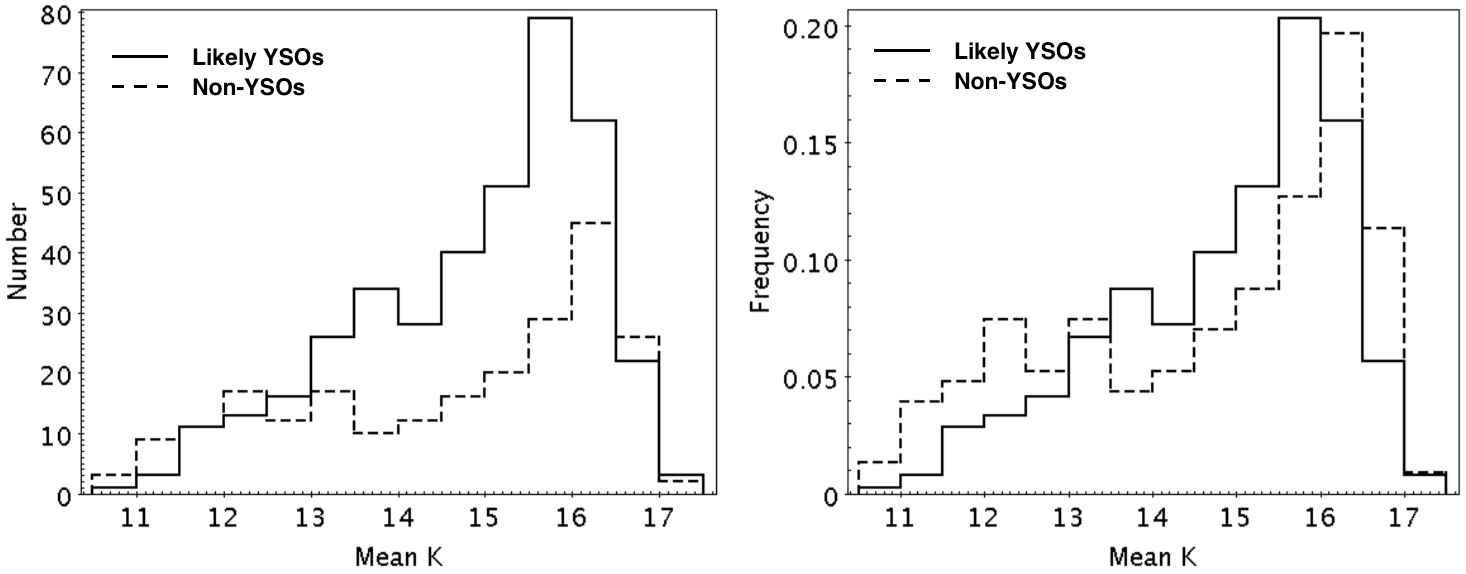}
    \caption{Mean UGPS $K$ magnitude distributions for likely YSOs and non YSOs. {\it (left)} Absolute numbers {\it (right)} Normalised distributions, each with unit area.}
    \label{fig:maghist}
\end{figure*}

In SIMBAD we searched for objects generally accepted as star formation indicators such as HII regions, YSOs and YSO candidates, molecular clouds, IRDCs, Herbig-Haro objects, class II methanol masers, mm and submm sources and clusters with ages $\le$10~Myr. We required a minimum of 5 SIMBAD indicators within a 5$\arcmin$ radius to classify a variable star as a YSO, unless the variable itself was listed as a YSO or YSO candidate or it was located within 5$\arcmin$ of a pre-MS cluster. In the WISE 3-colour images (blue, green and red corresponding to the W1 (3.3~$\mu$m), W2 (4.6~$\mu$m and W3 (12~$\mu$m) filters respectively) we required either a bright red or pink nebula to be present (indicating a likely HII region) or at least 5 red, yellow or green stars within 5\arcmin (normal stars generally appearing blue in WISE). Examples are given in figure \ref{fig:yso_select}. In the great majority of cases the designation as a likely YSO was clear from the apparent spatial association, such sources typically appearing as red, yellow or green stars themselves if they were detected by WISE.
(Sources fainter than W3=11 are often undetected in the W3 passband so faint YSOs with red SEDs typically appear green (due to their large W1-W2 colour) rather than red in the three colour images). Additional care with inspection of WISE images was required on inner galaxy mid-plane sightlines, defined $l<65^{\circ}$, $|b|<1^{\circ}$, owing to (i) the presence of numerous bright AGB stars and (ii) numerous spatially extensive IRDCs. The O-rich AGB stars common in the inner Galaxy (excepting unusually red dusty types) typically have 0$<$W1-W3$<$1.25, W3$<$7 (e.g. \citealt{tu13}). They appear slightly red around the edges of the stellar profile in WISE three colour images but they are noticeably less red than YSOs, see figure \ref{fig:yso_select} (lower left panel). We therefore did not consider such slightly red and unusually bright stars as an indication of star formation, given the notable lack of any accompanying population of fainter red, yellow or green stars in the field. IRDCs tended to cause all faint background stars in an area of a few square arcminutes to appear green in WISE three colour images (see  figure \ref{fig:yso_select}, lower right panel), similar to many bona fide YSOs. IRDCs are often but not always associated with active star formation so without more detailed analysis it was not always possible to assign YSO status to sources in fields with extensive IRDCs using the WISE and UGPS images alone. Consequently a small number of possible YSOs are not identified as such in the "Source Type" column of the catalogue but we note the possibility in the "Association" column.

The red colours of the likely YSOs support this classification, with most showing signs of ($H-K$) colour excess in the near infrared two colour diagram (figure \ref{fig:X}, lower panel)\footnote{The reddening vector displayed in figure \ref{fig:X} is based on the \citet{rieke85} extinction law. This is essentially indistinguishable from the very slightly curved reddening tracks derived by \citet{stead09} for UKIDSS data, after allowing for the effects of spectral type and changes in the effective wavelength of the broad band filters with extinction} 
or a location within the YSO selection region of \citet{koenig14} in the W1-W2 vs. W2- W3 diagram (figure 5, solid lines in the lower right panel). The YSO selection region in the latter plot is divided into class I (upper part) and class II (lower part) with flat spectrum systems spanning the division. \citet{koenig14} show that this selection region is not exclusive: it includes AGN and AGB stars (the latter marked as blue data points en- closed in black boxes in the class I YSO region) but it does at least serve to show consistency with our position-based classification. Sources that lie outside the YSO selection region are not excluded as YSOs since the selection was designed to include most but
not all YSOs. Class III YSOs have WISE colours near zero on both axes: there appear to be very few in the catalogue.

The comparison of the W1-W2 colours with those of normal stars in the Galactic plane is stark, e.g. 233 of the 390 YSO candidates are detected in W1 and W2 and almost all (231/233) have W1-W2 $>0.25$. By contrast, in a random sample of UGPS stars in the two-epoch footprint, we found that only 4\% have W1-W2 $>0.25$ (based on a sample with W2$<$14, a limit that includes the bulk of the YSO candidates). The 37\% of stars not spatially associated with SFRs tend to have bluer near infrared colours (figure \ref{fig:X}, upper left panel), except for the brighter stars that are likely to be dusty AGB stars in many cases (see section 3.4.1). 

To quantify the incidence of chance associations with SFRs, we randomly selected 200 of the 62 million stars that passed all our selections save variability from the UGPS two-epoch sky at $30<l<230^{\circ}$. We inspected SIMBAD and the WISE three colour images for these in the same manner as the variables and found that 13\% were spatially associated with SFRs by chance. However, since non-YSOs account for no more than half the catalogue, this means that the "likely YSO" proportion of 64\% is reduced by only about 6.5\%, leaving 57--58\% as YSOs. Five likely YSOs were subsequently re-classified as AGB stars or evolved stars, see section 3.4.1 below, but this does not significantly affect the proportion.
We should also allow for the presence of some relatively isolated YSOs with too few star formation indicators within 5$\arcmin$ to pass our selection. This seems likely given the existence of a substantial group of faint stars with W1-W2 colours and ($H-K$) colour excesses similar to the likely YSOs in the two colour diagram, see figures \ref{fig:X} and \ref{fig:nonYSO}. Therefore we round up the figure and quote the proportion of YSOs in the catalogue as 60\%, to 1 s.f.

We note that that catalogue includes a small number of published YSO candidates not yet listed as such in SIMBAD, typically relatively isolated objects that did not pass our position-based selection. E.g. sources 286 and 291, located about 0.5$^{\circ}$ away from the W51 complex and source 171, located about 0.4$^{\circ}$ from the W43 cluster \citep{saral17}. Similarly, the recent machine learning-based all-sky selection of YSO candidates by \citet{marton16} (based on AllWISE and 2MASS photometry) includes 28 members of the catalogue, six of which failed our YSO selection due to their relative isolation and a further two of which we classify as dusty AGB stars in section 3.4.1. For the sake of consistency (since we cannot identify all such omissions from SIMBAD) we do not include these candidates in our total of 390 likely YSOs but we note their candidacy in the "Association" column of the catalogue. The nature of high amplitude infrared variability is such that it often involves sources with infrared excess due to circumstellar matter, sometimes rare objects that would not be considered as likely contaminants even in the very careful colour-based YSO searches mentioned above. By retaining the "unclassified" status of these relatively isolated YSO candidates we aim to assist follow-up searches for rare objects in the catalogue.

The 60\% proportion of YSOs  is similar to the proportions found in VVV (c. 50\%, see CP17a) and our initial UGPS study (66\%, see Paper I). This confirms our earlier conclusion that they 
dominate the high amplitude near infrared variable sky on Galactic disc sightlines and extends the previous result to a much wider range of longitudes. The observed mean surface density of variable YSOs (0.25/deg$^2$) is very similar to that seen in Paper I and the magnitude distribution rises towards our sensitivity limit even in nearby SFRs. In Paper I we argued that the mean space density of high amplitude YSOs is higher than that of Mira variables in the Galactic disc (Miras being the commonest type of high amplitude variable seen in the optical waveband, see e.g. the General Catalogue of Variable Stars \citealt{samus10}). Our new results support this, particularly given that high amplitude YSO variability extends down to low mass systems
(see section 4.1.3) and the rising magnitude distribution. We refer to Paper I for the details of the argument.

\subsection{Non-YSO variables}

The distributions of mean UGPS $K$ magnitudes are a little different for the "likely YSO" and "non-YSO" subsets, see figure \ref{fig:maghist}. (We use "non-YSO" to
refer to the subset that did not pass our YSO selection even though there will be some YSOs in that group, see section 3.4.3.) The proportion of non-YSOs is relatively
high at $K<13.5$ and at $K>16$. 

\begin{figure*}
    \hspace{-4mm} \includegraphics[width=\textwidth]{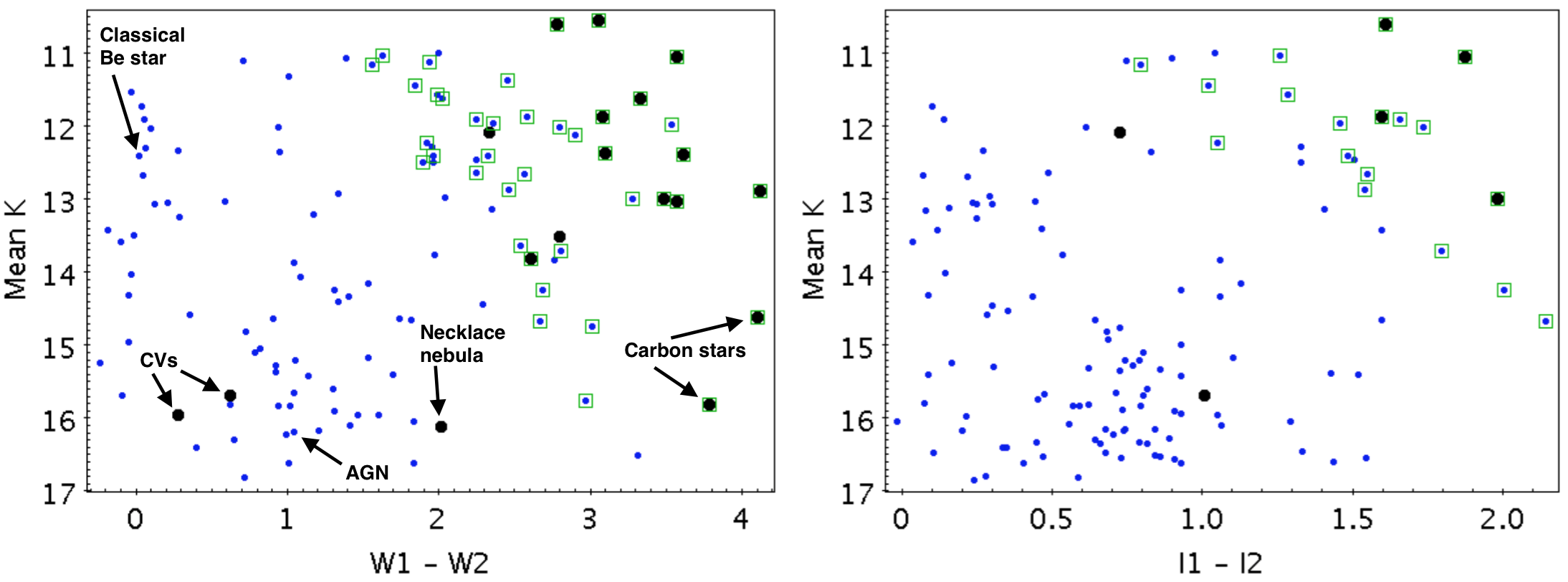}
    \caption{Colour magnitude diagrams for non-YSOs. {\it (Left):} Mean UGPS $K$ vs WISE W1-W2. {\it (Right):} Mean UGPS $K$ vs {\it Spitzer}/IRAC I1-I2. The IRAC
    I1 and I2 data are a little deeper than WISE W1 and W2 but they do not cover the full area of the UGPS catalogue. Known variables are marked in black and variables in our dusty 
    AGB star selection region in figure \ref{fig:AGB} are enclosed in green squares. Dusty AGB stars mostly found at the upper right. The nature of the small group of bright blue variable      
    stars in the left panel is unclear, see text. The larger group of faint variables includes both faint red stars (some of which will be relatively isolated YSOs) and faint blue variables such 
    as CVs, EBs and AGN, seen in  greater numbers in the upper left panel of figure \ref{fig:X}.}
    \label{fig:nonYSO}
\end{figure*}

\begin{figure}
    \hspace{-4mm} \includegraphics[width=0.5\textwidth]{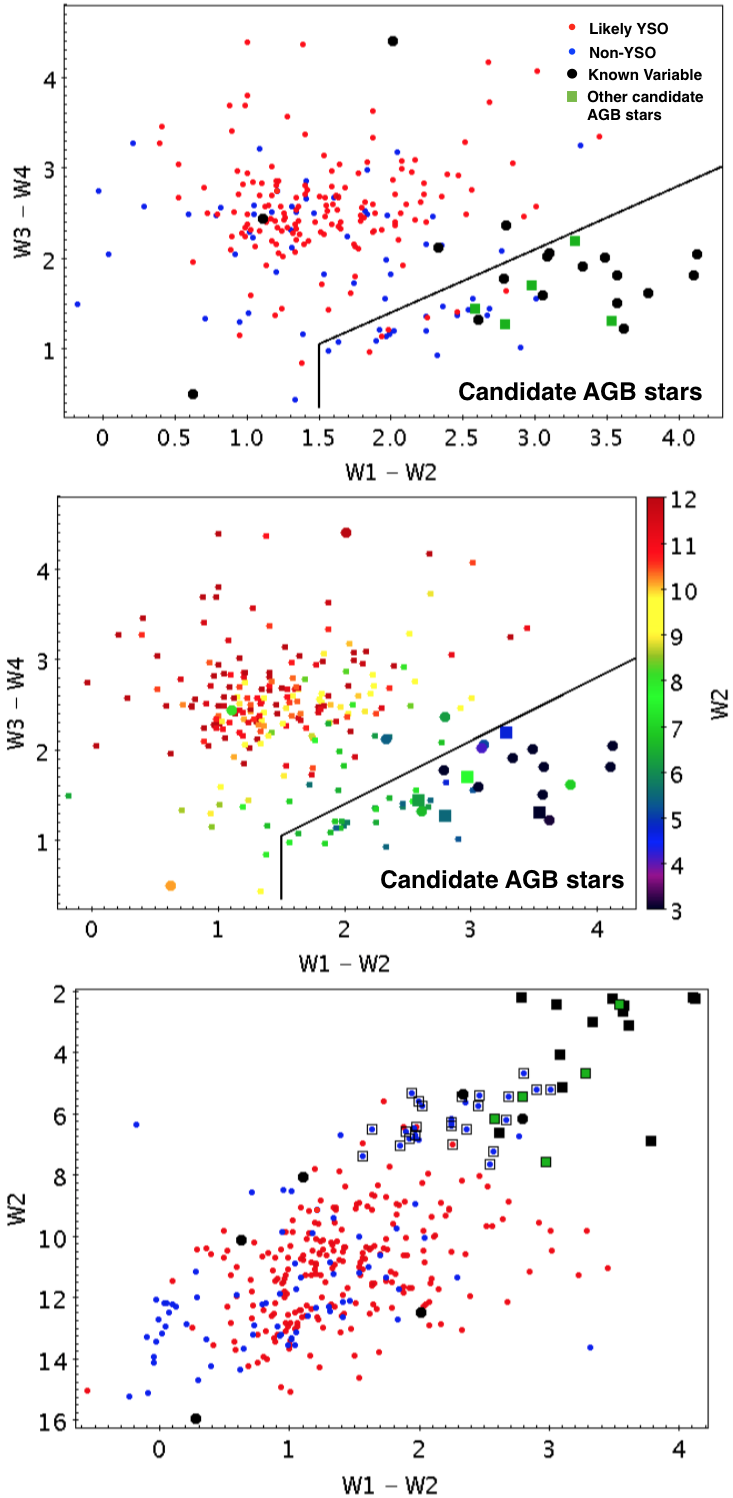}
    \caption{Selection of candidate AGB stars. {\it (Top):} WISE two colour diagram with likely YSOs in red, non-YSOs in blue, known variables as large black points and other candidate AGB
    stars from the literature as large green squares. The known variables are mainly dusty AGB stars located at lower right, save for a CV at lower left, a PN central star at top and
    a candidate eruptive YSO at centre left. The dividing line encompasses most of the known AGB stars while including only five sources projected
    against SFRs. {\it (Middle):}  as upper panel but with W2 mag colour coded to illustrate how the simple colour selection picks out the brightest mid-infrared stars. {\it (Bottom):} W2 vs. 
    W1-W2 colour magnitude diagram with the same colour coding as the upper panel. The 42 sources in the AGB selection region are enclosed in squares.}
    \label{fig:AGB}
\end{figure}

To aid our discussion of non-YSOs, in figure \ref{fig:nonYSO} we plot mean UGPS $K$ vs. W1-W2 and mean UGPS $K$ vs. I1-I2 colour magnitude diagrams for the non-YSOs only, with stars in our dusty AGB star selection region (see figure \ref{fig:AGB} indicated with green squares. The various IRAC I1 and I2 datasets are generally slightly deeper than the WISE All-Sky catalogue W1 and W2 data but cover only part of the area. The I1-I2 ([3.6]-[4.5]) and W1-W2 ([3.4]-[4.6]) colours are based on similar passbands but while the two colours correlate closely and agree for blue stars, W1-W2 has progressively larger values than I1-I2 for redder stars in our catalogue. (This is true for both non-YSOs and likely YSOs). 
\citet{cutri12} noted that W1-I1 becomes progressively larger with increasing I1-I2 for red sources at high Galactic latitude, with considerable scatter about the mean trend, but they did not discuss the cause. They found W2-I2 to be colour-independent: we see the same trend results for these two pairs of filters in our sample of highly variable sources. The W1:I1 differences for sources with red SEDs may in part be due to the slightly shorter effective wave- length of W1 than I1 but we note that it could be a signpost of
strong 3.1~$\mu$m absorption by molecules such as water ice in the W1 passband, which is centred at 3.4~$\mu$m but has significant response down to 2.8~$\mu$m. 

Circumstellar water ice would only be present in systems with cold circumstellar matter such as YSOs and very red O-rich AGB stars (OH/IR stars) with high mass loss rates (e.g. \citealt{meyer98}, \citealt{sylvester99}). However, two other strong 3.0--3.1~$\mu$m absorbers, hydrogen cyanide (HCN) and acetylene (C$_2$H$_2$), are commonly observed in dusty AGB stars, arising either from the circumstellar shell (in O-rich AGB systems) or in the photospheres of carbon stars, see \citet*{groenewegen94}. These molecular features should often play the same role in reddening W1-W2 more than I1-I2, with acetylene becoming more important in low metallicity stars (\citealt*{yang04}; \citealt{matsuura05}; \citealt{vanLoon08}). We note that \citet{antoniucci14} provide IRAC to WISE photometric transformations for YSOs (using a mostly low amplitude sample). While they did not discuss molecular absorption, their transformations indicate a similar trend for W1-W2 to be substantially redder than I1-I2, mainly due to differences between W1 and I1.

\subsubsection{Bright red variables: AGB stars}

The relatively high proportion of non-YSOs at $K<13.5$ very likely reflects the presence of many dusty AGB stars, as indicated by the extremely red colours of many bright non-YSOs 
in the colour-magnitude diagrams (figure \ref{fig:X} and figure \ref{fig:nonYSO}) and supported by the locations of several known OH/IR stars found at the upper right of the mean $K$ vs. W1-W2 plot
in figure \ref{fig:nonYSO}). One known AGB variable star system with similar mean $K$ and W1-W2 colour to the OH/IR stars is source 389, which is thought to be a D-type symbiotic system, as
noted in section 3.2. The mean magnitude histogram of the sample of 816 VVV variables in CP17a was bimodal, due to the large number of AGB variables found at bright magnitudes. 
That is not the case here because most AGB stars, especially normal AGB stars, are too bright to appear in our search, owing to shorter distances to the edge of the Galactic disc and lower extinction on most sightlines in the two epoch UGPS area, compared to the VVV disc region. (Mira variables typically have absolute $M_{K} = -6$ to -8, e.g. \citet{knapp03}, so $d>25$~kpc would be required
in the absence of extinction).

However, the two known carbon stars in the sample are somewhat fainter than the other known AGB stars in the sample (mean $K= 14.6, 15.8$) indicating that these typically less luminous AGB stars can contribute to the faint end of the distribution (see e.g. \citealt{ishihara11}).

In figure \ref{fig:AGB} we show that the W3-W4 vs. W1-W2 two colour diagram can be used to identify many of the AGB stars. The known AGB stars are mostly found at the lower right of this diagram and there is a substantial group of non-YSOs (i.e. sources not satisfying our YSO selection) that extends to the left and down from the group of known AGB stars
(black circles) and candidate AGB stars (green squares) clearly separated from the majority of YSOs. (The five candidate AGB stars were identified by \citealt{lewis90}, \citealt{chengalur93}, \citealt{kwok97} and \citealt{yung14}). Sources in this region of the plot, as far to the left as W1-W2=1.5, are in every case amongst the brightest mid-infrared sources in the catalogue, as shown in the middle and lower panels of figure \ref{fig:AGB}. All of these satisfy the criterion W2$<$7.8 that was used by \citet{robitaille08} to statistically distinguish "extreme" dusty AGB stars with high mass loss rates from YSOs and normal AGB stars (that study actually used IRAC I2, which gives almost identical magnitudes to W2 due to the similar bandpass.

Only six sources initially classified as likely YSOs are found in this "dusty AGB star" region of the W3-W4 vs. W1-W2 diagram, all of which have W2$<$7.8. These can readily be explained as chance projections of AGB stars against SFRs (see section 3.3) with the exception of sources 143(=GPSV3) and 239(=GPSV15). Source 143 was classified as an eruptive variable YSO in Paper I on the basis of an emission line spectrum, location in the Serpens OB2 association and luminosity-based problems with an AGB star classification. A similar classification was given for 
source 239 in Paper I for similar reasons, though this object is isolated rather than located in an SFR. For the present, we classify it as a likely YSO rather than an AGB star but we note that its
nature is uncertain. With these two exceptions, we classify 40 sources as likely AGB variable stars that satisfy:\\

$\mathrm{W3-W4 < 0.7}$($\mathrm{W1-W2}$)

$\mathrm{W1-W2 > 1.5}$\\

as indicated by the solid lines in figure \ref{fig:AGB}. Of the 40 stars, 22 were previously known as AGB stars or candidate AGB stars, though four of the candidates were identified
only by the selection of \citet{robitaille08}, which we do not plot as candidates because it has much more scatter in colour and magnitude when their "standard AGB stars" are included
 and these were not identified as variable. The sources in our selection region generally follow the $K$-[12] vs. $H$-$K$ colour-colour locus defined by \citet{vanloon98} for O-rich AGB stars, though source 301 is closer to the carbon star locus shown in that work. (The two known carbon stars in the sample are among 15/40 stars in the region that were undetected in 
 $H$).

The cut at W1-W2=1.5 was imposed because the two stars just to the left of this boundary are less luminous in the mid-infrared (W2$>$7.8), though one of these, source 203, has W2 = 7.87, I2 = 7.77 and therefore might be an AGB star despite being classified as a likely YSO due to its location in an IRDC that appears to host several YSOs. 
Two known AGB stars, source 351(=GLMP 939 = IRAS 19374+1626) and source 216(=IRAS 18569+0553), lie a little way outside our AGB region in figure \ref{fig:AGB}, closer to the main group of YSOs. Several other stars located close to the AGB region are bright mid-infrared stars that may well be AGB stars but our selection appears to capture the majority of the dusty variable AGB population with sufficient colour separation to suggest that it might be useful in other studies. 
We note that several authors have previously published colour selections for AGB stars or OH/IR stars (e.g. \citealt{veen90}, \citealt{lewis90}, \citet{vanloon98}, \newline \citealt{ishihara11}) but our selection differs in that it is defined for the high amplitude variable sky.
In particular, \citet{koenig14} proposed a very similar scheme to separate AGB stars from YSOs in the W3-W4 vs. W1-W2 two colour diagram. However, their template sample of dusty AGB stars did not include any sources with W1-W>3, with the result that their proposed selection would mis-classify some of the reddest known variable AGB stars in our catalogue as YSOs (these having W3-W4>2 by a small margin). Our selection therefore appears to be better for the reddest AGB stars, which lie in a region of colour space where there are no YSO candidates. The selection of \citet{koenig14} would be better for bluer, more typical AGB stars with less circumstellar matter but these are generally saturated in UGPS, as noted above. 

An additional evolved star, of uncertain nature, is source 507(=GPSV34, see Paper I and \citealt{cp15}). This bright and fairly red source (mean UGPS $K$=12.35, W1-W2=0.95, $J$-$H$=2.74, $H$-$K$=1.82) is projected in the Cygnus X star forming complex near the DR17 HII region. It lies outside our AGB star selection region (since it is not red enough in W1-W2) and \citet{cp15} showed that a mass-losing AGB star that followed the typical colour-luminosity relation of \citet{ishihara11} would be at a distance of $\sim$35~kpc, far outside the Galactic disc. The spectrum presented in \citet{cp15} shows strong $\Delta v=2$ $^{13}CO$ absorption lines, indicating an evolved star that has been through at least the first dredge up to increase the $^{13}$C/$^{12}$C ratio. Some remaining possibilities may be a mass-losing AGB star in the Galactic halo or a relatively low luminosity AGB star such as a J-type carbon star \citep{Morgan03}. We note that there are also a variety of the rare R Cor Bor stars, thought to be formed from binary white dwarf mergers, and related cooler objects (DY Per stars), in which large optical variations are caused by dust formation events and semi-regular pulsations (e.g. \citealt{tisserand09}, \citealt{otero14}. While these extinction events do not typically have high amplitude in the infrared, we should not discount the possibility.




\subsubsection{Bright blue variables}
Amongst the bright, non-YSO population there is a fairly distinct blue group of seven stars in the left panel of figure \ref{fig:nonYSO} with $0 < W1-W2 < 0.1$, $11<K<13$. These sources all have relatively blue near infrared colours ($J-Kc<1.25$), relatively low amplitudes ($1< \Delta K<1.5$, and $1< \Delta K_{all}<1.5$). One of this group, source 234 (=GPSV13), is an H$\alpha$ emitter previously discussed in Paper I as a likely classical Be star with extreme variability for such objects. However, none of the other six members of this bright blue group appear to be 
H$\alpha$ emitters (based on IPHAS photometry, see Appendix \ref{B} and section 4.1.4). Balmer line emission is variable in classical Be stars so absence of H$\alpha$ excess means little in individual cases but it would be somewhat surprising if all members of this group are classical Be stars. The locations of these stars in the IPHAS $r-$H$\alpha$ vs $r-i$ two colour diagram are consistent with moderately reddened B-type main sequence stars or lightly reddened G to early K-type dwarfs. Their locations in the UGPS $J-H$ vs. $H-K$ diagram favour the B-type option, with extinction values $A_V \approx 2$ to 9, consistent with the IPHAS data for each star. 

The nature of this small group is presently unclear but the simplest explanation is that most of them are eclipsing binaries (EBs) in which a smaller but more luminous early-type star is eclipsed by a less luminous giant star. In CP17a we examined the results of \citet{armstrong14} for the Kepler EB sample and found that EBs with amplitudes in $K_s$ above 1~mag should be dominated by systems with F- and G-type primaries because early type primaries emit less of their flux in the infrared. However, the one previously known EB in the CP17a sample is the detached (Algol-type) system PT Cen, an A2V+G6IV system \citep{budding04} with $\Delta K_s = 1.2$. This suggests that EB configurations involving early type primaries and subgiants, for example, are a possible explanation for the small group of bright blue variables.

\subsubsection{Faint red variables}
The higher proportion of sources classified as non-YSOs at $K>16$ (see \ref{fig:maghist}) appears to be due to a combination of one or more faint, relatively blue populations adding
to a redder population that has the same UGPS, WISE and IRAC colours as the YSO population, see figure \ref{fig:X}. It is likely that many of the faint red sources are indeed YSOs that were not 
selected in our search for sources spatially associated with SFRs, as indicated by the fact that a few of them are found within the areas occupied by the spatially extensive Cygnus X and 
Gemini OB1 groups of variable YSOs (see section 4.2). However, other types of faint high amplitude variable star can be red sources. 
The left panel of Figure 7 shows a few faint, very red AGB stars and new candidate AGB stars. We have also mentioned source 366, the binary central star of a PN. Source 366 is located in the region at W2-W3$>$4 that includes many PNe, as shown in the W1-W2 vs W2-W3 plot in \citet{koenig14} (figure 10 of that work). However, in our catalogue it is the only non-YSO in this sparsely populated colour space (see figure 5) that is also an H$\alpha$ emitter (see section 4.1.4). Other PNe may however be present that are too faint for a detection in W3 (see section 3.2) or in IPHAS or VPHAS+ H$\alpha$ images.

Source 83, identified in section 3.2 as a likely blazar is also a faint red source (mean GPS $K$=16.19, $W1-W2=1.04$, $J-H$=0.84, $H-K$=0.91). 
It is well established that blazars, including both radio loud optically violent variable (OVV) quasars and radio quiet BL Lac objects are highly variable sources in the infrared and across the electromagnetic spectrum, e.g. \citet{takalo92}, \citet{webb88}. In these rare types of AGN the consensus model involves inverse synchrotron emission by a highly beamed relativistic jet oriented very close to the line of sight, enabling variability of several magnitudes on a range of timescales from days to years, ultimately attributable to accretion variations in the black hole disc, (\citealt{blandford78}, \citealt{ghisellini93}). Blazars have near infrared colours in the range $1 < J-K < 2$, in the absence of Galactic extinction \citep{raiteri14}. While this is bluer than most YSOs in the catalogue, near infrared colours can of course be reddened by extinction and the WISE colours strongly overlap with those of YSOs \citep{raiteri14}.

Some low luminosity symbiotic stars and some CVs are also red sources (e.g. \citealt{phillips07}, \citealt{rodriguezflores14}, \citealt{hoard02}). It would therefore be unsafe to classify faint red variables as likely YSOs without spectroscopic data. We note that our SIMBAD and WISE-based selections of YSOs may reasonably be expected to be a little less complete at faint magnitudes, given that UGPS contains many distant low mass star formation regions that are unstudied (Lucas et al., in prep) and may not be identifiable by simple inspection of WISE or UGPS images.

\subsubsection{Faint blue variables}
The faint blue non-YSO variables with mean UGPS $K > 14.5$ typically have  $J-K<1.5$ (see figure \ref{fig:X}) and these have $J-H <1$, $H-K<0.72$. A total of 45 objects satisfy this colour selection, though there is no clear colour separation from the redder population. This faint blue component might plausibly be composed of CVs, eclipsing binaries (EBs) and AGN. Of the known variables with blue $J-K$ colours and mean UGPS $K>15.5$, 5/6 are CVs. The sixth, source 335, is the PN binary central star mentioned above, a source distinguished by its red mid-infrared colour (W1-W2=2.0). It is plausible that previously undetected CVs are numerous amongst the faint blue population. 

In Paper I we noted that a small proportion of AGN can vary by $>1$~mag in $K$, see \citet{kouzuma12}. The only known AGN in our sample is a fairly red object, 
a member of the rare blazar subclass often characterised by variability of a few magnitudes, see section 3.4.3. However, the broader sample of variable AGN studied by
\citet{kouzuma12} have relatively blue near infrared colours and the very few with $\Delta K>1$ would have relatively low amplitudes within our catalogue ($1 < \Delta K < 1.5$), similar to most of the faint blue variables. Most of the faint blue sources are located on sightlines with relatively low near infrared extinction ($|b| > 2^{\circ}$) so the colours are also consistent. In the mid-infrared, AGN typically display redder colours fairly similar to those of YSOs \citep{koenig14} but unfortunately the sensitivity of WISE is insufficient to detect most of the faint blue variables and aid classification. Scaling up the calculation in Paper I to the area of the present study, we might expect $\sim$30 variable AGN to be included in the sample, in the absence of extinction.
The five known CVs all displayed higher amplitudes so the data suggest it is indeed possible that AGN are contributing significantly to the faint blue membership of the catalogue.

In CP17a we found that EBs may have a space density comparable to that of YSOs amongst the variable population with $1<\Delta K_{s} < 1.6$, though not at higher amplitudes. The VVV sample of CP17a contained only 72 EBs and EB candidates (9\% of that sample). By randomly sampling 10 000 pairs of data points separated by at least 1.8~yrs from the light curves of each of the 72 VVV sources available in CP17a (with 45--50 data points each) we estimate that only 7\% of all EBs would be recovered in a two epoch search, which compares with 25\% of YSOs (Paper I). EBs and YSOs made up 
9\% and about 50\% of the VVV sample, respectively. If we assume they would be present in the same relative proportions in the area surveyed here and correct for the lower EB proportion recovered by two epoch sampling, we estimate EBs should make up $\sim$3\% of the catalogue, or $\sim 18$ objects. Alternatively, if we simply scale the 72 EBs found amongst 12 million VVV stars to the 62 million UGPS stars selected for the present search and then correct for the 7\% recovery fraction then we expect $\sim 24$ EBs in the present dataset. Both estimates are crude, owing to differences in Galactic populations on different sightlines and the fact that the search of CP17a was incomplete due to being based on typically only 17 early data points. (This is rather low for detection of eclipses covering perhaps $\sim$10\% of orbital phase, a typical fraction for deep eclipses in the Kepler Eclipsing Binary Catalogue, \citealt{prsa11}; \citealt{kirk16}). Nonetheless it appears that a mix of EBs and AGN can explain most of the faint blue population with amplitudes ($1<\Delta K<1.5$), with CVs likely comprising most of the rest of this population at $\Delta K > 1.5$. 

We refer the reader to section 4.1.4 for a discussion of the H$\alpha$ emission properties of the catalogue. We see there that three faint blue variables with $\Delta K_{all} > 2$ are also H$\alpha$ emitters, consistent with a CV interpretation. In section 5.2 we discuss the most intriguing faint blue object in the catalogue, source 363, which is also an H$\alpha$ emitter.

\section{The variable YSO sample}

\subsection{General properties}
\subsubsection{Spectral indices and amplitudes}

\begin{figure}
	\includegraphics[width=\columnwidth]{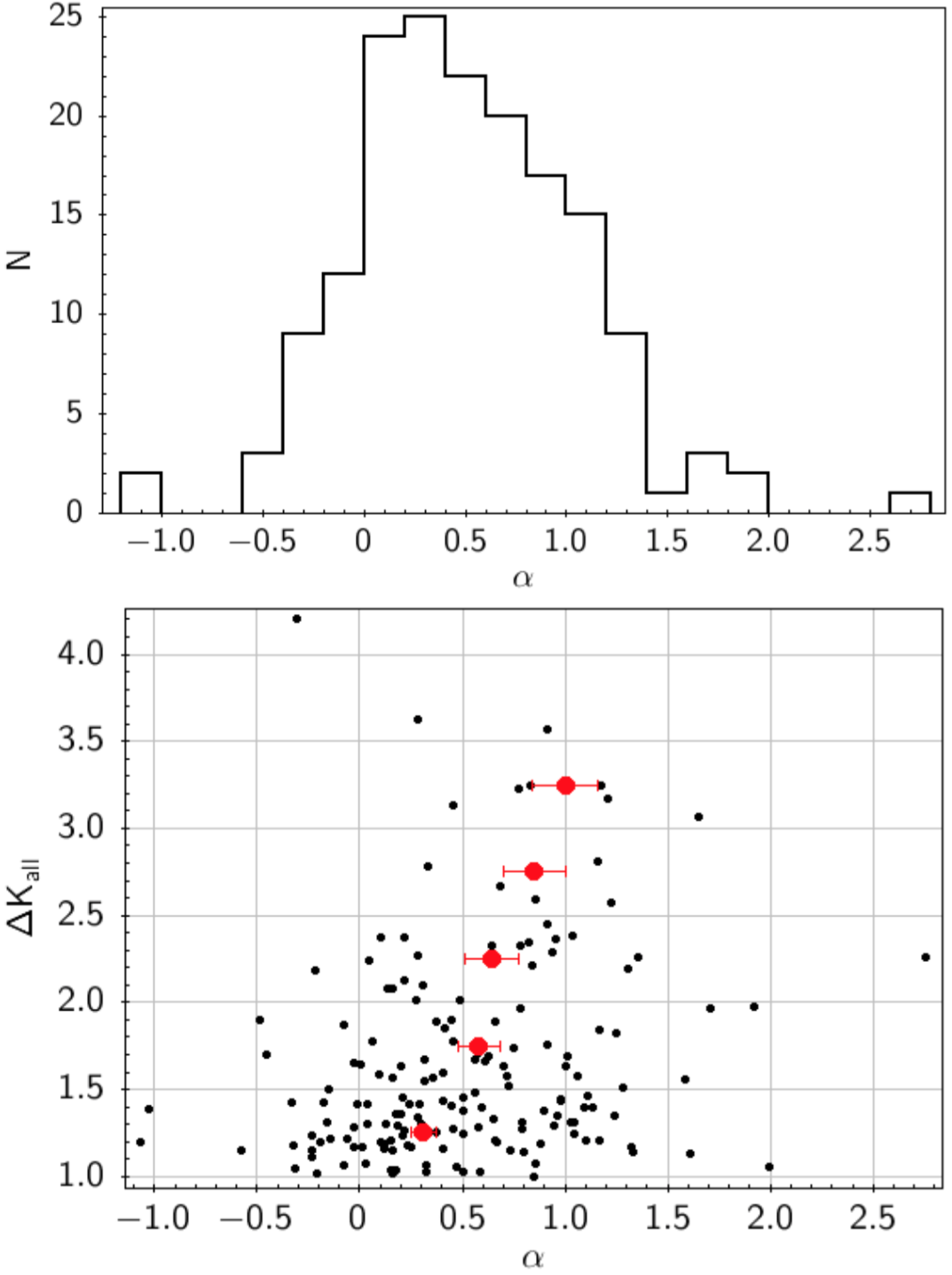}
   \caption{{\it (Upper panel):} histogram of the distribution of 2-22~$\mu$m spectral index, $\alpha$. {\it (Lower panel).} 
   Amplitude $\Delta K_{all}$ vs $\alpha$. YSOs with larger amplitudes show a subtle tendency to have redder spectral indices, as illustrated by the large red
  points corresponding to the median of $\alpha$ in amplitude bins of width 0.5~mag, up to $\Delta K_{all}$=3.25. Error bars represent the standard error, $\sigma$/$\sqrt{N}$.}
    \label{fig:alpha}
\end{figure}

In figure \ref{fig:alpha} (upper panel) we plot the distribution of spectral index for the likely YSOs (165/390) that have a 22 or 24~$\mu$m detection in WISE or MIPSGAL.
Using the divisions of \citet{greene94}, the overwhelming majority of these sources (95\%) are either class I systems, defined by $\alpha>0.3$, or 
flat spectrum systems, defined by $-0.3<\alpha<0.3$.  This distribution is of course influenced by selection effects given that class II and class III
YSOs will typically be too blue to be detected in the WISE W4 or MIPSGAL 24~$\mu$m passband. In addition, the gradual decline in the distribution at $\alpha>0.4$
is influenced by the fact that sources with redder SEDs are more likely to fall below the $K \approx 16$ magnitude limit of the UGPS variable catalogue.  

The 156 systems with $\alpha > -0.3$ make up 40\% of the full sample of likely YSOs. We can gain a better idea of the proportion of class I or flat spectrum systems by considering 
the I1-I2 and W1-W2 colours of faint YSOs lacking a 22 or 24~$\mu$m detection. Inspection of our sample of likely YSOs and the "Cores to Disks" sample of \citet{evans09} indicates 
that YSOs satisfying the condition "W1-W2 > 1.4 OR I1-I2 > 1" almost always have $\alpha > -0.3$. Moreover, at least half of YSOs that satisfy a second condition 
"0.8$<$W1-W2$<$1.4 OR 0.6$<$I1-I2$<$1" also have $\alpha > -0.3$. Considering likely YSOs in the catalogue that lack a measurement of spectral index, 52 satisfy the first condition and 80 satisfy the second. If we add the 52 and half of the 80 to the 156 likely YSOs with measured $\alpha > -0.3$ we find that $\sim$64\% are class I or flat spectrum systems. The true proportion is likely to be higher because some faint YSOs were not detected by Spitzer or WISE in any passband. We note that while a few distant YSOs suffer sufficient foreground extinction to influence the calculation of spectral index, this appears to be a very minor effect for the sample of 390 likely YSOs, most of which have a $J$ detection and relatively low 
extinction in comparison to the VVV sample of CP17a (see section 4.1.2). We see no significant correlation between distance and spectral index.

The high proportion of class I or flat spectrum systems supports the result of CP17a that these SED classes are found more frequently amongst high amplitude YSOs than in the YSO population as a whole. E.g. \citet{dunham14} found that class I and flat spectrum YSOs represent only 28--31\% of YSOs in nearby SFRs, being heavily outnumbered by class II systems. Since we have photometry at only two or three epochs we cannot test the result of CP17a that the trend towards red SEDs is strongest amongst YSOs with eruptive light curves.

In figure \ref{fig:alpha} (lower panel) we plot amplitude against spectral index. There is a subtle but clear tendency for YSOs with higher amplitudes to have redder spectral indices,
as illustrated by the large red points that give the median spectral index for bins of width $\Delta K_{all} = 0.5$, in bins from $\Delta K_{all} = 1.25$ to 3.25. Error bars show the standard error for each bin. The Pearson's correlation coefficient, $r$, for these 165 data points is only 0.221 but this is a significant correlation given the size of the dataset: the probability that 
$|r| > 0.2$ for a set of 100 randomly generated points is only 4.6\%, falling to 1.3\% for $|r| > 0.25$ \citep{taylor97}, so we can be confident in this result.
Relatively few YSOs with $\Delta K_{all}>1.5$ have $\alpha<0$ and all YSOs with amplitudes $\Delta K_{all}>2.5$ are class I systems, save for one system with $\Delta K_{all} > 4$ located on the class II/flat spectrum boundary.

\begin{figure}
	\includegraphics[width=\columnwidth]{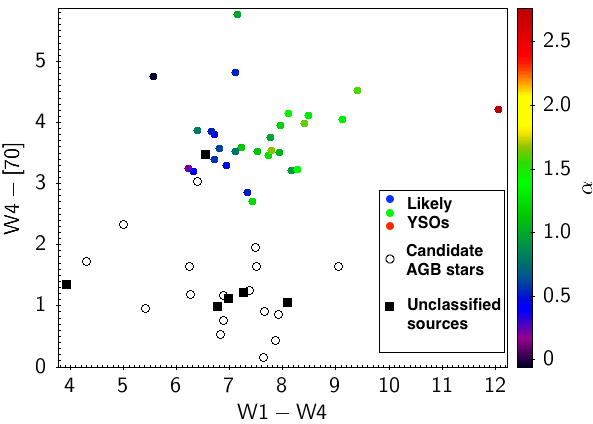}
   \caption{Separation of variable YSOs and dusty Mira variables in the W4-[70] colour. The W4-[70] vs. W1-W4 two colour diagram shows a fairly clean separation of 
   YSOs (filled circles) and AGB stars (open circles), supporting our position-based YSO selection method and our mid-infrared selection method for dusty AGB stars. 
   Unclassified sources (filled black squares) mostly lie amongst the AGB stars. The YSOs are colour-coded by spectral index, which is closely correlated with the 
   W1-W4 colour.}
    \label{fig:pacs}
\end{figure}

Our cross match against the {\it Herschel}/PACS 70~$\mu$m data (see Appendix \ref{B}) recovered 52 members of the catalogue, see figure \ref{fig:pacs}. These comprise 29 likely YSOs (27 class I and two flat spectrum systems), 6 unclassified sources and 17 sources classified in section 3.4.1 as dusty AGB star candidates via the W3-W4 vs. W1-W2 two colour diagram. The YSOs and the AGB stars are quite well separated in the W4-[70] colour, owing to the relative lack of dense cold dust around the AGB stars. This further verifies our WISE-based colour selection of the AGB stars. Moreover the plot shows that the likely YSOs, identified mainly by our position-based selection, have the colours expected for class I and flat spectrum systems. Comparison with figure 2 of CP17b shows that the W4-[70] and W1-W4 colours are very similar to those of spectroscopically confirmed high amplitude YSOs in the VVV sample as well as the MIPS [24]-[70] and I1-[24] colours of class I or flat spectrum YSOs in the Perseus cloud complex \citep{young15}.

Of the six unclassified sources, five have W4-[70] colours similar to the likely AGB stars. This is unsurprising given that four of these five have WISE colours that place them in the region between our fairly conservative AGB star selection region and the main body of YSOs in figure \ref{fig:AGB}. The fifth object, source 183, appears in SIMBAD as an unstudied IRAS source (IRAS~18492-0148). With W1-W2 = $-0.18$, it is in a sparsely populated part of figure 8 but inspection of the WISE images and GLIMPSE photometry shows that this colour is erroneous (caused by a rogue W1 datum) and this source is indeed likely to be a dusty Mira variable.

While all the likely YSOs in figure 10 have the expected colours, one candidate AGB star (source 158) and one unclassified source (source 278) are located amongst the YSOs. Source 278 was unclassified despite the presence of several adjacent faint green sources in the WISE three colour image, due to concerns that these might be normal background stars reddened by an IRDC (see section 3.3). The WISE colours in figure \ref{fig:AGB} are similar to those of YSOs so we can be fairly confident that this source is actually a YSO, though we note that the PACS detection has a signal to noise ratio of only 2. Source 158 (a high signal to noise ratio PACS detection) was classified as a candidate AGB star by \citet{robitaille08}, using their simple probabilistic colour-magnitude selection of YSOs and AGB stars. The WISE colours also place it firmly in the dusty AGB star region of figure \ref{fig:AGB}. The WISE three colour image shows no sign of additional YSOs in the field but there is some fairly bright 12--22~$\mu$m nebulosity a few arcminutes east (not unusual given the inner Galaxy mid-plane location) and a bright 22~$\mu$m nebula immediately adjacent that might relate either to previous mass loss or to star formation activity. Given the unusual SED, this source is not readily classified.

\begin{figure}
	\includegraphics[width=\columnwidth]{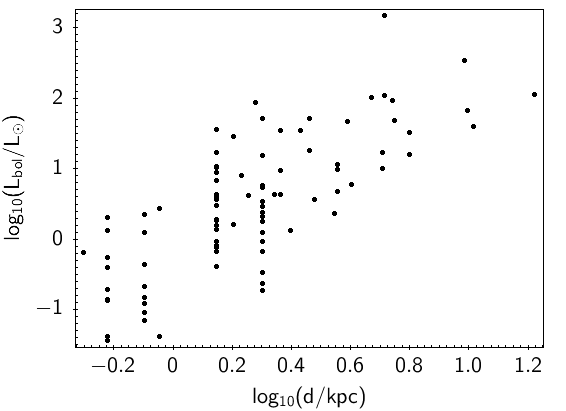}
	\includegraphics[width=\columnwidth]{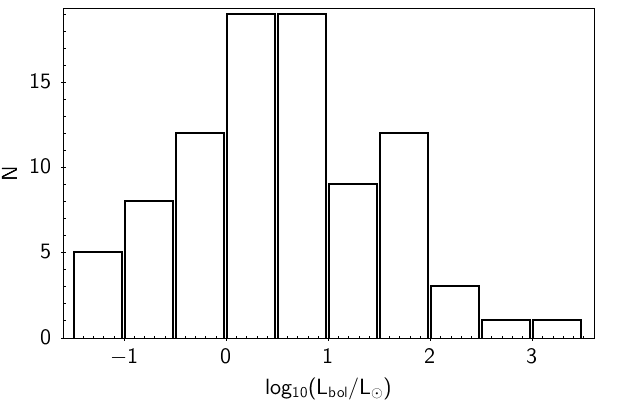}
   \caption{{\it (Upper panel).} Estimated bolometric luminosity vs distance for likely YSOs with 2--22~$\mu$m
   detections. Luminosities are based mainly on the 2010 epoch of 3--22~$\mu$m data from the WISE All-Sky catalogue.
   {\it (Lower panel).} Histogram of estimated bolometric luminosity.}
    \label{fig:Lbol_d}
\end{figure}

\subsubsection{Luminosities and distances}

The bolometric luminosities span a wide range, from $L_{bol} \sim 0.1 L_{\sun}$ to $\sim$$10^3 L_{\sun}$, see figure \ref{fig:Lbol_d}, though only rarely exceeding $10^{2.5} L_{\sun}$.
(Distances of likely YSOs were taken from the literature via spatial association with individual SFRs, see Appendix \ref{B}).
The plotted luminosities are estimated only for sources with 2 to 22~$\mu$m photometry, using the ratio of measured luminosity in this range, $L_{2-22}$, to $L_{bol}$ based on data for 
nearby YSOs from the "Cores to Disks" project \citep{evans09} and WISE photometry. The plotted luminosities do not include the effect of foreground extinction, which the data in table 9 of that study show would increase the $L_{bol}$ by a factor from 1 to 4 for YSOs with typical extinction and spectral indices in the range $0<A_V<12$, $0<\alpha<1$. (This assumes that the measured
extinction is all in the foreground, rather than the YSO disc and envelope, thereby indicating the maximum possible correction). Individual luminosities are uncertain
(and not given in Table 1) especially for sources with $\alpha >0$ where there is substantial scatter in the  $L_{bol} $ vs. $L_{2-22}$ ratio. However we were able to 
fit this ratio as a function of $\alpha$ and $F_{22}$/$F_{12}$ (or very similarly $F_{24}$/$F_8$) with a standard deviation of 0.15\footnote{The equations used were:\\ 
$L_{2-22}$/$L_{bol}=0.39$ \hspace{4cm} ($\alpha < -1.3$)\\
$L_{2-22}$/$L_{bol}=1.01 + 0.4 \alpha - 0.047 F_{22}$/$F_{12}$ \hspace{8mm} ($-1.3 < \alpha < -0.5$)\\
$L_{2-22}$/$L_{bol}=0.71$ \hspace{4cm} ($-0.5 < \alpha < 0$)\\
$L_{2-22}$/$L_{bol}=0.83-0.275\alpha - 0.217 log$($F_{22}$/$F_{12}$) \hspace{2.5mm} ($\alpha > 0$)}, 
such that luminosities should typically be correct within a factor of 2 (0.3 dex).
The least luminous variables with $L_{bol} \la 0.1 L_{\sun}$ include some nearby sources with low extinction ($A_V$=1--4 for sources 594, 555 and 101 at $d$=800--900~pc) 
and several additional GLIMPSE sources with similar absolute I1 to I4 magnitudes that lack 22~$\mu$m or 24~$\mu$m detections.
Therefore the presence of YSOs with $\sim$0.1~$L_{\sun}$ is reasonably secure, limited only by the sensitivity of the search. This suggests that the episodic accretion phenomenon 
extends down to the low mass peak of the Initial Mass Function, though this is of course hard to quantify given the debate over the existence of the stellar birthline (e.g. \citealt{baraffe09}).

Previous examples of very low luminosity eruptive variable YSOs include V1180 Cas \citep{kun11}, with $L \approx 0.07 L_{\sun}$ in the low state, ASASSN-13db \citep{holoien14} with
$L \approx 0.1 L_{\sun}$ pre-outburst and two eruptive candidates with high 
3.5~$\mu$m amplitudes identified by \citet{antoniucci14}, referred to as sources 828 and 1247 in that work. A few others (e.g. EX Lup) have $L$ = 0.1--1~$L_{\sun}$ when in the low state, see 
\citet{audard14}. Our calculated luminosities are dominated by the WISE data from 2010 so they will correspond to a mix of low state, high state and intermediate values.

The two most luminous sources plotted, with $L_{bol} =10^{2.5}$ to $10^3 L_{\sun}$, are both sources at large distances with very red SEDs ($d$>5~kpc, $\alpha \ga 0.7$).
While a few known optically bright eruptive YSOs in a nearby SFR are saturated in UGPS images (see section 4.2.2) neither of these two embedded YSOs (sources 325 and 229)
are close to the catalogue bright limit in UGPS $K$ and all other sources have $L_{bol} \la 10^{2.0}$. The surveys have sufficient sensitivity to detect luminous
YSOs across the Milky Way, as is evident from figure \ref{fig:Lbol_d}. Hence the rarity of luminous variables appears to be genuine, 
even allowing for a factor of 3 or 4 underestimate of luminosity due to extinction. The rarity may be due to the
fact that the photospheres of luminous YSOs typically produce more luminosity than the accretion disc even at very early evolutionary stages, e.g. \citet{calvet91}. One
might expect variable extinction to influence the light curve more often than variable accretion amongst luminous YSOs. Only a few
luminous eruptive variables ($>10^3~L_{\sun}$) have been previously identified: the class I YSO V723 Car \citep*{tapia15} in NGC~3372 and several distant embedded 
candidates identified by \citet{kumar16} from the sample of CP17a as (usually) the most luminous YSOs in their respective SFRs. In addition, a very luminous YSO
undergoing episodic accretion, S255IR NIRS3, was recently found by \citet{caratti16}.

We also plotted amplitude against luminosity (not shown) but found no correlation between these two variables. This suggests that in YSOs similar processes occur to cause high amplitude variability across the measured range of luminosities.

\begin{table}
	\centering
	\caption{Expected proportions of YSO light curve types as a fn. of UGPS amplitude. Each column is normalised to $\sim$100\%}
	\label{tab:proportions}
	\begin{tabular}{lcccc}
		\hline
		Light Curve Type & \multicolumn{4}{c}{$\Delta K$ (mag)}\\
		                            &               >1     &    >1.5    &     >2    &   1.0--1.5\\ \hline
		Short term variable	&    11\%  & 4\%  &  2\%   & 15\%      \\
		Eruptive 			& 36\% & 38\%  &  36\%     &  34\%       \\
		Fader			& 24\% & 32\%  &  40\%     &  19\%      \\
		Long term periodic variable & 15\% & 13\%  &  14\%  & 17\%      \\
		Dippers			& 12\% & 13\%  &  8\%   &  12\%     \\
		EBs				& 1\% & 0\%  &  0\%    &   2\%  \\   \hline
	\end{tabular}
\end{table}

\subsubsection{Expected light curve types}

In Paper I we used the two year $K_{s}$ time series dataset of \citet{carpenter01} for the Orion Nebula Cluster to show that high amplitude YSOs discovered by our two 
epoch search should mostly have long term variability on timescales of years rather than short term variability on timescales of days or weeks. The recent CP17a sample of
441 high amplitude YSO candidates allows us to revisit that calculation with a longer time series (4 years), better sampling of long time baselines and far more high amplitude 
variables. We randomly sampled 10 000 pairs of points from all the CP17a light curves in each of the YSO light curve categories defined therein (short-term variable, eruptive, 
fader, dipper, long period variable YSO and EB), with a minimum time baseline of 1.8 yrs. We find that short term variables should constitute only 11\% of a high amplitude 
($\Delta K>1$) YSO sample selected using two widely separate epochs. This fraction falls to 4\% if a $\Delta K>1.5$ threshold is used, and 2\% above a $\Delta K>2$ threshold. 
These results should be applicable to the YSOs in the present UGPS catalogue, allowing for minor caveats such as a small amount of contamination by non-YSOs
in the CP17a YSO sample and the fact that UGPS time baselines range from 1.8 to 6.3 years. We conclude that variability in the overwhelming majority of YSOs in our catalogue 
is predominantly on timescales of years.  The proportions in the different light curve categories given by our sampling experiment are listed in Table \ref{tab:proportions} for various amplitude thresholds
and ranges. We see that eruptive light curves and faders should dominate the UGPS sample, especially at higher amplitudes. Eruptive light curves are thought to be 
caused by episodic accretion in most cases (see CP17b). The same may also be true for the faders and long period YSOs, though evidence
in CP17b was more limited for those light curve types.

Note that the table does not include the effects of contamination by non-YSOs projected against star forming regions, such as pulsating AGB stars, AGN, CVs and EBs
(see section 3.3). While we have attempted to identify most of AGB stars in the sample via colour selection (see section 3.4.1) a small number will remain. Late M-type AGB 
stars can have infrared spectra that somewhat resemble those of FUors and MNors (e.g. CP17b) so light curves may be required to identify them. 
Similarly, the proportions of EBs listed in the table are based on the numbers projected against SFRs in the study of CP17a, neglecting the fact that not all of these
would be YSOs.


\begin{figure}
	\includegraphics[width=\columnwidth]{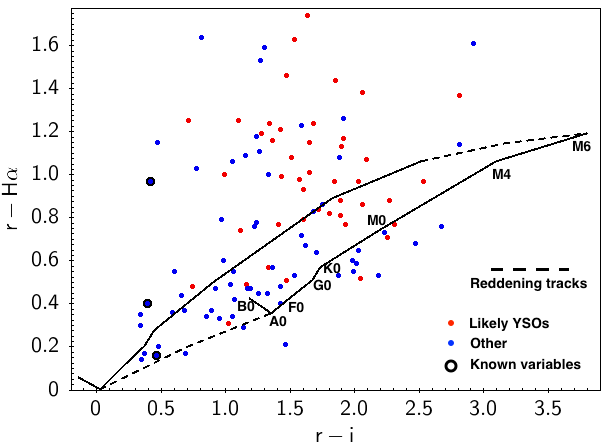}
   \caption{IPHAS $r - $H$\alpha$ vs. $r-i$ two colour diagram. Solid lines show the main sequence tracks from B0 to M6, with
   zero reddening (left track) and {\it E(B-V)}=2 (right track). Red points are likely YSOs, most of which show clear evidence for
   H$\alpha$ emission, based on location above both tracks. Blue points are other variables, almost half of which are also clearly H$\alpha$ emitters.
   The three known variables plotted are enclosed large black circles: the central star of the Necklace nebula, dwarf nova EG Lac and Nova Cyg 2008,
   in descending order of $r-$H$\alpha$. The lower and upper dotted lines are the reddening tracks for A0V and M6V stars respectively.}
    \label{fig:iphas}
\end{figure}

\subsubsection{H$\alpha$ emission}

In figure \ref{fig:iphas} we plot the IPHAS $r$-H$\alpha$ vs. $r$-$i$ two colour diagram for the 108 stars detected in all three optical passbands of the IPHAS 
2nd Data Release. Main sequence tracks are overlaid for zero extinction (left track) and for {\it E(B-V)}=2 (right track) using data from \citet{drew05}. 
Stars located above the zero reddening track are identified with high confidence as H$\alpha$ emitters, given that the photometric uncertainties are nearly
always smaller than the distance above the track. Three known variables are marked with large black points: the central star of the Necklace nebula,  
dwarf nova EG Lac and Nova Cyg 2008, in descending order of $r-$H$\alpha$. A further ten sources are detected in all three passbands of the VPHAS$+$
survey. These are not plotted in figure \ref{fig:iphas} because the $r$, $i$, and H$\alpha$ passbands are sufficiently different to shift the main sequence tracks
appreciably \citep{drew14}.

At least 64\% of the likely YSOs in the plot (red points) are H$\alpha$ emitters, with equivalent widths up to 250\AA~ (based on comparison with figure 6 of
\citealt{drew05}). All of the other YSOs may also be H$\alpha$ 
emitters (if we allow for extinction up to {\it E(B-V)}=3) but this cannot be ascertained from a 2D plot in which locations are determined by three variables (spectral type, extinction
and H$\alpha$ emission). Interestingly, almost half 
(41\%) of the sources not classified as YSOs or known variable stars (blue points) are also clearly identified as H$\alpha$ emitters. Inspection of the near infrared two 
colour diagram for these sources (not shown) indicates that H$\alpha$ emitters with $r$-H$\alpha > 1$ typically have locations consistent with reddened YSOs, 
i.e. they deredden to the red end of the classical T Tauri locus (CTTS) or they appear in the "protostar" region with large $H$-$K$ colour excesses.
Moreover, the subset of these sources with W1, W2 and W3 detections (eight objects) are all located in the YSO selection region of the W1-W2 vs. W2-W3 plot (figure \ref{fig:X}). It therefore seems very likely that many of these are YSOs with insufficient adjacent indicators of star formation to pass our selection.
We note that the IPHAS and near infrared colours of the more common H$\alpha$ emitting populations are illustrated in \citet{corradi08}.

The bluer H$\alpha$ emitters (with $r$-H$\alpha < 1$ and $r$-$i<1.5$) are typically located below the CTTS locus in figure \ref{fig:X}, though many of them have redder
$H$-$K$ colours than would be expected for normal main sequence stars or giant stars. Some of these blue H$\alpha$ emitters are faint infrared sources
with similar optical/infrared colours and magnitudes to the dwarf nova EG Lac. Most CVs are H$\alpha$ emitters \citep{witham08} and, as noted in section 3.4.4, they are one of the three main populations expected to contribute faint blue
variables to the catalogue. In support of this, we see that the two bluest unidentified H$\alpha$ emitting variable stars (sources 359 and 445, with $r-i<0.4$) both have unusually high 
amplitudes ($\Delta K_{all}=$2.84, 2.30, respectively) which supports a CV interpretation. Source 127, a VPHAS$+$ detection not plotted in figure \ref{fig:iphas},
also has blue colours ($r-i=0.39$, $J-K=0.04$), H$\alpha$ emission and a high amplitude $\Delta K_{all}=3.44$, so it can also be considered a candidate CV.

Other blue H$\alpha$ emitters may be unusual sources such as source 366, the binary central star of the Necklace PN. A new discovery of interest is source 363, discussed
separately in section 5.2. Sources lacking clear evidence for H$\alpha$ emission and non-YSOs are often relatively bright and blue
infrared sources (mean UGPS $K<14$, $J-K<2$) whose nature is unclear (see section 3.4.2). However, some of these are red enough to be on the unreddened CTTS locus.


\begin{figure}
	\includegraphics[width=\columnwidth]{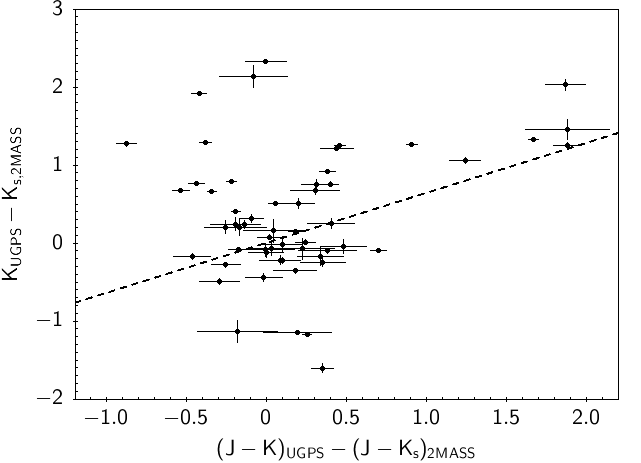}
	\includegraphics[width=\columnwidth]{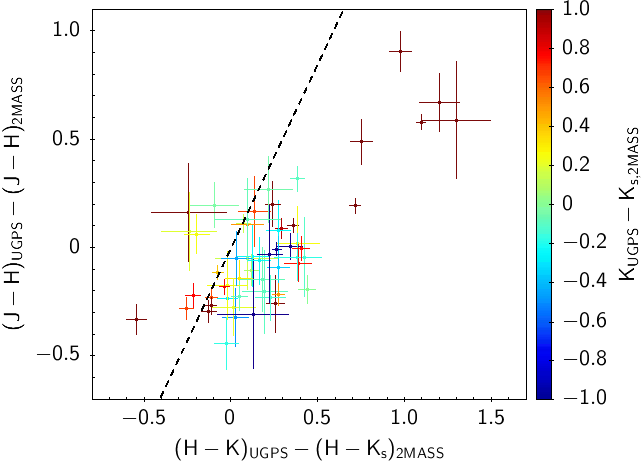}
   \caption{Colour variability. {\it (Upper panel):} plot of change in flux vs. change in colour between UGPS and 2MASS. The 
   reddening track is overlaid as a dashed line. Most sources are fainter in UGPS than 2MASS due to greater survey sensitivity.
   A few sources with large flux changes lie close to the reddening line but most do not. {\it (Lower panel):}  two colour diagram of 
   colour changes between UGPS and 2MASS, with extinction track overlaid and the change in flux colour coded. Sources with a large 
   2MASS:UGPS reddening increase show larger changes in $H-K$ than would be expected from extinction alone. Some sources at the 
   lower left also became fainter while becoming bluer on both axes.}
   \label{fig:colvar}
\end{figure}

\subsubsection{Colour variability}


Some information on the physical origin of the variability can be gleaned from the near infrared colour changes between
UGPS and 2MASS. In figure \ref{fig:colvar} we show these changes for the 56 likely YSOs with $JHK_s$ fluxes from 2MASS that
have "ph\_qual" quality grades A, B or C in all three 2MASS passbands. Most sources are fainter in UGPS than 2MASS
due to greater UGPS sensitivity and the bright limit of the catalogue set by UGPS saturation near mean $K$=11.  A linear regression fit 
shows that, on average, flux changes in the $J$ and $H$ passbands are only slightly greater than the change between $K_s$ and $K$, 
though there is considerable variation for individual sources. Defining $\Delta J$ and $\Delta H$ respectively as $J_{UGPS} - J_{2MASS}$ 
and $H_{UGPS} - H_{2MASS}$, the equations of best fit are $\Delta J = 1.19 \left(K-K_{s} \right)+0.13$  and 
$\Delta H = 1.09 \left(K-K_{s} \right)+0.17$.

Considering sources with a large positive or negative change in $K$ (upper panel), we 
see that only a minority lie close to the reddening track (dashed line). Looking at the lower panel, we see that the six sources
at the upper right (sources that faded between 2MASS and UGPS) have larger changes in $H-K$, relative to $J-H$, than would be 
expected from an increase in reddening alone. These may be cases where an acccretion-driven outburst led to a large reduction in 
extinction at the 2MASS epoch (common in eruptive variables) that then reversed as the outburst faded. In the lower panel the data 
are colour-coded by flux change: we see that some sources close to the reddening track at the lower left actually faded whilst becoming bluer 
on both axes. These correspond to sources at the upper left in the upper panel, far from the reddening track. As noted by \citet{lorenzetti07} 
the wide range in measured colour changes may be due to differing locations and temperatures of the part of the disc that has 
increased in brightness. Those authors found that EXors often become bluer when brighter if the full amplitude of variability is 
observed but a wider range of behaviour is observed if only part of the outburst is sampled.

\begin{figure*}
	\includegraphics[width=\columnwidth]{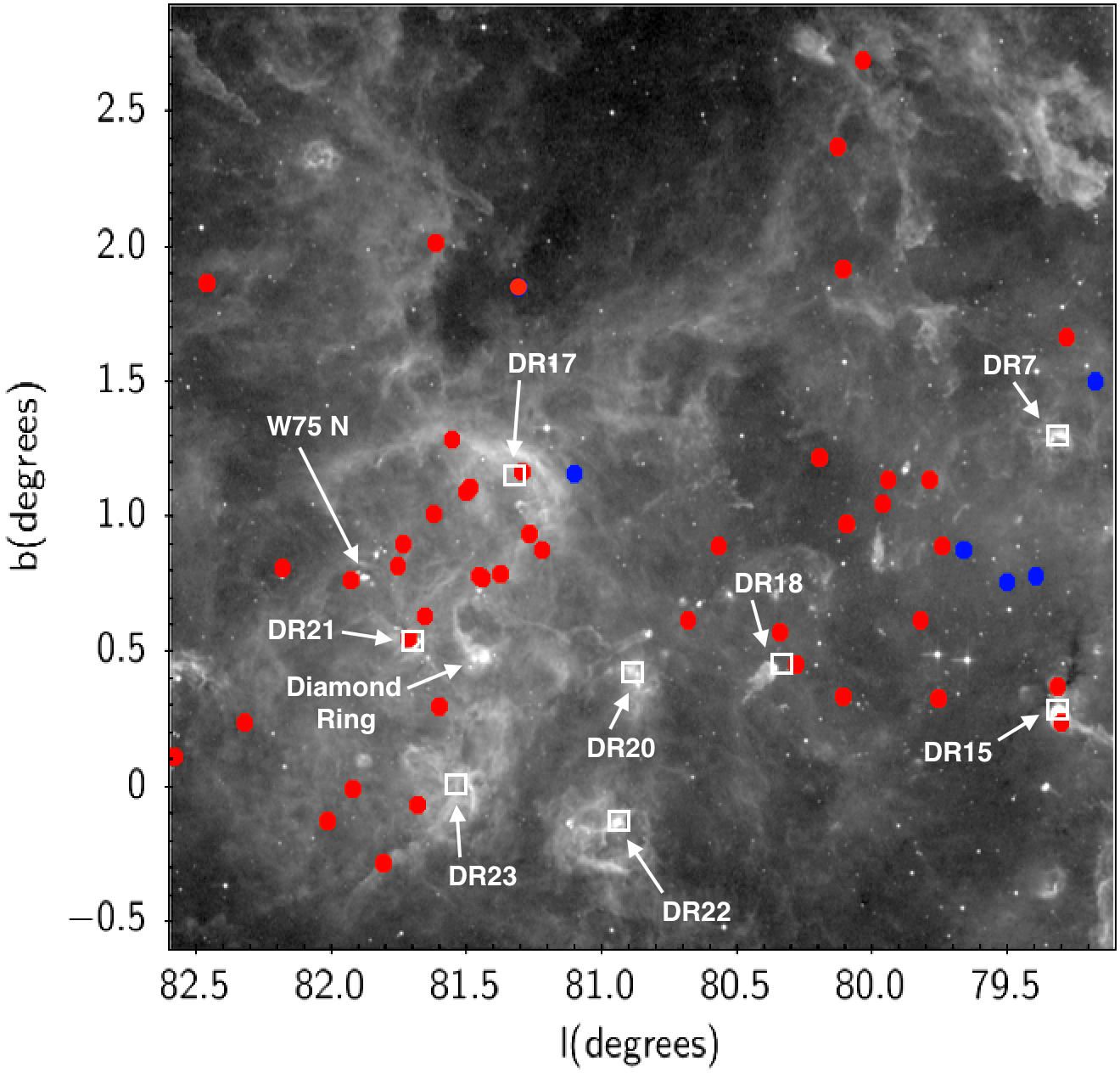}
	\includegraphics[width=1.005\columnwidth]{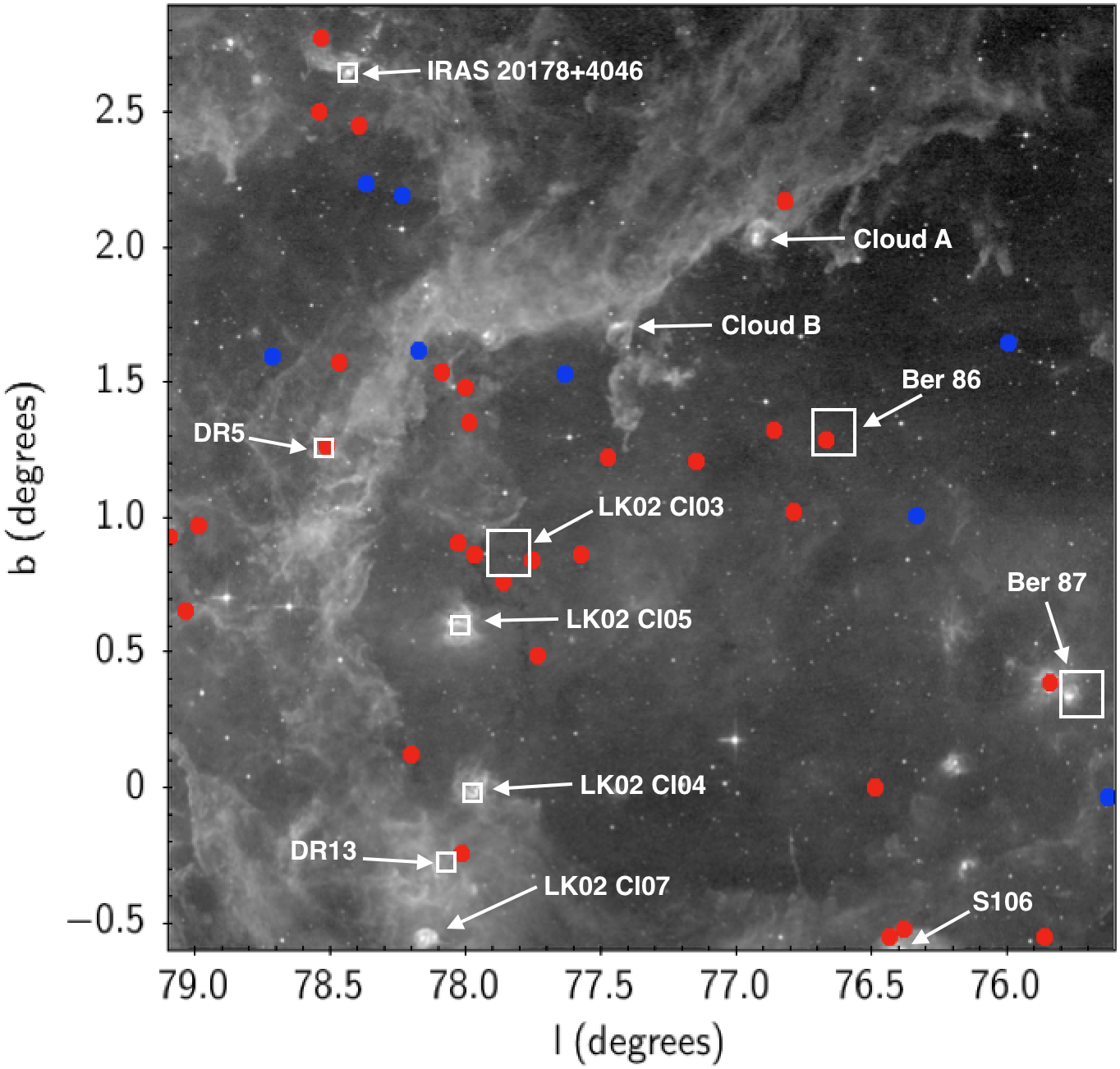}
   \caption{The Cygnus X spatial group. The greyscale background is the 8.3~$\mu$m image data from the MSX satellite. The two panels include almost
   all of the group. Red circles are likely YSOs and blue circles are other sources. Infrared-bright SFRs are labelled,
   along with selected well known HII regions and pre-MS clusters from \citet{downes66} and \citet{leduigou02} that are located close to variable sources. 
   (Clusters and radio sources are marked with squares, with larger squares for the larger clusters). In the left panel we see that most of the likely YSOs are located either in the active 
   region of star formation around W75~N, DR21, DR17, DR23 and the Diamond Ring or within the Cyg OB2 association centred near $l=80^{\circ}$. In the right panel there 
   is some concentration of likely YSOs near the LK02 Cl03 cluster and near the upper left portion of the large IR-bright ring that extends from Clouds A and B through DR5,
   DR13 and down to S106 (see  \citealt{schneider07}). However, some likely YSOs are associated with smaller SFRs that are not apparent on this scale.
} 
   \label{fig:cygx}
\end{figure*}

\subsection{Spatial groups}

In figure \ref{fig:cov} we show the spatial distribution of the likely YSOs and indicate the locations of over-densities that can be identified with particular well-studied star forming complexes.
Several over-densities are apparent. Below we discuss the well-studied regions containing at least four likely YSOs, except the group in the Serpens OB2 association that was 
described in Paper I. We also discuss the set of three likely YSOs possibly associated with the massive W43 complex near $l=31^{\circ}$. Another set of 3 likely YSOs (sources 57--59) is located in the S241 HII region. These three all have relatively low amplitudes ($\Delta K_{all}<1.65$).

An apparent over-density in the outer galaxy near $l=217^{\circ}$ is spatially correlated with two separate little-studied groups of molecular clouds, one at $d \approx$~2--3~kpc in the Perseus arm and another at $d \approx$~6--7~kpc in the Norma-Cygnus arm, see \citet{kaminski07}. Another overdensity of seven sources at $l$=171--174$^{\circ}$, $b \approx +2.5^{\circ}$ corresponds to a large, fairly bright $^{13}$CO emission feature visible in the low resolution maps of \citet*{dame01}, with a velocity of -17 to -18 km/s. Two sources in this "L171-174 region" are in the vicinity of the S231/S232/S233/S235 group of HII regions at $l\sim$~173--174$^{\circ}$ and $d$=1.8~kpc \citep{evans81}; two more at $l\sim$~171--172$^{\circ}$ are associated with other molecular clouds in Auriga that appear to be at a similar distance \citep{liu14}. While the CO emission breaks down into separate sections in higher resolution maps \citep{kawamura98} it may be that the seven variable sources in the L171-174 region are part of a single large star formation complex. We leave such discussion to detailed studies of large scale structures, e.g. \citet*{kiss04}, \citet{liu14}.

\subsubsection{Cygnus X}
A large, spatially extended group is located at $75.6<l<82.6^{\circ}$, $-1.5<b<3.8^{\circ}$ corresponding to the massive Cygnus X star forming complex at $d=1.4$~kpc 
\citep{rygl12}. This region contains 79 likely YSOs and 14 other variables, some of which were listed in Paper I and \citet{cp15}. Most of the other variables are relatively faint 
and unusually red stars (e.g. in the IRAC I1-I2 colour provided by the {\it Spitzer} Cygnus X project) and some are therefore quite likely to be YSOs with too few adjacent star formation 
indicators to meet our selection criteria. Figure \ref{fig:cygx} shows the locations of the variable stars in this region superimposed on Band A (8.3~$\mu$m) images from the Midcourse Space Experiment (MSX, \citep{egan98})\footnote{The MSX data are shown as two images with slightly different contrasts due to limits on the gif-format image sizes available at the NASA Infrared Survey Archive}.
The major sites of active star formation are labelled (following \citealt{schneider06} and \citealt{schneider07}), along with well studied HII regions and any pre-MS clusters located near individual variables. We see that most of the likely YSOs (red points) are located either near these star formation indicators or in the Cygnus OB2 association at $l=79$--$81^{\circ}$ or in bright mid-infrared nebulosity (attributable to hot dust or PAH emission in photo-dissociation regions). 

Projected within the bounds of the Cygnus X complex are the more distant region AFGL~2590 at $d=3.3$~kpc \citep{rygl12} and separately the HII region [WC89] 075.83+0.40 at $d=5.5$~kpc
\citep{wood89}. Only one variable YSO appears to be associated with each of these regions: sources 486 and 450 respectively. Chance projections of other more distant YSOs against the extensive
Cygnus X complex are of course possible. Many of the 77 likely YSOs that we have associated with the Cygnus X group in column 24 of the catalogue were already identified as YSOs, e.g. 20 stars from the catalogue of \citet{kryukova14} and two in the vicinity of the S106 HII region from the catalogue of \citet{gutermuth09}.

While we have adopted a common distance of 1.4~kpc for all other likely YSOs in Cygnus X, we note that \citet{maia16} have recently disputed this (see also \citealt{gottschalk12}), finding that star formation towards this complex is split into layers at $d \sim 0.7$, 1.5 and 3~kpc. Some care regarding this thorny issue is therefore warranted for future studies of these variable stars, though we note that clusters in the sample of \citet{maia16} have some spread in age and the older ($\sim$10~Myr) clusters are less likely to host the class I and flat spectrum systems that dominate the YSO sample. The forthcoming 2nd Data Release of the GAIA mission will of course assist with this and many similar issues concerning other relatively nearby star formation regions.

Ten stars were detected in all three passbands of the IPHAS survey: 9/10 show clear evidence of H$\alpha$ emission due to location above the main sequence tracks plotted in figure \ref{fig:iphas}. Separately, five stars (sources 40, 80, 93, 109 and 110) are identified as molecular outflow driving sources in Makin \& Froebrich (submitted) utilising data from
the UWISH2 survey \citep{froebrich11}.
The amplitudes and mean UGPS $K$ magnitudes of the 77 likely YSOs have a similar distribution to that of the full YSO sample (shown in figures \ref{fig:hist_ampl} and \ref{fig:maghist}). 

Two sources have amplitudes above 2.5 mag: source 537 (=GPSV35 from Paper I) and source 470. Source 537 has $\Delta K_{all}$=2.81. It faded from $K_{s}$=14.04 in June 1998 (2MASS data) to $K$=15.06 in June 2006 and then faded further to $K$=16.84 in June 2009. It was undetected in the $J$ and $H$ images (from June 2009) but the WISE and IRAC data both show that it is a very red source. (W1-W2=2.11, spectral index $\alpha$=1.16 and I1-I2=1.76). 
Source 537 is one of two variables located in the recently discovered embedded pre-MS cluster [SUH2012] G079.852-01.507 \citep{solin12}, located within the LDN 896 dark cloud. This cluster was identified in the UGPS dataset and it is very obvious in the WISE three colour images, where an HII region can be seen at the northeastern end of the cluster. The other variable in this cluster, source 538, is similarly faint and red but has a lower amplitude ($\Delta K$=1.16) and was undetected by 2MASS.

The other very high amplitude variable, source 470, has $\Delta K$=2.64. It rose from $K$=18.51 in September 2008 to $K$=15.88 in October 2011. It was undetected by 2MASS.
This source is located in the DR5 HII region \citep{downes66} which includes several YSOs visible within the brightest part of the 12 and 22~$\mu$m nebulosity in the WISE three colour images.

A further eight sources have amplitudes $\Delta K_{all}>2$. However, the highest $H$ band variability between 2MASS and UGPS is seen in source 521, for which $\Delta H=4.18$ and $\Delta K_{all}=1.97$. This source faded from $K_s=11.90$, $H=14.27$ ($H-K_s=2.37$) in November 1998 to $K=13.87$, $H=18.45$ ($H-K=4.58$) in July 2006 before partly recovering to $K=12.44$ in 
October 2010. The change in colour is more than would be expected from a change in extinction alone. The object is located in the dense dust clump BGPS G081.218+00.872 \citep{rosolowsky10}.




\subsubsection{North America and Pelican Nebulae}
Adjacent to the Cygnus X group are 17 likely YSOs at $83< l <85.5^{\circ}$, $-1 \la b \la 1^{\circ}$ associated with the North America nebula (NGC7000) and the Pelican nebula (IC5070), at a distance of only 520--600~pc from the sun (\citealt{laugalys02}; \citealt{laugalys06}; \citealt{guieu09}). One other variable is located in this area, source 560(=MSX6C~ G085.3935+00.1268). We classified this very bright mid-infrared source (W2 = 4.69, W4 = -0.93) as a likely AGB star via the W3-W4 vs. W1-W2 diagram (see section 3.4.1) and it was 
previously selected as a candidate AGB or post-AGB star by \citet{yung14} on the basis of {\it Akari} colour criteria. Four known highly variable YSOs are located in the region: the
classical FUors V1057 Cyg and V1515 Cyg \cite{herbig77}, the unusual eruptive variable HBC~722 (=V2493~Cyg, \citealt{semkov10}) and VSX~J205126.1+440523 (=V2492~Cyg,
\citealt{itagaki10}; \citealt{covey11}), a source that may have both variable extinction and variable accretion \citep{kospal11}. These four YSOs are all too luminous to be included in our
catalogue owing to saturation.



\begin{figure*}
	\includegraphics[width=1.6\columnwidth]{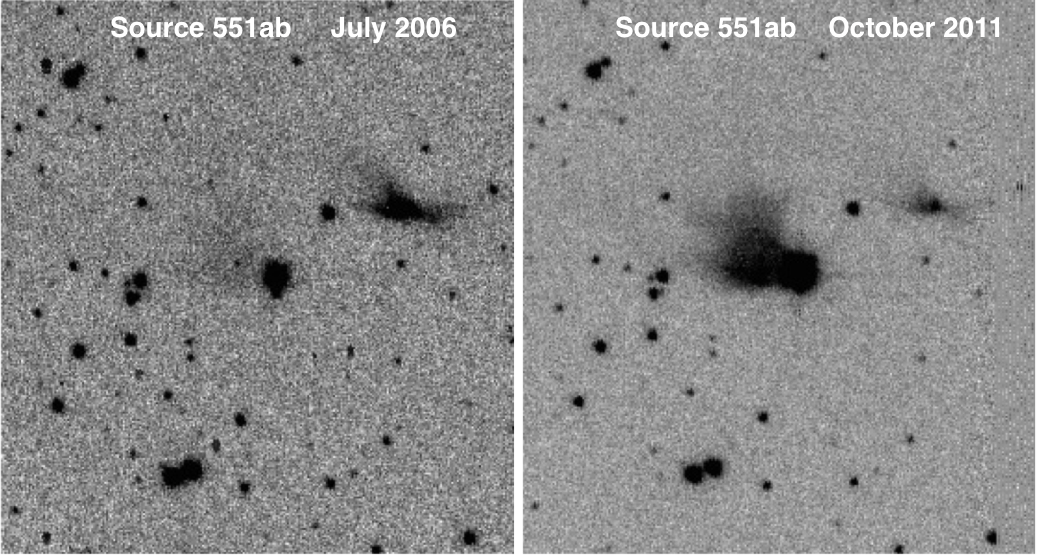}
   \caption{UGPS $K$ images of source 551ab (=2MASS J20500940+4426522). This pair of highly variable YSOs within the North America nebula have unusually extensive nebulosity and
   they may be a bound system. The primary at the centre is the only variable YSO in the catalogue that was identified as an eruptive YSO candidate prior to our UGPS search, see 
   \citet{scholz13}. The images are 1 arcminute across and they have conventional equatorial orientation.}
   \label{fig:551}
\end{figure*}

This region contains two new sources with amplitudes above 2.5 mag, sources 551 and 558. Source 551, with $\Delta K_{all}$=3.62, is the central star of a cometary nebula that appears much brighter at the second UGPS epoch than the first, see figure \ref{fig:551}. As noted earlier, it was previously identified by \citet{scholz13} as a candidate eruptive variable YSO via two-epoch photometry from {\it Spitzer}/IRAC and WISE: it had amplitudes 1.70 and 2.00 mag in the $\sim$3.5~$\mu$m and $\sim$4.5~$\mu$m passbands respectively. The central point
source rose in brightness by 1.6~mag from $K_{s}$=14.46 in June 1998 (2MASS data) to $K$=12.86 
in July 2006 and then rose by a further 2.0~mag to $K \approx 10.83$ 
in October 2011. The 2nd UGPS epoch has an uncertainty of 0.1 mag in the point source flux due to the combination of saturation and the (relatively faint) surrounding nebulosity.  
Cometary nebulae in YSOs are interpreted as reflection nebulae caused by light scattering on paths through a low density cavity in the circumstellar envelope that has been cleared
by an outflow. Scattering may occur at the walls of the cavity (e.g. \citealt{whitney93}) or in the dusty outflow itself \citep{lucas97}. Assuming the variability of the point
source is due to episodic accretion,
the large rise in brightness of the cometary nebula may be due both to the increased flux from the central star and a reduction in extinction on paths from the star to the cavity during
the outburst. The central point source became redder as it became brighter ($H-K_{s}=1.55$ in 2MASS in 1998, $H-K=1.82$ in UGPS in 2006) so its variability cannot be explained
as a simple reduction in extinction.

Intriguingly, there is a second spatially extended and highly variable YSO, UGPS J205008.1+442659.9, just 16$\arcsec$ northwest of source 551. In UGPS images, this YSO displays a bipolar nebula around a relatively faint point source. It appeared much brighter than in 2MASS ($K_s$=14.19 in June 1998) than in UGPS ($K$=15.26 in 2006, $K$=15.71 in 2011).
N.B. We give UGPS point source magnitudes in the default 2$\arcsec$ diameter aperture, which should include a contribution from the surrounding nebula in a broadly similar
manner to the 2MASS point source flux, which is based on data with $\sim$2$\arcsec$~spatial resolution. \footnote{Photometry from both "filtered" and "unfiltered" images are available 
for this source in the WSA. In filtered images, extended nebulosity on scales larger than 5--10$\arcsec$ is removed to aid point source photometry. For this source we give unfiltered
magnitudes in order to better compare with 2MASS but effect of filtering was small in any case (0.04 mag).} 
The large change in flux is not in doubt because UGPS~J205008.1+442659.9 is clearly brighter than source 551 in 2MASS images, whereas the reverse is true in the UGPS images.
UGPS J205008.1+442659.9 is not in our variable catalogue because of the smaller variation in UGPS and the relative faintness of the point source compared to the surrounding nebula, which led to a profile $class=+1$ (i.e. resolved) at both epochs.

Source 551 and UGPS J205008.1+442659.9 have a projected separation of just 8000--10000~au for the distance range quoted above so it is likely that they are a weakly bound
binary pair that formed at the same time. The companion may therefore be referred to as source 551b. Spatially resolved binaries are likely to be rare in the catalogue, given that most sources are more distant and most binaries have smaller orbital radii. Observation of high amplitude variability in both members of an embedded pair of supports our view that episodic accretion is quite common in embedded YSOs. Source 551 has spectral index $\alpha =0.28$ and while the companion is not included in the WISE or {\it Spitzer}/Cygnus~X source catalogues it is clearly an embedded system.

These two variable YSOs with circumstellar nebulae are in a small minority amongst likely YSOs in the catalogue. \citet{connelley10} found that most class I YSOs with FUor-like spectra
in nearby SFRs tend to display resolved nebulae in conventional ground-based near infrared images. However, our search method required the candidates to have point source image profiles
at the central flux peak, in order to minimise false positives caused by blended stars. The combination of this selection and the large distance to many of the likely YSOs reduces the number
with resolved nebulae in the UGPS images.

The other very high amplitude YSO in this region, source 558 (=2MASS J20524822+4435182) was identified as a YSO by \citet{guieu09} and \citet{rebull11}. It is a point source with $\Delta K_{all}$=2.67 and $\alpha=0.68$ that faded monotonically from $K_s=13.26$ in May 2000 to $K$=14.51 in July 2006 and $K$=15.92 in October 2011. It became much redder as it faded: $J-K_s = 2.27$ in 1998 and $J-K=4.15$ in 2006. These changes are consistent with a change in extinction but the $J-H$ and $H-K$ colours became redder by similar amounts. This is probably not
consistent with a change in extinction alone, though we caution that scattered light or large dust grains could cause colour changes arising from variable extinction to depart from the
interstellar reddening law.


\subsubsection{Gemini OB1 molecular cloud}
A spatially extended group of 15 likely YSOs is located in the mid-plane at $188<l<195^{\circ}$, coincident with the Gemini OB1 molecular cloud at $d=2$~kpc (\citealt{carpenter95a}; \citealt{carpenter95b}). A further four faint red variable stars in the region (sources 68, 69, 71, 76) are also plausible YSO candidates, although there is less evidence within 5$\arcmin$ of each of these sources. The Gemini OB1 molecular cloud complex is centred on the S254--S258 HII regions at $l$=192.5--192.7$^{\circ}$ (see \citealt{chavarria08}) and four high amplitude YSOs are located in that vicinity (sources 67, 72, 73 and 75).
None of the 15 sources classified as YSOs in our catalogue have amplitudes higher than $\Delta K_{all} = 1.75$ but one of the other four YSO candidates, source 68, has $\Delta K = 2.08$. These
19 sources are all relatively faint (mean $K>15$) except for sources 61 and 62, which each have mean $K<13$. Source 61 shows a large H$\alpha$ excess in the IPHAS $r$-H$\alpha$ vs. $r-i$ two colour diagram.

\subsubsection{Rosette complex}
A somewhat more compact group of 10 likely YSOs at $206.2 < l < 207.3^{\circ}$, $-2.6 < b < -1.7^{\circ}$ is associated with the Rosette complex at $d \approx 1.6$~kpc  (e.g. \citealt{perez87}). Only three of these (sources 85, 86 and 88) are located within 10~arcminutes of NGC~2244, the ionising cluster of the Rosette Nebula. The rest are spread out to the south and east 
amongst the other embedded clusters, $^{13}$CO clouds and dark clouds that define the complex, see \citet{romanzuniga08}; \citet{phelps97}; \citet{poulton08}; \citet{blitz86}; \citet{cambresy13}. The REFL08 cluster hosts two variables (sources 93 and 94) and a third (source 95) is located amongst the extensive group of YSOs with near infrared excess that surrounds the REFL08, PL04 and PL05 clusters (\citealt{romanzuniga08}; \citealt{phelps97}). This larger group corresponds approximately to the "cluster E" identified by \citet{poulton08} using {\it Spitzer} data.
Of the other four variables, one (source 91) is associated with the PL02 cluster \citep{phelps97}, and another (source 87) is in the small [SUH2012] G207.312-02.538 cluster identified by \citet{solin12}. The other two are somewhat removed from any of the YSO groups and clusters in the region. Half of the 10 Rosette variables are relatively bright, with mean UGPS $K$=12.5--14.5. 

Source 94 in the REFL08 cluster has the highest amplitude, $\Delta K_{all}=3.17$. It was previously identified as a likely class I or stage I YSO (WISE J063419.49+041747.9) by \citet{cambresy13} using WISE colours, and independently by the MYStIX team (\citealt{povich13}; \citealt{broos13}) via near to mid-infrared excess.
It rose by 1.32~mag from $K_s=13.45$ in November 1999 (2MASS data) to $K=12.13$ in November 2007 and then fell by 3.17~mag to $K=15.30$ in March 2012. While there is no 2MASS $J$ detection, the $H-K_s$ and $H-K$ colours agree very closely between the 1999 2MASS epoch and the 2007 UGPS first epoch (values of 1.95 and 1.99 respectively).
This would appear to rule out reduced extinction as an explanation for the rise in flux.

Only one likely YSO, source 88 (=CXOU J063211.9+050030), is detected in all three of the IPHAS filters: this source ($\Delta K_{all}=1.79$) shows a large H$\alpha$ excess in the IPHAS $r$ $-$ H$\alpha$ vs. $r-i$ two colour diagram. This is one of two sources in the NGC~2244 cluster that were detected in the X-ray catalogue of \citet{wang08}. Source 88 was clearly detected in both the 0.5-8~keV and 2.5-8~keV ranges, whereas the other (source 89 = CXOU J063205.98+045334.4) had only a 2$\sigma$ detection.

\subsubsection{W51}
A group of nine likely YSOs is associated with the W51 star forming complex at at $d \approx 5.6$~kpc (e.g. \citealt{parsons11}). Owing to the large distance this group
is compact, the sources all lying within the area $48.72 < l < 49.45^{\circ}$, $-0.39 < b < -0.18^{\circ}$. They are spread out in longitude along the central spine of this massive 
complex where most of the class I and class 0 protostars are located (\citealt{kang09}; \citealt{kang10}). None of them are located in the cluster of low mass YSOs discovered
by those authors on the Galactic equator at $l=49.4$, $b=0.0$. The nine sources are relatively faint, as might be expected from the large distance, with 8/9 having
mean UGPS $K$>15; source 293 was somewhat brighter at all three of the UGPS and 2MASS epochs. None of the nine show variations above 2.5~mag but two stars (sources 296 and 
297) have $\Delta K_{all} \approx 2.2$. None were detected by the IPHAS survey, which is unsurprising given the substantial extinction toward the complex ($A_V \ga 10$, see \citealt{kang10}).

Two additional YSO candidates, sources 286 and 291, were recently identified about 0.5$^{\circ}$ from the complex by \citet{saral17}, see section 3.3. While these relatively isolated sources did not pass our YSO selection, they are still candidates for association with the region.

\subsubsection{NGC 2264 / the Cone Nebula}
A group of six likely YSOs is associated with the NGC~2264 cluster (the region of the Cone nebula) at $d = 740$~pc \citep{kamezaki14}. These six are located within the area $202.6 < l < 204.1^{\circ}$, $1.7<b<2.4^{\circ}$, aligned along the north-south axis that connects the Cone nebula and S Mon. All six are relatively bright (mean UGPS $K<15.1$) but five have amplitudes 
$\Delta K_{all}<1.5$~mag. Source 106 has $\Delta K_{all}=2.03$ and $\Delta J=3.90$, the latter being the largest change in $J$ amongst all the likely YSOs in the catalogue.
This system fell from $K_s =14.08$ in 1998 to $K=16.11$ in early April 2011 and recovered to $K=14.10$ in late March 2013. The source became much redder in $H-K$ as it faded between 1998 and 2011, far more than would be expected from variable extinction: $H-K_s=0.46$ and $J-H=0.6$ in 1998, rising to $H-K_s=1.66$ and $J-H=1.27$ in 2011.
The falls in flux in $J$ and $H$ were 3.90~mag and 3.23~mag, respectively. 

The spectral indices are in the range $\alpha$=0.0--0.4 for four of the six sources; sources 103 and 106 were not detected in W4 but their colours W1-W3 = 4.45 and 4.24, respectively, also suggest class I or flat-spectrum systems, given that W1-W3 and $\alpha$ are well correlated amongst the likely YSOs in the catalogue. The IPHAS data and previous studies (\citealt{reipurth04}; \citealt{dahm05}) show that four sources are H$\alpha$ emitters: sources 106, 100 (=ESO-HA~400),  103(=ESO HA~414 = NGC 2264 DS 159), and 101(=CSIMon-000959). The first three of these have 
$r-$H$\alpha$ above the unreddened main sequence locus in figure \ref{fig:iphas}. The last, source 101, is one of the reddest CTTS in the optical study of \citet{venuti14}, with an estimated spectral type M2 and a bolometric luminosity of 0.34~$L\sun$. Adopting the M2 type allows us to confirm H$\alpha$ emission despite the lower $r-$H$\alpha$ value of 0.77. We estimate an H$\alpha$ equivalent width of $\sim$10~\AA~ from figure 6 of \citet{drew05}. Given the faint $u$ band magnitude ($u \approx 23.3$) the accretion estimates reported by \citet{venuti14} should be treated with caution for this star. Our computed spectral index, $\alpha$=0.11, is consistent with a flat-spectrum CTTS.



\subsubsection{S86 HII region}
A compact group of seven likely YSOs at $59.1 < l < 59.8^{\circ}$, $-0.2 < b < 0.2^{\circ}$ is associated with the large S86 HII region \citep{billot10} at $d \approx 2.3$~kpc \citep{chapin08}. This is
part of the Vul OB1 association. Three of these stars (sources 360, 362, 364) are in the vicinity of the well-studied NGC~6823 cluster (e.g. \citealt{riaz12}; \citealt{xu12}) which 
has an age of $\sim$3~Myr. Source 358 is in the pre-MS cluster Collinder 404 (= far IR cluster [BSP 2011] 19, see \citealt{bica08}; \citealt{billot11}. Three other objects (sources 355, 357 and 361) are located on the outskirts  of the HII region. Source 361 is furthest from NGC~6823, in the massive SFR IRAS~19410+2336 (=cluster G3CC 73), see \citealt{qiu08}) This cluster has an independently determined maser parallax distance $d=2.2$~kpc \citep{xu09} so it seems certain that it is part of the same complex, in which star formation is thought to have been triggered by the G59.5+0.1 supernova (\citealt{billot10}; \citealt*{taylor92}).

Sources 355, 362 and 364 have relatively high amplitudes, $\Delta K_{all}$ = 2.17, 2.08 and 1.98, respectively. Source 355 rose from $K$=16.69 in August 2008 to $K$=14.52 in September 2011
and was undetected in 2MASS and WISE and MIPSGAL. The GLIMPSE colour, [3.6]-[4.5]=0.39, suggests a class II YSO. Source 362 is a fairly bright flat spectrum system ($\alpha$=0.16) that rose from $K_s$=13.95 in 1998 in 2MASS to $K$=11.87 at the first UGPS epoch in August 2008 and then fell back to $K$=13.69 by the second UGPS epoch in September 2011. The (H-K) colour changed only slightly from 1.43 to 1.58 (a 1$\sigma$ reddening) between 1998 and 2007 despite the $\sim$2 mag flux increase, which indicates that the variability was not caused by a reduction in 
extinction. Source 364 is a faint source ($K$=15.69 in 2007, $K$=17.67 in 2011) that was undetected in 2MASS and WISE and MIPSGAL; the GLIMPSE colour, [3.6]-[4.5]=0.47, suggests a class II YSO.


\subsubsection{S236 / NGC 1893}
A very compact group of four likely YSOs (sources 29--32) is associated with the S236 HII region and NGC~1893 cluster at $d$=3.6~kpc (\citealt{prisinzano11}, see also \citealt{caramazza08}). 
These four are located within the area $173.57 < l < 173.68^{\circ}$, $-1.85 < b < -1.57^{\circ}$. None show variation higher than $\Delta K_s$=1.76 and these are relatively faint sources in the near infrared (mean UGPS $K>14.8$). Spectral indices are available for two objects (sources 29 and 31), both of which have steeply rising SEDs, $\alpha$=0.88 and 1.82 respectively. All four sources were previously identified as YSO candidates by \citet{prisinzano11}. They were undetected in the IPHAS H$\alpha$ data. The additional epoch of $JHK$ photometry from \citet{prisinzano11}
does not significantly increase the amplitude of variation of any of these sources. Similarly, the WISE W1 and W2 fluxes show no large differences from the IRAC I1 and I2 fluxes in \citealt{caramazza08}.




\subsubsection{Possible W43 group}

The W43 complex at $d$=6.5~kpc \citep{morales13} is a very massive SFR but only three likely YSOs (sources 165--167) may be associated with it, located at $30.74 < l < 31.00^{\circ}$, $-0.05 < b <  0.07^{\circ}$. Sightlines in the mid-plane region near $l=31^{\circ}$ pass along the Scutum-Centaurus spiral arm and the region contains several high amplitude sources classified as likely YSOs, so this set of three sources does not stand out as an overdensity. The most noteworthy is source 167 (=SSTGLMC G030.9948-00.0384), which has a very high amplitude ($\Delta K$=3.42) and a very red 
SED ($\alpha=2.02$). It was identified as a red YSO candidate in the \citet{robitaille08} list of red {\it Spitzer} sources and it rose from $K$=18.02 in June 2005 to $K$=14.60 in August 2011.
It was undetected in 2MASS and the UGPS $J$ and $H$ images but it brightened by 1.73 and 1.89~mag at $\lambda \sim 3.5 \mu$m and $\lambda \sim 4.5 \mu$m respectively, between the GLIMPSE epoch in 2003 and WISE data in 2010. Source 167 is associated with the HII region [KB94] 9 at approximate distance $d$=8.0 kpc \citep{kuchar94} but the authors noted the
region may in fact be part of the W43 complex. A similar possible association with W43 applies to source 165 in the HII region [KB94] 8. The last of the 3, source 166, is in the W43 cluster.
A fourth YSO candidate, source 171, was identified in the general vicinity of W43 by \citet{saral17}, though it is a relatively isolated source not included in our selection of likely YSOs, see section 3.3.


\subsection{New pre-MS clusters and groups}

Many of the likely YSOs in the catalogue were identified by spatial association with five or more red sources in the WISE three colour images. While
most such cases involve known clusters or diffuse groups on the outskirts of large SFRs, simple visual inspection of the UGPS and WISE images
indicates that some are substantial pre-MS clusters not identified as such in the SIMBAD database or the papers listed therein. In table \ref{tab:newclus} we list the 18 cases where
variable sources are associated with pre-MS clusters, SFRs or compact stellar groups that appear to be new discoveries. The epoch J2000.0 coordinates
are estimated from visual inspection of the WISE and UGPS images. Many of them are located in little-studied parts of the outer Galaxy. For more details of associated 
mm and submm sources and references for the distances see the individual catalogue entries for each variable source.

\begin{table*}
	\centering
	\caption{New pre-MS clusters associated with individual variable stars}
	\label{tab:newclus}
	\begin{tabular}{lccccl}
		\hline
		Cluster & Source no. & RA & Dec & d/kpc & Comment\\ \hline
		  1     &  3 & 52.517  & 55.772   & & see figure \ref{fig:yso_select}  \\
		  2	&  4 & 52.701 & 54.817    &  & a compact little cluster, members blended in WISE, see figure \ref{fig:yso_select}   \\
		  3	&  13 & 60.935 & 51.451  & 2.5 & a small embedded group near Camargo 441 and S206 HII region (=NGC 1491) \\
		  4	&  36 & 81.735  & 38.751  &   &   \\
		  5    &  45 & 84.094  & 36.661 &   &  \\
		  6	& 62 & 92.091 & 21.349   & 2 & in the Gem OB1 molecular cloud  \\
		  7	& 63 & 92.181 & 31.415   & &  a filamentary SFR   \\ 
		  8    &  66 &  92.849 & 16.549 & 2 & compact cluster in Gem OB1, largely unresolved in WISE\\		  		  
		  9	& 70 & 93.035 & 20.247   & 2 & in the Gem OB1 molecular cloud \\
		  10	&  92 & 98.297 & 8.564   &     &  \\
		  11  &  118 & 104.765 & -4.889  & 2 & connected to pre-MS cluster [FSR2007] 1129 = [KPS2012] MWSC 1042  \\
		  12  &  122 & 104.985 & -3.934  &    &  \\
		  13  &  190 &  283.639 & 7.948  &   & \\
		  14  &  250 &  286.441 & 5.022  &   & an embedded cluster candidate associated with IRDC G39.046-0.877\\
		  15  &  370 &  297.008 & 26.098 &   & a cluster with a tightly-packed core, unresolved in WISE.\\
		  16	&  374 & 297.547 & 23.918   & 0.8  &    \\
		  17  &  469 &  306.415 & 36.985  & 1.4  & in the Cygnus X complex, associated with BGPS G075.861-00.551 mm core.\\
		  18	&  599 &  325.504 & 52.699  &  &  see figure \ref{fig:yso_select} \\  \hline
	\end{tabular}
\end{table*}

\section{Other sources of interest}

\subsection{The highest amplitude sources}

Of the 43 sources in the high amplitude tail with $\Delta K_{all}>2.5$, only 20 are classified as likely YSOs. However, that classification
is given to five of the seven sources with the very highest amplitudes, $\Delta K_{all}>3.5$. We are unable to determine the nature of most of the very 
high amplitude sources not classified as likely YSOs, though some may certainly be relatively isolated YSOs. Below we briefly discuss a few of the 
highest  amplitude systems.

\begin{itemize}
\item Source 282 ($\Delta K_{all} = 4.60$) has the highest amplitude in the catalogue. It is classified as a likely YSO due to location in the HII region 
GRS G048.60+00.20 at $d$=10.6~kpc \citep{kuchar94}, seen clearly as an SFR in the WISE three colour image. The source brightened from
$K$=18.02 in August 2007 to $K$=13.42 in August 2011. It was undetected in all other optical to mid-infrared bandpasses (adjacent bright sources
inhibit detection in the relatively low resolution WISE images).

\item As noted earlier, source 391 (Nova Cyg 2008) has the second highest amplitude, $\Delta K_{all}=4.40$.

\item Source 266 ($\Delta K_{all} = 4.20$) is located in molecular cloud GRSMC 43.30-0.33 at $d$=3~kpc \citep{simon01}. It is unique amongst the
YSOs in the high amplitude tail in having a negative spectral index ($\alpha = -0.31$), corresponding to a class II YSO. It rose very slightly from 
$K_s$=15.06 (2MASS data) in 1999 to $K$=14.86 in April 2008 but then rose to $K$=10.86 in July 2011. The UGPS near infrared colours from
2008 place it in the region of reddened CTTS. It was undetected in IPHAS.

\item Source 570 ($\Delta K_{all} = 3.80$) is an isolated source not classified as a YSO. However, the {\it Spitzer}/IRAC colours (I1-I2 = 1.44) indicate a red 
object with circumstellar matter. This is supported by UGPS near infrared colours consistent with a lightly reddened CTTS or perhaps a more strongly
reddened CV. Source 570 faded from $K=14.70$ in July 2006 to 18.50 in September 2011. It was undetected in IPHAS and the WISE images.

\item Source 133 (=GPSV1, $\Delta K_{all} = 3.75$) is a likely YSO located in the Serpens OB2 association. It was discussed in Paper I as a possible
FUor, based on the large rise in flux and slow rate of decline. CO absorption was not detected in the low quality spectrum presented in that work.

\item Sources 551 and 167 ($\Delta K_{all} = 3.62$ and 3.42 respectively) have been mentioned earlier as members of the North American/Pelican group
and the possible W43 group respectively.

\item Source 190 ($\Delta K_{all} = 3.57$, $\alpha$ = 0.97) is a likely YSO located in a previously uncatalogued group or small cluster of YSOs apparent in 
the WISE three colour images (see below).  Unusually for a class I system, it is detected in all three of the IPHAS optical passbands. However,  the colours 
$r-i=2.27$ and $r-$H$\alpha=0.88$ do not clearly establish an H$\alpha$ excess. It faded from $K$=13.87 in June 2010 to $K$=17.44 in August 2012 and 
it was undetected in 2MASS.

\item Source 127 ($\Delta K_{all} = 3.43$) was mentioned earlier as one of three faint blue H$\alpha$-emitting sources that we regard as candidate CVs. 

\item Among the remaining likely YSOs with amplitudes above 3~mag, only source 615 ($\Delta K_{all} = 3.25$, $\alpha$ = 1.18) is notable as having a well-determined 
distance. It is a member of the embedded pre-MS cluster no. 43 identified by \citet{kumar06}, near the S138 HII region and S138 IR Cluster at $d$=5.1~kpc. 
Source 615 faded from $K$=15.14 in July 2006 to $K$=18.39 in October 2011. It was undetected in 2MASS.

\item Source 156 (= V1360 Aql or IRAS 18432-0149) is notable as the highest amplitude OH/IR star in the catalogue, with $\Delta K_{all} = 3.80$. Other 
previously known variable stars with amplitudes above 3~mag are source 389 ($\Delta K_{all} = 3.20$, mentioned earlier as a likely D-type symbiotic star, and 
source 301 ($\Delta K_{all} = 3.30$) which we classified as a likely carbon star (see section 3.4.1).
\end{itemize}

\subsection{Source 363}

\begin{table}
	\begin{centering}
	\caption{Optical photometry of source 363 (Vega system).}
	\label{tab:363}
	\begin{tabular}{lccccc}
		\hline
		Data & i   &  H$\alpha$   &  r    &     g    &   u   \\ \hline
		IPHAS July 2005 &19.27 & 19.42 & 20.54 & N/A  & N/A\\ 
		IPHAS May 2007 & - & - & 21.14 & N/A & N/A \\
		UVEX July 2010 & N/A & N/A & 19.22 & 20.75 & 20.0: \\
		UVEX August 2011 & N/A & N/A & 19.93 & 20.50 & 19.63 \\ \hline
	\end{tabular}
	\end{centering}
	Note: The IPHAS data were taken in contemporaneous $r$, $i$, H$\alpha$ sets, and UVEX data in $u$, $g$, $r$ contemporaneous sets.
\end{table}

\begin{figure*}
	\includegraphics[width=\columnwidth]{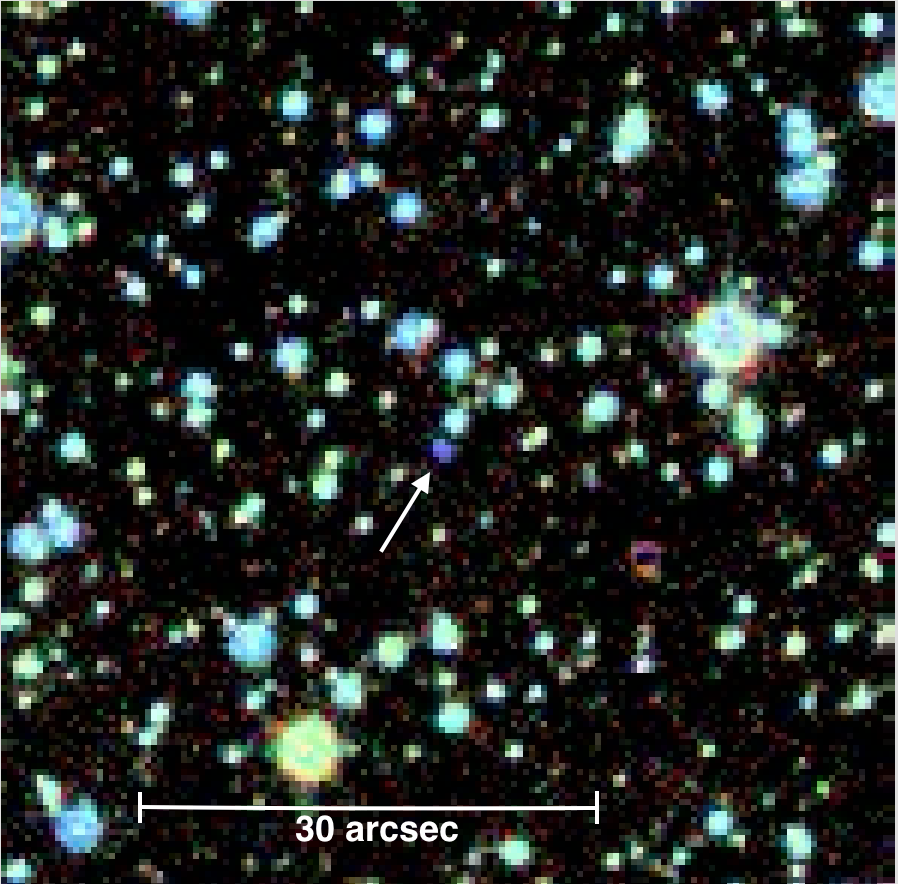}
   \caption{UGPS near infrared three colour image of source 363 (July 2010 data, with red, green and blue represent $K$, $H$ and $J$, respectively). The source is indicated with an 
   arrow and the image has conventional equatorial orientation. The deep blue colour of the object is reflects the extreme measurement $J-H = -0.6$.}
   \label{fig:363}
\end{figure*}

The strangest of the isolated H$\alpha$ emitters is source 363, with near infrared colours $J-H = -0.60$, $H-K = 0.67$, $J-K = 0.07$ measured in July 2010, see figure \ref{fig:363}. This source has a redder $r-i$ colour than the three CV candidates ($r-i = 1.26$, $r-$H$\alpha$ = 1.11 in July 2005) located closer to the YSOs in the IPHAS two colour diagram. It faded by 1.97 mag from $K$ = 15.72 to $K$ = 17.69 between July 2006 and July 2010. At $l=55.4^{\circ}$, $b=-2.5^{\circ}$ it lies outside the {\it Spitzer}/GLIMPSE footprint and the location is blended with adjacent stars at the 6$\arcsec$ resolution of WISE, preventing a mid-infrared detection. It was not quite bright enough to be included in the 2MASS Point Source Catalogue or the 2MASS Point Source Reject Table but a source is visible at the location in the 2MASS $J$, $H$ and $K_s$ images, apparently at an intermediate state somewhat brighter in all three passbands than in 2010. Precise measurement is problematic due to blending with the adjacent source 2$\arcsec$ north and slightly west, measured in UGPS with $J$ = 17.02. However, the peak of the image profile in the 2MASS images is coincident with the UGPS coordinates in all three passbands indicating that $J < 17$ in 2MASS, considerably brighter than the $J$ = 17.76 UGPS datum in 2010. It is therefore significant that the source does not appear unusually blue in a 2MASS three colour image but similar to the other stars in this lightly reddened field with typical colours $J-H$=0.4 to 0.8.

The optical flux in $r$, as measured by the UVEX survey \citep{groot09} and the IPHAS survey, has also shown variation of almost 2~mag, see Table \ref{tab:363}. The photon-noise uncertainties on the optical measurements are under 0.05~mag, though we note that the UVEX 2010 data have a significant calibration uncertainty due to off-axis location of the source in the focal plane. It is clear from blinking the images that the $r$ flux faded substantially between July 2010 and August 2011 while the $u$ and $g$ fluxes either remained constant or increased. The negative UGPS $J-H$ colour is unique in the catalogue, considerably bluer than typical CVs (\citealt{hoard02}, \citealt{corradi08}) or white dwarfs (\citealt{girven11}, \citealt{leggett11}). The very negative $u-g$ colour is also bluer than is typically found in white dwarfs, comparable to an early B-type star \citep{drew14}. Inspection of the UGPS stacked images and individual exposures shows no problem with the measurement and the large changes in $r$ magnitude are obvious in the IPHAS and UVEX images. The $J-K$ colour is also very blue for the field, though the $H-K$ colour is amongst the redder sources present. One possibility is that the apparently extreme $J-H$ colour is an illusion caused by fading of the source in the 7.5~minute interval between the $J$ and $H$ observations. Common varieties of EB are ruled out by the H$\alpha$ emission but interacting binary systems containing compact objects can show large changes in flux on this timescale. E.g. \citet{bogdanov15} describe high amplitude optical variations on timescale of minutes in PSR~J1023+0038, a millisecond pulsar that appears to alternate between accretion disc-dominated LMXB-like behaviour and discless pulsar-like behaviour. The available optical-infrared data do not allow us to determine the nature of this system; time series monitoring and pointed X-ray and radio observations would better constrain the possibilities.

\subsection{The nearest sources}

Several of the likely YSOs in the catalogue are located within 1.5 kpc, close enough for high resolution imaging studies to resolve structures such as envelopes, discs and jets.
This may help to determine the cause of variability, e.g. \citet{liu16}, \citet{dong16}. We have already described the North America/Pelican group at 520--600 pc, the 
NGC~2264/Cone nebula group at $\sim$800 pc and the Cygnus X group at 1.4~kpc. Below we list the other likely YSOs at $d \le 1$~kpc and one source projected in a
mature open cluster.

\begin{itemize}

\item Source 284 ($\Delta K_{all} = 1.02$) is close to the well-studied YSO V1352 Aql (=AS 353 A) and the bright HH32 outflow, located in the Lynds 673 dark cloud. \citep{herbig83} states that this cloud is in front of other molecular clouds in Aquila and gives the distance as 300~pc using a somewhat uncertain method based on velocity dispersion of foreground stars. While this
distance is considered uncertain, source 284 may well be the nearest member of the catalogue. Although the amplitude is relatively low, the IPHAS colours from August 2004 provide clear evidence for H$\alpha$ excess ($r-i =1.51$, $r-$~H$\alpha$ = 1.08), implying active accretion. The near infrared colours became slightly bluer as it faded from $K_s$ = 11.72 in July 1999 (2MASS data) to K=12.49 in August 2007 but they remained consistent with a lightly reddened CTTS. It then rose to K=11.47 in July 2011. It was undetected by WISE.

\item The W40 cloud in the Aquila rift is host to two or three catalogue members. Sources 148 and 149 (GPSV48 and GPSV42 respectively in \citealt{cp15} and Paper I) are located 
within the region studied by \citet{mallick13}. In addition, source 150 (= GPSV43 from Paper I) has mid-infrared WISE colours consistent with a flat spectrum YSO and may well be part of the same region, though it is located almost half a degree south of the centre of the young cluster. It is not classified as a likely YSO in the catalogue due to its relative isolation. This region is at $d \approx 500$~pc \citet{radhakrishnan72}. All three sources have amplitudes $\Delta K_{all} < 1.5$ and all were undetected by IPHAS and VPHAS+ due to their red SEDs.

\item Source 413 ($\Delta K_{all} = 1.40$) is located in the IRAS 20050+2720 pre-MS cluster at $d \approx 700$~pc \citep{guenther12} in the slightly part of this region which those authors referred to as cluster core E. The only detections are the UGPS near infrared data and the source is relatively faint (mean $K=$15.80).

\item Source 605 ($\Delta K_{all} = 1.15$) is a faint blue source ($J-H=0.25$, $H-K=0.1$, mean UGPS $K=15.34$) projected in the outskirts of the $\sim$250 Myr-old cluster NGC 7243 at $d \approx 700$~pc \citep{jilinski03}. It was not detected by WISE. Inspection of the catalogue of \citet{jilinski03} indicates that most sources outside the central part of the cluster are not members. Moreover, source 605 appears to lie below the cluster main sequence on a near infrared CMD. The relatively low amplitude permits an EB interpretation and the blue colours ($J-H=0.25$, $H-K=0.1$) are consistent with this and the fact that 2MASS $JHK_s$ fluxes are fairly consistent with the UGPS $JHK_c$ data (within 2$\sigma$) tends to support this. An AGN or CV interpretation is also possible, see section 3.4.4. Unfortunately the source lies just outside the IPHAS area ($b = -5.24$). Panstarrs/PS1 mean optical colours are consistent with a blue source but they cannot be used to estimate a spectral type without information on the observing dates.

\item Source 374 ($\Delta K_{all} = 2.33$, $\alpha=0.78$) is located in the Planck core PLCKECC G060.75-01.23 ($d=800$~pc). This core is associated with IRAS 19480+2347, seen in the UGPS and WISE images as a new embedded cluster (no.16 in Table \ref{tab:newclus}). Source 374 faded from $K_s$=12.06 in April 2000 (2MASS data) to $K=14.39$ in July 2006 and then recovered to $K=12.37$ in September 2011. The near infrared colours were similar in the 2000 and 2006 observations (becoming slightly bluer in $J-H$ and slightly redder in $H-K$),
indicating that the drop in brightness was not caused by extinction. The source was not detected by IPHAS.

\item Source 555 ($\Delta K_{all} = 1.11$, $\alpha=-0.23$) is located in the [DBY94] 089.2+03.6 molecular cloud ($d=800$~pc, \citealt{dobashi94}). This lightly reddened flat spectrum system had similar colours and brightness ($K \approx 14.4$) in 2MASS in June 2000 and UGPS in September 2011 but was fainter in August 2013. The IPHAS photometry from November 2003 
($r-i = 1.10$, $r -$~H$\alpha$ = 1.25) showed a large H$\alpha$ excess, indicating active accretion.

\item Source 581 ($\Delta K_{all} = 1.17$, $\alpha=0.02$) is located in the [DBY94] 092.9+01.7 molecular cloud ($d=800$~pc, \citealt{dame85}). This bright system had similar colours and fluxes ($K \approx 12.5$) in 2MASS in November 1998 and UGPS in October 2010 but was fainter in October 2012. The IPHAS photometry from July 2005 ($r-i = 2.04$, $r -$~H$\alpha$ = 0.52) are consistent with a highly reddened A-type star with no H$\alpha$ excess at that time.

\item Source 586 ($\Delta K_{all} = 1.19$, $\alpha=0.12$) is located in molecular cloud [DBY94] 090.3-02.3 ($d=800$~pc, \citealt{dobashi94}, specifically in bright nebula [B77] 27, on 
the edge of the Lynds 989 dark cloud. This source is highly reddened but bright in $K$, rising from $K$=13.06 in August 2011 to $K$=11.87 in July 2013.

\item The AFGL 490 region at $d$=900~pc in the outer galaxy hosts two variable YSOs. Source 1 ($\Delta K_{all} = 2.28$, $\alpha=0.94$) is YSO no. 25 in the catalogue of 
\citet{gutermuth09} and source 2 ($\Delta K_{all} = 1.21$) is YSO no. 337 in the same catalogue. The region is further analysed in \citet{masiunas12}. Source 1 rose from
$K_s$=15.26 in December 1998 (2MASS data) to $K$=12.99 in October 2005 and then faded to $K$=14.25 in October 2008. This very red source was undetected in $J$. Source 2 has
near infrared colours indicating a CTTS. It dipped between December 1998 and October 2005 before recovering to its original brightness in October 2008.

\item Source 594 ($\Delta K_{all} = 1.22$, $\alpha=-0.14$) is located in IC 1396A (the Elephant Trunk nebula), a bright-rimmed cloud on the periphery of the giant HII region IC 1396 
ionized by the Trumpler 37 cluster ($d \approx 0.9$~kpc, \citealt{contreras02}). It rose from $K=15.66$ in September 2011 to $K=14.44$ in September 2013.

\item Source 168 ($\Delta K_{all} = 1.41$) is a relatively faint, very red source (mean UGPS $K$=15.28) located in a cluster of far infrared sources, no. 12 in the list of \citet{billot11}, at $d \approx 900$~pc.

\item Source 602 ($\Delta K_{all} = 1.48$) is located in the IC 5146 pre-MS cluster (the Cocoon nebula) at $d$=1~kpc, \citep{harvey08}). Its near infrared colours are consistent with a CTTS. 
It is a relatively faint source (mean UGPS $K$=15.57). 
\end{itemize}

\section{Summary and Conclusions}

We have searched the UGPS two-epoch dataset for highly variable stars across some 1470 deg$^{2}$ of the Galactic plane and compiled a catalogue
of 618 sources with amplitudes $>1$~mag in $K$, nearly all of which are new discoveries.  About 60\% of these sources are YSOs, based on spatial association with SFRs 
at distance ranging from $\sim$300~pc to 10~kpc or more, identified using either the SIMBAD database or inspection of the WISE three colour images. YSO luminosities (determined 
mainly from WISE data taken in 2010) range from 0.1~$L_{\odot}$ to 10$^3$~$L_{\odot}$, though only rarely exceeding 10$^{2.5}$~$L_{\odot}$.

There is also a population of dusty AGB stars with high mass loss rates, typically very red sources at the bright end of the magnitude distribution, some of which are identified here as 
new candidates. Most are likely to be O-rich systems. 
Our WISE-based colour selection of dusty Mira variables differs slightly from the AGB star selection of \citet{koenig14} and it appears to be better for selecting the very reddest 
AGB stars. However, their system is better for bluer, more typical AGB stars, which are generally absent from UGPS due to saturation.

Amongst the fainter sources with relatively blue near infrared colours there may be comparable proportions of CVs, EBs and AGN, as well as rarer variables such as
the binary PN central stars. Within this group we identify three new candidate CVs by their H$\alpha$ emission, high amplitudes (2 to 3.5~mag) and blue colours. There is also an 
H$\alpha$ emitting source of unknown nature with $J-H = -0.6$ and $\Delta K \approx 2$, an outlier with respect to the rest of the catalogue.
A small population of bright blue variable sources of unknown nature is present: these present easy targets for follow-up work.
 
More than half of the likely YSOs in the sample are class I or flat spectrum systems, a higher proportion than in the YSO
population in general (e.g. \citealt{dunham14}). Simulations of two-epoch sampling of the VVV light curves for YSOs from CP17a indicate that variations 
will usually be due to long term flux changes on timescales of years. Many of these can reasonably be expected to be examples of episodic accretion, often
located in relatively nearby SFRs with less foreground extinction than the VVV sample. This is supported by evidence for active accretion from the H$\alpha$
emission detected in many catalogue members by the IPHAS survey. The catalogue should therefore provide a rich sample of eruptive 
variable YSOs for investigation by the community. The substantial groups of variable YSOs in well-studied regions such as Cygnus X, the North America/Pelican 
nebula region, the Rosette complex, the NGC~2264/Cone nebula region and the Gemini OB1 cloud are statistically-valuable samples.
This panoramic two-epoch study complements recent more focussed high cadence studies, e.g. the infrared study of Cygnus OB2 by \citet{roquette17} 
and the optical study of NGC~2264 by \citet{venuti17}.

The amplitude distributions for both the YSOs and the remainder of the sample decline steeply from 1 to 2.5~mag before flattening off, with a low level tail
extending up to 5~mag. However, the much better sampled VVV sample of CP17a does not show this break in behaviour so we caution against over-interpretation,
given the small number statistics at the highest amplitudes and the wide variety of YSO light curves seen in VVV. The higher amplitude YSOs have a redder distribution
of SEDs, as in the VVV sample. While most pre-MS catalogue members have lower amplitudes than classical FUors and EXors, it is only by studying the
full range of amplitudes and timescales attributable to accretion-driven variability that we are likely to come to understand the underlying physical causes. E.g. photometric
monitoring of the variables in this catalogue would allow us to test predictions of unsteady accretion models such as those of \citet{dangelo10} and \citet{dangelo12},
in which cycles of eruption and quiescence can emerge naturally in discs that are truncated close to the co-rotation radius by the stellar magnetosphere. While those
models were proposed as an explanation for EXors, the range of amplitudes and timescales would include the new "MNor" class suggested by CP17b to describe
variations in more embedded systems that vary on slightly longer timescales, often with lower amplitudes than the classical EXors and a diverse mixture of EXor-like
and FUor-like spectra. A difficulty with this model (see \citealt*{hartmann16}) is that the measured accretion rates in EXors do not drop to zero between 
outbursts but it is still worthy of consideration.

The large proportion of YSOs confirms our earlier finding in Paper I and CP17a that they dominate the high amplitude near infrared variable sky on Galactic disc sightlines, 
extending the previous result to a much wider range of longitudes. The observed surface density of variable YSOs is similar to that seen in Paper I and the magnitude 
distribution rises towards our sensitivity limit even in nearby SFRs. This confirms our initial result in Paper I that the mean space density of high amplitude YSOs is higher than that 
of Mira variables, the commonest type of high amplitude variable seen in the optical waveband, see e.g. the General Catalogue of Variable Stars \citep{samus10}.

We note that our selection of point sources will tend to exclude YSOs with bright circumstellar nebulosity, commonly seen in systems with FUor-like spectra
\citep{connelley10}. This may lead to a bias against systems with near edge-on discs, particularly in nearby SFRs, that should be taken account of in follow-up
work. 

We encourage more detailed investigation of the sources in this catalogue by the astronomical community. 



\vspace{1cm}

\section*{Acknowledgements}

This work was supported by the UK Science and Technology Facilities Council (STFC), grant numbers ST/J001333/1, ST/M001008/1 and ST/L001403/1.
We made use of data products from the Cambridge Astronomical Survey Unit and the Wide Field Astronomy Unit at the University of Edinburgh.
L. Smith and C. Contreras Pe\~{n}a were supported by an STFC PhD studentship and a University of Hertfordshire PhD studentship, respectively, in the earlier stages of this research.
Support for DM is provided by the Ministry of Economy, Development, and Tourisms Millennium Science Initiative through grant IC120009, awarded to the Millennium Institute of Astrophysics, MAS. DM is also supported by the Center for Astrophysics and Associated Technologies PFB-06, and Fondecyt Project No. 1130196. This research has made use of the SIMBAD database, operated at CDS, Strasbourg, France. We also used the NASA/ IPAC Infrared Science Archive, operated by the Jet Propulsion Laboratory, California Institute of Technology, under contract with the National Aeronautics and Space Administration, and the SAO/NASA Astrophysics Data System (ADS).




\bibliographystyle{mnras}
\bibliography{extbib} 



\appendix

\section{Details of the Variable Star Search}
\label{A}


%

\subsection{Stage 1: searches of SQL data tables}

The 45 variables in Paper I and the 26 variables in \citet{cp15} were found by searching the 5th, 7th and 8th 
UKIDSS data releases. In each release we selected candidates for visual inspection having $\Delta K \ge 1$~mag 
in the two epochs of $K$ photometry given in the $gpsSource$ table as $k\_1apermag3$ and $k\_2apermag3$, 
which provide photometry in the default 2\arcsec diameter aperture. We selected only stars having $K<16$~mag in at 
least one of the two measurements and for which no serious post-processing photometric warnings are found 
($k\_1ppErrBits < 256$, $k\_2ppErrBits < 256$). One effect of the latter cut was to exclude stars that were 
flagged as close to saturation at either epoch. Further quality constraints were to require a stellar image profile 
({\it mergedClass} = -1), low ellipticity ($k\_1Ell<0.3$, $k_2Ell<0.3$) and coordinates that agreed within 0.5\arcsec 
at the two epochs. These last three cuts were designed to minimise the number of false positives caused by blended stars. 
Finally, we rejected any candidates located in bad fields, identified by having more than 10 candidates in
one $13 \times 13$\arcmin~array.

\subsection{Stage 2: initial search of FITS catalogues}

The early searches were very inefficient due to bad pixels and severe source confusion in the more crowded
parts of the Galactic plane. As the survey approached completion, we realised that searches for both variable stars and 
high proper motion stars would be easier in less crowded parts of the Galactic plane. A dataset covering 
900~deg$^2$ of two-epoch sky at longitudes $60<l<230^{\circ}$ was compiled at Hertfordshire for the purpose of finding 
high proper motion stars, see \citet{smith14b}. That dataset included all data taken up 31 March 2013. The number of 
false positive high proper motion stars was shown to be far lower at $l>60^{\circ}$ (due to fewer blends) than at smaller
longitudes so we decided to search that the dataset for high amplitude variables. 

This search was based on the publicly available FITS catalogues for each UGPS stacked ``multiframe'' image 
in the K filter alone (multiframes consist of four images for the four array detectors in WFCAM. Source 
magnitudes were extracted from the FITS catalogues with the {\sc fitsio\textunderscore cat\textunderscore list.f} 
program that forms part of the public software release of the Cambridge Astronomical Survey Unit (CASU)
(http://casu.ast.cam.ac.uk/surveys-projects/software-release).
This search differed slightly from the earlier searches of the SQL database. 

\begin{itemize}
\item First, sources with potentially bad pixels in the photometric aperture were excluded from the search, since 
the {\sc fitsio\_cat\_list.f} program sets the image profile-based $class$ parameter to -7 in these cases. This flag is not 
set in the data releases at the WSA, which is why bad pixels were a serious issue in our earlier searches.

\item Second, the stellar profile cut used $class=-1$ at each epoch rather than $mergedclass=-1$, i.e. it was based 
on the profile classification in the two $K$ images rather than the merged classification from all available passbands and 
epochs.

\item Third, the FITS catalogues do not include the $ppErrBits$ flags so out initial selection included stars that would 
have been flagged as close to saturation (or even in some cases clearly saturated, since visual inspection
shows that not all saturated stars have the profile $class=-9$ that is applied when saturation is detected by the
CASU data reduction pipeline.) 

\item Fourth, not all the FITS catalogues had been through the various quality control procedures implemented for
public SQL data releases. These quality issues would be expected to further increase the number of false positive 
candidates arising from various causes including poor weather or gross electronic defects. We did however gain 
greater completeness because high amplitude variables can be readily detected even in poorer quality data. 
\end{itemize}

Following visual inspection, this search of a $\sim$900~deg$^2$ area at $l>60^{\circ}$ yielded 297 genuine
variables from only 501 candidates, including 32 discovered in our earlier searches. This 59\% success rate 
was due to the rarity of blends in less crowded parts of the Galactic plane and the far smaller number of
sources with bad pixels. The bona fide variables include some sources that are flagged as close to saturation
in the DR10 $gpsSource$ table with the $k\_1ppErrBits$ or $k\_2ppErrBits$ parameters. 
However, comparison with 2MASS $K_s$ data across a large area shows that effects of saturation on UGPS 
photometry are typically small ($<0.05$~mag) for stars with 2MASS magnitudes $11<K_{s,2MASS}<12$ that
are not blended in 2MASS. The photometry can be useful as bright as $K=10$, depending on seeing, sky 
background and source location on the WFCAM arrays. In order to ensure that at least one of the two 
epochs was always well measured, we excluded all candidates that were obviously saturated
at both epochs. This done by simple visual inspection but no source in our catalogue of variables has $K<11$ at 
both epochs. A close-to-saturation warning flag is set in the catalogue for all measurements where the
issue may be significant. This flag was set either if the relevant measurement had bit 16 of the $ppErrBits$ flag set in 
the WSA (see wsa.roe.ac.uk/ppErrBits.html) or, if the data were not in DR10, the flag was set for 
measurements brighter than $K=11.5$, $H=13$, $J$=13. We note that bona fide variables were always very 
easy to identify by visual inspection of the FITS images even if one epoch was saturated.

\subsection{Stage 3: final searches of FITS catalogues}

Final searches were undertaken when the complete UGPS dataset became available, again using the 
publically available FITS catalogues since data taken after 31 Jan 2012 are not available in the most 
recent SQL data release (DR10) at the WSA. An additional 375 deg$^2$ of two epoch sky at 
$60<l<230^{\circ}$ were available, based on a 2nd epoch obtained between 1 April 2016 and the end of 
UKIDSS operations on 31 December 2016. We further expanded the search of the $60<l<230^{\circ}$ area by 
exploring the 3rd epoch of data that were taken for a small proportion of UGPS fields due to doubts about 
the data quality at one of the two earlier epochs. The database created for the proper motion search had 
used the first and last image in the sequence to maximise time baseline; we now used the 2nd and 3rd 
image in the sequence, provided the 2nd image was separated from the other two by at least 1.5~yrs.
We chose not to use 1st and 3rd image for additional searches, owing to the low ratio of genuine to false 
candidates found with the 2nd and 3rd images. 
In addition, we modified the {\sc fitsio\textunderscore cat\textunderscore list.f} program to stop it from overwriting 
the stellar profile classification when potentially bad pixels were included in the photometric aperture. By recording 
the information separately, we were able to search for highly variable sources classified as stellar 
($k{_1,2}class=-1$) even if potentially bad pixels were present. We found that candidate variables with the bad pixel 
flag were mostly located at the edges of the field and were not real variables, but the minority of 
candidates further from the edges were often genuine. A threshold 
distance of 83 pixels from the array edges in X and Y was applied. 

These searches yielded only 86 real variables from 350 new candidates, most of them arising from the new area and only 
a small proportion arising from the 3rd epoch or sources flagged as having a bad pixel. We note that a single bad pixel 
appeared to have a very small effect on the photometry (smaller than the recorded uncertainty) for the relatively
bright variable stars in our catalogue, based on comparison of source magnitudes in different photometric 
apertures. In some cases no identifiable bad pixel was visible in the images, presumably because the effect was small or 
intermittent. Consequently we flag the nine affected sources in our catalogue (see Appendix \ref{B}) but we have not attempted 
to correct the photometry. 

Lastly, we expanded the search to the two-epoch sky at $30<l<60^{\circ}$, finding 202 real 
variables from 595 candidates. 
This very reasonable ratio of real variables to candidates shows that high stellar density has much less effect on two-epoch 
variability searches than on our past two-epoch high proper motion searches. The ratio becomes much worse at 
$l=<30^{\circ}$, see section 2, so that region was not searched due to lack of available time (though some of this area
was  included in the initial SQL searches). For similar reasons we did not attempt to mine any 3rd epoch of data in the 
$30<l<60^{\circ}$ region, though the few candidates with the bad pixel flag were included.

\section{Catalogue description}
\label{B}

Columns 1 and 2 give a running number and the UGPS designation; column 3 gives any other name by which a source is already known.
Equatorial and Galactic coordinates are given in columns 4 to 7. Columns 8 to 13 give the contemporaneous
UGPS $J$, $H$ and $K$ magnitudes, the latter  denoted "$K_{c}$", and their the associated errors.
The other UGPS $K$ magnitude is denoted "$K_{o}$", given in columns 14, with the error in column 15. 
We note that the $J$, $H$ and $K_{c}$ fluxes given here
were taken from the $gpsSource$ or $gpsDetection$ tables in DR10 at the WSA (carefully avoiding the occasional
non-contemporaneous entries in $gpsSource$) rather than the FITS catalogues, in order to benefit from small ($\le 0.03$~mag) 
improvements to the calibration of the four WFCAM arrays that are not available in the FITS catalogues. The K$_{o}$ photometry 
were left unchanged because many of these data are not available in the DR10 $gpsSource$ table. We impose a floor of 0.02~mag
on the errors, representing the typical calibration uncertainty for well-measured sources.
Column 16 contains a flag ("$a$" or "$b$") indicating whether the "$K_{c}$" corresponds to the first UGPS epoch ($a$) or the second ($b$)
\footnote{Our alphabetical notation of epochs $a$ and $b$ is used to distinguish our catalogue from data in the $gpsSource$ table in the UKIDSS public data 
releases at the WSA, which typically uses $K_{1}$ and $K_{2}$ in the way that we use $K_{c}$ and $K_{o}$ respectively. 
Our selection of measurements occasionally differs from those used for the primary detection of each source in DR10.}. In three cases the flag is "c", indicating that the
contemporaneous $JHK_{c}$ photometry were taken at a third epoch when the $K_{c}$ magnitude was between the other two $K$ magnitudes.  
In these cases $K_{o}$ is left blank and $K$ data from all three epochs are given in Table B1.

Columns 17 and 18 give the modified Julian date of epochs $a$ and $b$.  
Column 19 indicates whether the source had a "possible-saturation" warning in each of $J$, $H$, $K_{c}$ and $K_{o}$ in sequence (see Appendix A). E.g. "0010" would 
indicate that the $K_{c}$ measurement might be influenced by saturation but $J$, $H$ and $K_{o}$ were not. Column 20 flags any
cases where bad pixels may be present that might possibly reduce the precision of photometry in $K_{c}$ or $K_{o}$ by listing the affected measurement (see Appendix \ref{A}).
Column 21 gives the amplitude of variation, $\Delta K$,  based on the two UGPS fluxes in Table 1 (or the maximum and minimum fluxes for the three sources in Table 2). 
$\Delta K$ is marginally below unity in six sources despite our initial selection requirement, owing to the small calibration differences between data in the FITS catalogues and the $gpsSource$ table noted above.
Column 22 gives the amplitude $\Delta K_{all}=K_{max}-K_{min}$ after taking account of a 3rd epoch in $K_{s}$ from the 2MASS Point Source Catalogue \citep{skrutskie06}, where available. We note that measurements in the UKIRT MKO $K$ filter and 2MASS K$_{s}$ filter typically agree within a few hundredths of a magnitude for most non-variable stars. Transformation between systems is generally not possible for individual stars of unknown spectral type, on Galactic plane sightlines with unknown extinction.

The later columns in Table 1 contain information based mainly on other public datasets and searches of the literature via the SIMBAD database, see section 3.
Column 23 indicates the likely type of astronomical object, if known. Column 24 gives distances to likely YSOs derived from the associated SFR, based on references in the literature that
are indicated in column 25. In column 26 we identify the SFR or cluster associated with each YSO, provided there is significant literature mentioned in SIMBAD. If no cluster or SFR is listed, we name any millimetre source, submillimetre source, or infrared dark cloud (IRDC) that is coincident with the source. Information from the WISE colour images

\begin{table*}
	\centering
	\caption{$K$ fluxes for Table 1 entries where $JHK$ photometry were taken at a 3rd epoch.}
	\label{tab:proportions}
	\begin{tabular}{lcccccc}
		\hline
		No. &  $K_{a}$ & $K_{b}$ & $K_{c}$   &   MJDobs a    &   MJDobs b    &  MJDobs c\\ \hline
		243	&  11.98 & 13.36  &  12.96 & 53638.24097 & 55776.41255 & 53503.58199 \\
		244	&  15.78 & 17.22 &  16.24  & 53638.24097 & 55776.41255 & 53503.58199 \\
		572	&   16.11 & 14.61 &  15.42  & 53915.50097 & 56208.25244 & 55463.36109 \\ \hline
	\end{tabular}
\end{table*}

\noindent is included if a substantial uncatalogued cluster is present.
References to the named region and/or the source itself are given in column 27. 
These searches also reveal whether the source was previously known as a variable star or other source of interest, named in column 3.
In column 28 we indicate whether the source is associated with any of the nine well-studied spatial groups of four or more pre-MS variables listed in
section 4.2, provided it is classified as a likely YSO in column 23.\\
\newline

In column 29 we give the 2--22~$\mu$m spectral index, $\alpha=d \left(log \left( \lambda F_{\lambda} \right) \right)$/$d\left( log \left( \lambda \right) \right)$. In almost all cases it is based on the 
average of the two UGPS K fluxes and the four contemporaneous W1, W2, W3 and W4 fluxes from the WISE All-Sky data release which was based on data obtained in 2010. 
We found that using either the brighter or fainter of the two UGPS $K$ fluxes instead of the average typically changed $\alpha$ by 0.2, which indicates the uncertainty caused by this non-contemporaneous passband. 
The spectral index is given only for the 57 sources classified as YSOs that were measured in W4 and a further 8 sources for which 24~$\mu$m photometry is available from MIPSGAL (\citealt{rieke04}; \citealt{carey09}; \citealt{gutermuth15}) but not in WISE. In these 8 cases we combined MIPSGAL photometry with {\it Spitzer}/IRAC photometry \citep{fazio04}, rather than WISE. Column 30 gives the discrete SED index classification for the YSOs as class I, flat spectrum or II, based on the spectral index.

Columns 31 to 33 give contemporaneous optical photometry in the $r$, $i$ and narrowband H$\alpha$ filters taken either from the IPHAS 2nd data release (\citealt{drew05}; \citealt{barentsen14}), for sources at Declination $\delta > +1.3^{\circ}$, or from the VPHAS$+$ dataset \citep{drew14} for more southerly sources. Column 34 gives the modified Julian date of the IPHAS or VPHAS$+$ observations, as appropriate. 2MASS photometry in the $J$, $H$ and $K_{s}$ passbands is given in columns 35 to 37 and the WISE W1 to W4 photometry is given in columns 38 to 41. MIPSGAL 24~$\mu$m data are given in column 42, and IRAC photometry are given in columns 43 to 46 for bands I1 to I4. The Spitzer/IRAC data are from the GLIMPSE, GLIMPSE3D, GLIMPSE360, Deep GLIMPSE, Cygnus X and SMOG public surveys (\citealt{benjamin03}; 
\citealt{churchwell09}) which were not contemporaneous with MIPSGAL or UGPS.

Inspection of the WISE images indicated that some sources with a W3 or W4 flux in the All-Sky catalogue were actually not detected or poorly measured in that passband, due to the effect of bright nebulosity on the background estimation and the use in WISE catalogues of list-driven photometry applied to all passbands for detections in any passband. \citet{koenig14} investigated this issue in detail for SFRs in the outer Galaxy for the AllWISE dataset, which has similar characteristics to the All-Sky dataset for this issue. They found that in every passband the genuine and false WISE detections have different but overlapping distributions in plots of the reduced $\chi^2$ statistic for a point source detection vs. the signal to noise ratio. We used quality cuts on the W3 and W4 photometry similar to those proposed in that work to identify sources for visual inspection in all four WISE passbands: we checked $\sim$90 sources located above the diagonal lines in their figure 1 (the $w4rchi2$ vs. $w4snr$ and $w3rchi2$ vs. $w3snr$ plots) that also had $w4rchi2 > 2$ or $w3rchi2 > 2$, as appropriate. WISE fluxes in one or more passbands were removed for slightly under half of these. Our adopted reduced $\chi^2$ thresholds are slightly more relaxed then those adopted for a reliable selection in \citet{koenig14} since we saw a slightly broader distribution of bona fide sources in the reduced $\chi^2$ vs. signal to noise ratio plane (perhaps due to increased source confusion in the inner Galaxy) and we aimed more for completeness than reliability. Only one source had photometry removed in the W1 or W2 passbands, due to blending rather than poorly measured background. We did not go so far as to visually check all WISE detections in all passbands, nor all optical, {\it Spitzer} and UGPS photometry in all pass-bands (only the K measurements, the few available MIPSGAL 24~$\mu$m measurements and $J$, $H$ in addition for sources with unusual near infrared colours). A small amount of erroneous photometry may therefore remain in the catalogue. We retained the small number of obviously blended WISE detections if the WISE active deblending appeared to be successful, checking the higher resolution Spitzer/IRAC images where possible for confirmation.

Matches to 2MASS and IRAC data used a $1\arcsec$ cross-match radius whereas matches to the lower resolution WISE and MIPSGAL datasets used a $2\arcsec$ cross-match radius. All matches
with separations over $0.4\arcsec$ were then matched back against the UGPS DR10 database in order to identify, inspect and remove any mis-matches to another UGPS source. One such mis-match occurred for source 273, for which the 2MASS counterpart with coordinates $0.77\arcsec$ north of the UGPS variable appears to be a blend of source 273 and an adjacent UGPS source 1.06 \arcsec to the north. All other mis-matches were to sources $>1\arcsec$ from the UGPS coordinates.
Three very bright WISE sources (sources 241, 273 and 301) required a larger cross-match radius of $3\arcsec$ due to slightly incorrect coordinates in the WISE All-Sky source catalogue (an issue that was fixed in the more recent AllWISE source catalogue, created from non-contemporaneous data). 

In column 47 we give the 70~$\mu$m flux in mJy from the PACS instrument \citep{pog08} on the {\it Herschel} satellite. These data were obtained by cross-matches to (i) the Hi-GAL first data re- lease (\citealt{molinari10}, \citealt{molinari16}) covering $-70 < l < 68^{\circ}$, $|b| < 1^{\circ}$ and (ii) the {\it Herschel}/PACS Point Source Catalogue \citep{marton17} which includes all PACS data from the satellite (though there is limited coverage outside the mid-plane region). Most matches were from the Hi- GAL dataset and we used the Hi-GAL data release as the first choice, though the two catalogues agree well for sources in common. Hi-GAL data were taken in 2010-2011 (typically about 1 year after the WISE All-Sky data). Initially we found 64 matches, using a 6$\arcsec$ matching radius (similar to the beam size). We found that matches with no corresponding WISE W4 detection almost always had separations over 3$\arcsec$ from the UGPS source while those with W4 detections had separations under about 3$\arcsec$. We therefore retained only the 52 matches with a W4 detection, leading to an effective cross-match radius of 3.1$\arcsec$. Most of these have a signal to noise ratio over 5 but we have included the small number of low significance detections. The conversion to a Vega magnitude system in figure \ref{fig:pacs} of the main text was done with the zero point given by \citep{nielbock13}.


\bsp	
\label{lastpage}
\end{document}